\documentclass[fleqn,usenatbib]{mnras}

\usepackage{amsmath}
\usepackage[T1]{fontenc}

\DeclareRobustCommand{\VAN}[3]{#2}
\let\VANthebibliography\thebibliography
\def\thebibliography{\DeclareRobustCommand{\VAN}[3]{##3}\VANthebibliography}

\usepackage{graphicx}	
\usepackage{amsmath}	
\usepackage{amssymb}	
\usepackage{comment}
\usepackage{csquotes}
\usepackage{xspace}
\usepackage{supertabular}
\usepackage{longtable}
\usepackage{graphicx}
\usepackage{pdflscape}
\usepackage{color}
\PassOptionsToPackage{normalem}{ulem}
\usepackage{ulem}
\usepackage{hyperref}

\usepackage{txfonts}
\def\br{B$_\text{R}$}
\def\bp{B$_\text{Plk}$}

\def\degr{$^{\circ}$}
\newcommand{\dgr}{$^{\circ}$}

\newcommand{\nthp}{N$_{2}$H$^{+}$}

\def\arcmin{^\prime}

\title[Core orientations and magnetic fields in isolated molecular clouds]{Core orientations and magnetic fields in isolated molecular clouds}

\author[E. Sharma et al.]{
Ekta Sharma,$^{1,2,3}$\thanks{E-mail: ektasharma.astro@gmail.com }
Maheswar Gopinathan,$^{1}$
Archana Soam,$^{1,4}$
Chang Won Lee$^{5,6}$
and T. R. Seshadri$^{7}$
\\
 $^{1}$Indian Institute of Astrophysics (IIA), Sarjapur Road, Kormangala, Bangalore 560034, India\\
    $^{2}$Physical Research Laboratory, Navrangpura, Ahmedabad 380009, India\\
    $^{3}$National Astronomical Observatories, Chinese Academy of Sciences, Beijing, China\\
    $^{4}$SOFIA Science Centre, USRA, NASA Ames Research Centre, MS-12, N232, Moffett Field, CA 94035, USA\\
    $^{5}$Korea Astronomy \& Space Science Institute, 776 Daedeokdae-ro, Yuseong-gu, Daejeon, Republic of Korea\\
    $^{6}$ University of Science and Technology, Korea (UST), 217 Gajeong-ro, Yuseong-gu, Daejeon 34113, Republic of Korea\\
    $^{7}$Department of Physics and Astrophysics, University of Delhi, Delhi 110007, India
    }

\date{Accepted XXX. Received YYY; in original form ZZZ}

\pubyear{2015}

\begin{document}
\label{firstpage}
\pagerange{\pageref{firstpage}--\pageref{lastpage}}
\maketitle

\begin{abstract}
{Molecular clouds are sites of star formation. Magnetic fields are believed to play an important role in their dynamics and shaping morphology. We aim to study any possible correlation that might exist between the magnetic fields orientation inside the clouds and the magnetic fields at envelope scales and their connection with respect to the observed morphology of the selected clouds. We examine the magnetic field orientation towards the clouds L1512, L1523, L1333, L1521E, L1544, L1517, L1780 and L183 using optical and \textit{Planck} polarization observations. We also found the correlation between the ambient magnetic field and core orientations derived using \textit{Astrodendrogram} on the \textit{Herschel} 250 $\mu$m data. We find that the magnetic fields derived from optical and \textit{Planck} agree with each other. The derived magnetic fields are aligned along the observed emission of each cloud as seen in \textit{Herschel} 250 $\mu$m data. We also find that the relative orientation between the cores and the magnetic fields is random. This lack of correlation may arise due to the fact that the core orientation could also be influenced by the different magnetization within individual clouds at higher densities or the feedback effects which may vary from cloud to cloud. The estimated magnetic field strength and the mass-to-flux ratio suggest that all the clouds are in a magnetically critical state except L1333, L1521E and L183 where the cloud envelope could be strongly supported by the magnetic field lines.}
\end{abstract}



\section{Introduction}
In molecular clouds, the filamentary structures are ubiquitous and some of them are believed to be created by compression of material due to multi-scale supersonic flows \citep{2014A&A...569L...1A}. Localised density enhancements will shrink under self-gravity to form prestellar cores where stars form through collapsing gas motions \citep{1987ARA&A..25...23S}. These prestellar cores are the preferential and ideal sites of star formation in nearby dark clouds like the Taurus and the Perseus \citep{1994MNRAS.268..276W, 2005A&A...440..151H}. Previous studies based on the optical inspection of the Palomar sky plate and molecular line observations in NH$_{3}$ and N$_{2}$H$^{+}$ showed that the dense cores are typically of size $\sim$ 0.1 pc and contain a few solar mass of subsonic material at temperatures of $\sim$ 10 K and with an average volume density $\sim$ 10$^{4}$ cm$^{-3}$ \citep{1983RMxAA...7..238M, 1983ApJ...270..589B}. The correlation of dense core positions with the locations of highly embedded young stellar objects (YSOs) led to an idea that some cores are forming stars currently or some of them have initiated such formation very recently. On the basis of association of YSO or IRAS point source, the cores are broadly divided into two categories: prestellar and protostellar. The morphology of their emission can be observed in dust continum and molecular line observations. The results from the \textit{Herschel} Gould Belt survey suggest that the cores are mostly being embedded in the filaments \citep{2010A&A...518L.102A, 2010A&A...520A..17K, 2014A&A...569L...1A}. The association is consistent with the theoretical prediction that the thermally supercritical filaments experience a longitudinal fragmentation into cores \citep{1992ApJ...388..392I}.

The significance of magnetic fields in the star formation process has been a subject of debate \citep[e.g., ][]{2012ARA&A..50...29C, 2014prpl.conf..101L}. It is widely believed that the magnetic fields along with the other physical factors like turbulence and gravity are the key factors affecting the star formation process at different scales and different evolutionary stages \citep{2007prpl.conf...63B,2007ARA&A..45..565M}. The magnetic fields can regulate the gravitational contraction of the cloud at higher densities (n $\geq$ 10$^{3}$ cm$^{-3}$) thus providing evidence of the field lines being dynamically important \citep{2011MNRAS.411.2067L}. Several studies have been carried out to understand the dynamical importance of magnetic fields \citep{2008A&A...486L..13A,2010ApJ...716..299S,2011ApJ...742L...9C,2013ApJ...762..120P,2016A&A...588A..45N,2018MNRAS.476.4782S,2020A&A...639A.133S}. 

The cloud-scale magnetic field at the lower densities can affect the compression in turbulence-induced shocks and channel the flows towards the high density regions \citep{2008ApJ...680..420H,2015ApJ...810..126C}. The core-scale magnetic field regulates the gas dynamics within the cores by removing angular momentum in collapsing cores and flattening the cores through the mechanism of ambipolar diffusion. The correlation between the magnetic fields in the dense cores and the core orientation has been studied in various starless and star-forming regions \citep{2000ApJ...537L.135W,2006MNRAS.369.1445K}. The magnetic fields at the intermediate scales connecting the large scale magnetic field to the field within cores can have a significant effect on their dynamics and the structure. Using the \textit{Planck} polarization observations, \cite{2017NatAs...1E.158L} suggested a bimodal distribution in the alignment between  the elongation of cloud (direction of major axes) and the direction of magnetic field using the \textit{Planck} observations  indicating that the alignment is more likely either parallel or perpendicular. Their study found the magnetic field as a primary regulator of the star formation rate. When the overall cloud evolves as a consequence of internal dynamics and stellar feedback effects, the clumps will also grow by accreting material turning into a non-spherical structures \citep{2014ApJ...785...69C,2018ApJ...865...34C,2011ApJ...728..123K}. Hence, clump shapes and orientation might be an indicator of formation mechanism. As a result of the interaction between magnetic field and core material through well defined orientation, the collapse may get inhibited and will result in reducing the star formation rate \citep{1995ApJ...453..271B,2011ApJ...740...36N,2017A&A...601A..18B}. Therefore, it is crucial to explore the effect of large-scale magnetic field orientation on the morphology of dense core structures.

Observational studies were conducted by \cite{2014ApJ...791...43P} on the cores extracted from 350 $\mu$m \textit{Herschel} observations to investigate the core orientations in the Lupus I cloud with respect to the large-scale magnetic field. The environment plays an important role in shaping the prestellar cores through ram pressure and magnetic pressure \citep{2018ApJ...865...34C}. Cores are usually either oblate or prolate shaped structures as suggested by MHD simulations \citep{2018ApJ...865...34C,2009MNRAS.398.1082B}. There are anisotropic gas flows along the magnetic field lines during the formation and evolution of cores which lead to the preferential orientation of the cores along the background magnetic field lines \citep{2014ApJ...785...69C,2015ApJ...810..126C}. \cite{2020MNRAS.494.1971C}, by combining simulations and observational results from the archive, studied a sample of cores in three clouds and investigated whether there is any systematic relation between the core orientation and the cloud-scale magnetic field. They found that the correlation has a regional dependence on the individual magnetic properties of the clouds.  
Optical polarization of starlight is used as a technique to trace the plane-of-sky magnetic field geometry which typically works in regions of low extinction (A$_{V}$ $\sim$1-3) \citep{1990ApJ...359..363G}. Thus, optical polarimetry is mainly applicable towards the low density parts of the clouds. This is because the non-spherical dust grains which are believed to be aligned with their major axes perpendicular to the magnetic field cause selective extinction making the starlight polarized. At higher extinctions, the background starlight gets extinguished completely making it impossible to measure any polarization \citep[e.g., ][]{1949Natur.163..283H, 1976AJ.....81..958V, 1990ApJ...359..363G, 2008A&A...486L..13A, 2013MNRAS.432.1502S}. The polarized thermal dust emission is used as a technique to trace the plane-of-sky magnetic field geometry of regions with high obscurations \citep{2009MNRAS.398..394W}. This is because the non-spherical dust grains that are believed to be aligned with their major axes perpendicular to the magnetic field emit the radiation which is polarized in a direction parallel to the major axis of the dust grains. The all-sky maps of polarized emission from dust using the \textit{Planck} satellite at 353 GHz gives a useful insight into the Galactic magnetic field in the star-forming regions \citep{2016A&A...586A.138P}.

\begin{figure*}
\centering
\includegraphics[height=4.4cm, width=4.5cm]{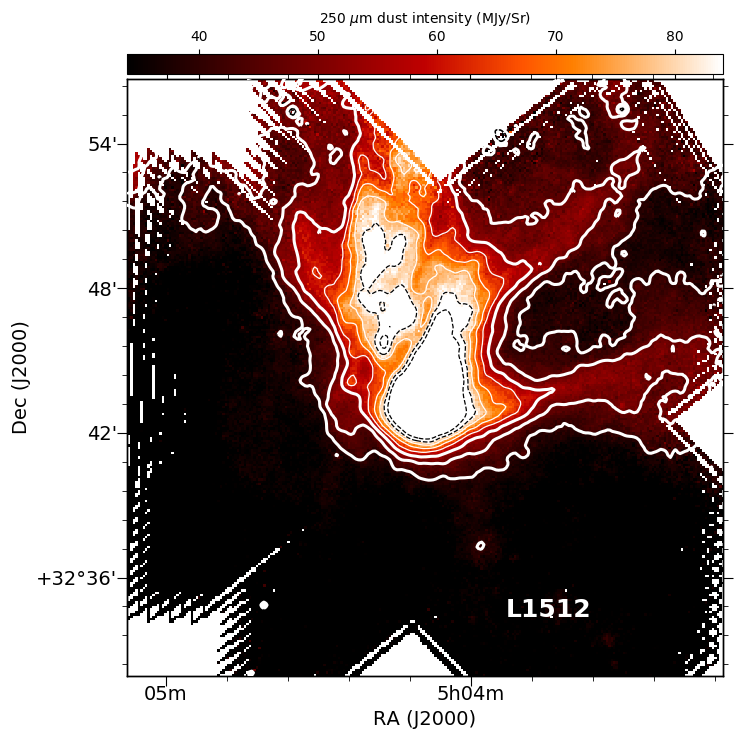}
\includegraphics[height=4.4cm, width=4.4cm]{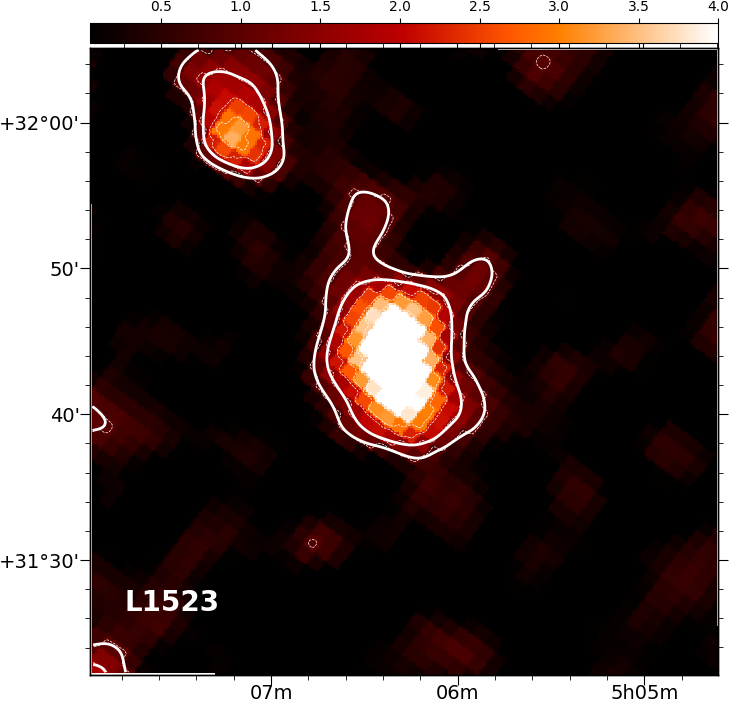}
\includegraphics[height=4.4cm, width=4.4cm]{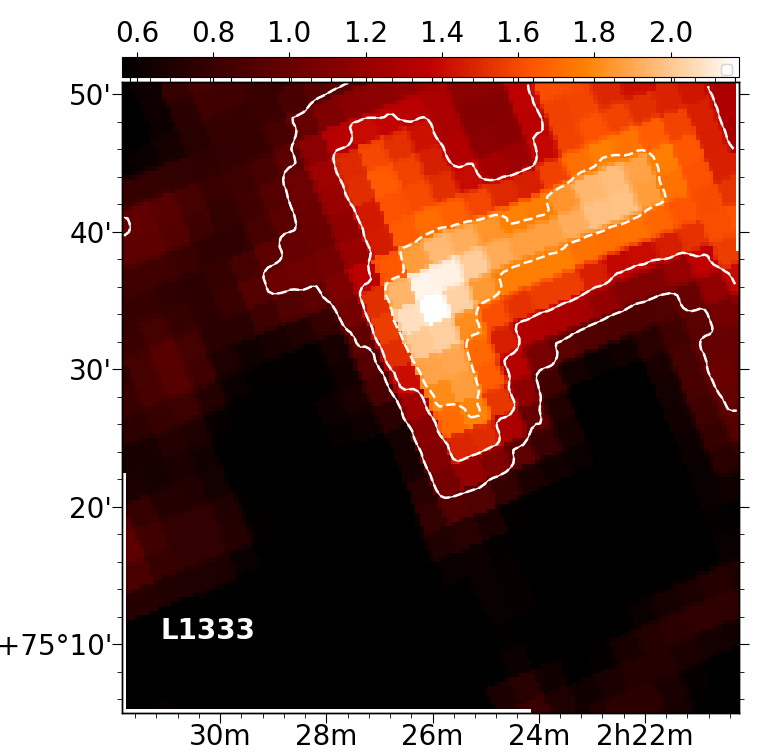}
\includegraphics[height=4.4cm, width=4.4cm]{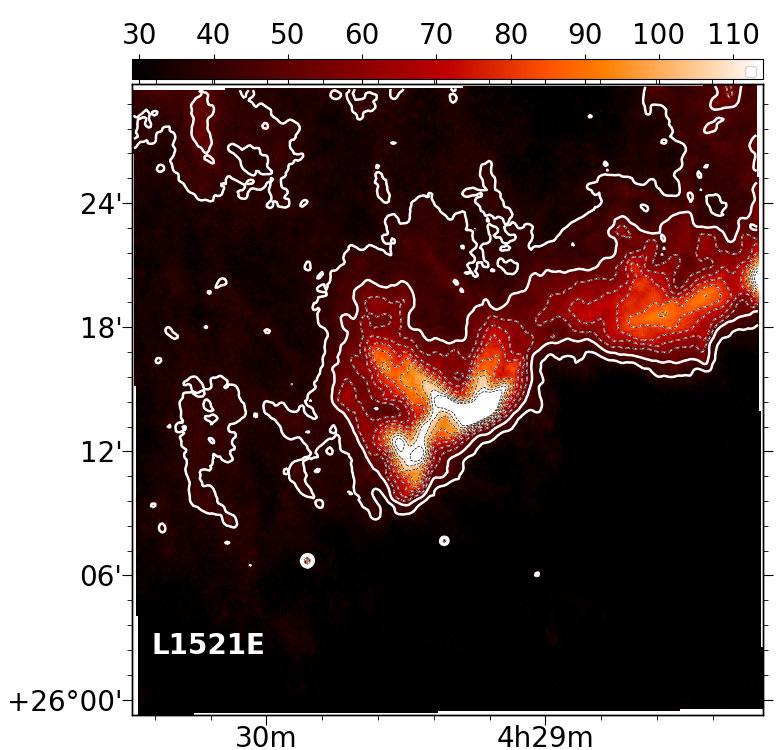}
\includegraphics[height=4.4cm, width=4.4cm]{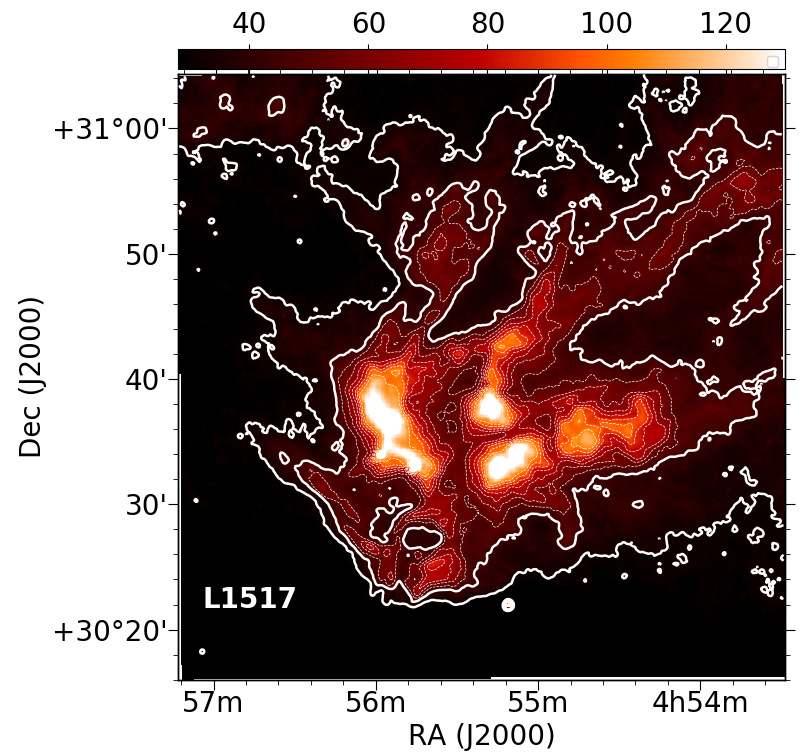}
\includegraphics[height=4.4cm, width=4.4cm]{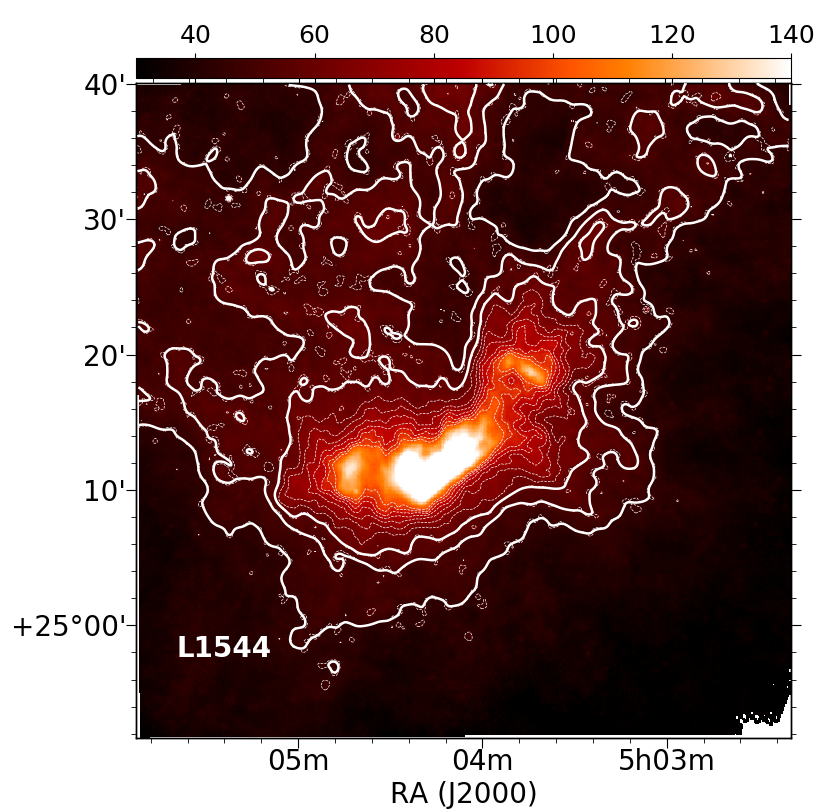}
\includegraphics[height=4.4cm, width=4.4cm]{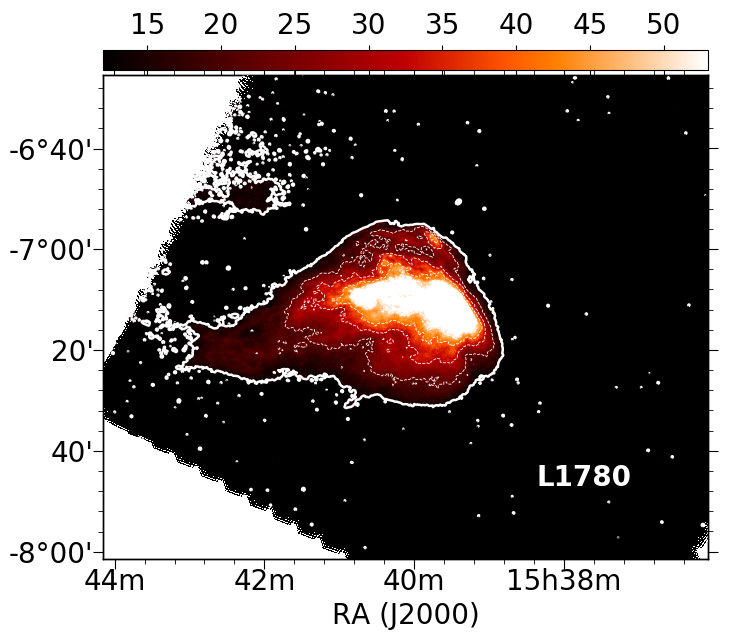}
\includegraphics[height=4.4cm, width=4.4cm]{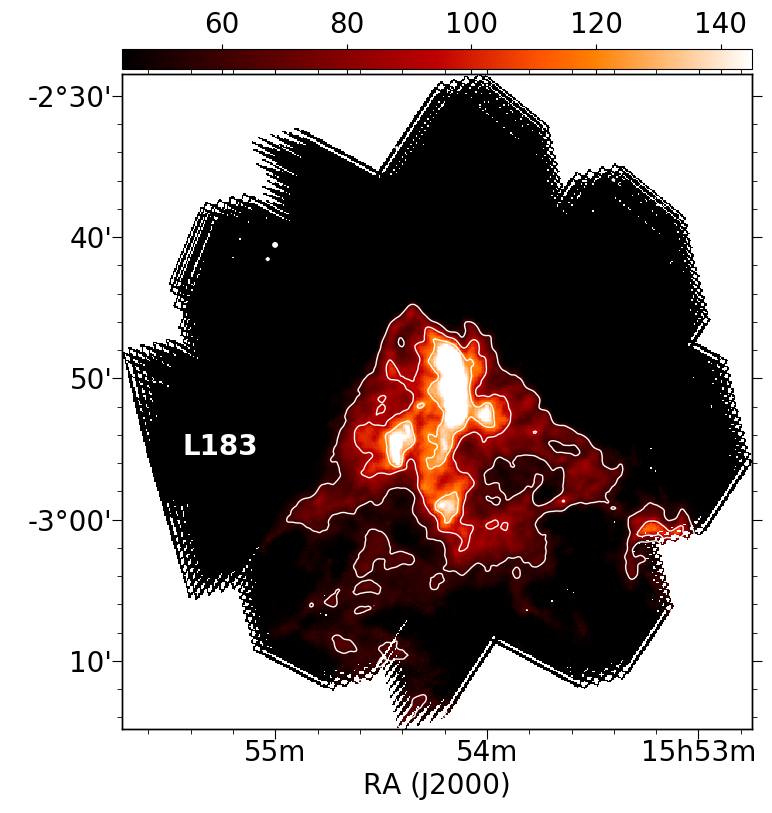}
\caption{{\bf Left:} \normalsize{The \textit{Herschel} 250 $\mu$m SPIRE image of size 0.5\degr$\times$0.5\degr for L1512, L1521E, L1544, L1780 and L183 cloud and 0.8$^{\circ}\times0.8^{\circ}$ for L1517 with intensity contours shown in white lines. The \textit{Dobashi} emission maps of size 0.5\degr$\times$0.5\degr for L1523 and 0.75\degr$\times$0.75\degr for L1333.}}\label{fig:l12_l13_el}
\end{figure*}

In this paper, we present results of a study conducted on eight clouds that are found to manifest their physical appearance in the form of sharp edges on one side and an extended diffuse emission of material on the opposite side of the clouds. The possible reasons behind this morphological emission could be the isotropic interstellar radiation field (ISRF) or the heating effect from nearby stars which may have interacted with the cloud material during the process of their evolution \citep{2012A&A...547A..11N,2009MNRAS.396.1851N,2013A&A...551A..98L,2016ApJ...824...85K}. In Fig. \ref{fig:l12_l13_el}, we show the fields contained the clouds using the \textit{Herschel} 250 $\mu$m SPIRE images of L1512, L1521E, L1517, L1544, L1780 and L183. The fields containing L1523 and L1333 are shown using the \textit{Dobashi} extinction maps since there is no \textit{Herschel} data available for them. The naming convention for the clouds given here are used to represent the whole cloud as shown in Fig. \ref{fig:l12_l13_el}. The white solid contours drawn on the images reveal the emission morphology of the clouds. Basic information of these eight clouds are given in columns 1-5 of Table \ref{tab:core_archive}. 

The questions that we are trying to investigate are the following: What is the relationship between the magnetic field orientations inside the clouds inferred from the \textit{Planck} polarization and the envelope-field inferred from the optical polarization measurements of background starlight? What is the relationship between the magnetic field orientation with respect to the elongated morphology of the whole cloud? How the core major axes are oriented with respect to both inside and outside magnetic field directions? This paper is organised in the following manner. We begin with the observations and the data reduction. Then, we present results from our R-band polarization measurements and describe the procedure adapted to extract the cores in each cloud. We discussed our main results for each cloud on magnetic field geometry, strength and conclude the results with a summary in last section.  
\begin{table}
\centering
\caption{Log of the observations.}\label{tab:core_log}
\begin{tabular}{lll} \hline
ID  & Cloud Id.  & Date of observations      \\\hline\hline
1   &  L1333    & 10 December 2015; 11,12 January\\
                & & 5,6 February 2016\\
2   &  L1521E   & 10, 16 January; 8 February 2013\\
3   &  L1517    & 28 November; 1, 28 December 2013 \\ 
4   &  L1512	& 26-30 October, 2017 \\ 
5   &  L1544    & 23-25 January, 2014 \\
6   &  L1523	& 28 October; 23, 24, 25 November 2016\\

 \hline
\end{tabular}
\end{table}


\section{Observations and data reduction}
The polarization measurements of stars projected on L1512, L1523, L1333, L1521E, L1517 and L1544 were carried out using ARIES Imaging Polarimeter \citep[AIMPOL;][]{2004BASI...32..159R} which is mounted on 1.04 m Sampurnanand Telescope (ST) at Nainital, India. The polarization measurements were done in the standard Johnson R-band filter. The details on the polarization measurements of clouds L1780 and L183 are given in \cite{2016A&A...588A..45N}. Table \ref{tab:core_log} shows the log of the observations for all six clouds. The AIMPOL has a combination of an achromatic half-wave plate (HWP) modulator and a Wollaston prism in form of a beam splitter. It splits the incoming light into two rays - ordinary (I$_{o}$) and extra-ordinary (I$_{e}$). The observations were made at $\lambda_{eff}$ = 0.63 $\mu$m. The plate scale of the CCD is 1.48$^{''}$/pixel and the field of view is $\sim$8$^{'}$ in diameter which covers 325$\times$325 pixel$^{2}$ area on the CCD. The full width at half-maximum of the stellar image varies from 2.9$^{''}$ to 4.4$^{''}$ (2-3 pixels) due to seeing of the atmosphere. The read-out noise and gain of CCD are 7.0 e$^{-1}$ and 11.98 e$^{-1}$/ADU, respectively. 

The standard aperture photometry available in the IRAF package was used to extract the fluxes of ordinary (I$_{o}$) and extraordinary (I$_{e}$) images of all the observed sources with a good signal-to-noise ratio. The ratio R($\alpha$) is given by,
\begin{align}
 R(\alpha) = \frac{\frac{I_{e}(\alpha)}{I_{o}(\alpha)}-1}{\frac{I_{e}(\alpha)}{I_{o}(\alpha)}+1} = P\cos(2\theta - 4\alpha),
\end{align}where P ($\%$) is the fraction of the total linearly polarized light, $\theta$ is the polarization angle in the plane-of-sky and the $\alpha$ is the position angle of the fast axis of the half-wave plate with respect to the axis of the Wollaston prism. The P and $\theta$ are calculated using the combination of the normalized stokes parameters q1, u1 and q2, u2 at angles 0\degr, 22.5\degr, 45\degr, 67.5\degr, respectively.

Instrumental polarization was evaluated by observing zero-polarization standard stars during every observation run. The typical instrumental polarization is found to be of the order of $\sim$0.1\% \citep{2013MNRAS.432.1502S,2015A&A...573A..34S,2016A&A...588A..45N}. The reference direction of the polarizer was determined by observing polarized standard stars taken from \cite{1992AJ....104.1563S}. The zero point correction in the polarization angle was derived using the offset obtained between the position value obtained in our observations and the standard position angles given in \cite{1992AJ....104.1563S}.

\section{The data from archive}
\subsection{The \textit{Planck} polarization}
We used the 353 GHz (850 $\mu$m) all-sky map of the polarized emission from dust provided by $\textit{Planck}$ Legacy Archive\footnote{http://www.cosmos.esa.int/web/planck/pla/}. The 353 GHz channel in \textit{Planck} satellite is the highest-frequency polarization-sensitive channel. We constructed the geometry of the magnetic fields for the whole sample based on the \textit{Planck} polarization maps. The gnomonic projection of HEALPix software \citep{Zonca2019,2005ApJ...622..759G} has been used to construct the stokes Q, U and I maps using all-sky \textit{Planck} data. Then, the stokes maps were smoothed down to the 8$\arcmin$ resolution to obtain a good signal-to-noise (SNR) ratio. We estimated linearly polarized intensity which is the quadrature sum of Stokes Q and U:
\begin{align}
    P = \sqrt{Q^{2} + U^{2}},   
\end{align}
The dust emission is linearly polarized with the electric vector normal to the sky-projected magnetic field. Therefore, the polarization position angles were rotated by 90\degr~ to infer the projected magnetic field as follows:
 \begin{equation}
   B_{pos} = \frac{1}{2} \arctan({-U,Q})+\pi/2.    
 \end{equation}
The positions angles have been taken in accordance with HEALPix standards \citep{2015A&A...576A.104P} and converted to IAU convention by multiplying stokes U with -1 as shown in above equation. The zero position angle is oriented along the Galactic North and increases towards East of North on the sky.

\subsection{The GAIA EDR3 data}
We have used \textit{Gaia} early data release 3 (EDR3) \citep{2021A&A...649A...1G} for the distance measurements of the stars used in the polarimetric observations. The \textit{Gaia} EDR3 data contains precise and updated astrometric as well as photometric parameters of $\sim$ 2 billion stars. The survey lists the magnitude values at the broadband photometric bands of G, G$_{BP}$ and G$_{RP}$ in the wavelength range of 330-1050 nm \citep{2021A&A...649A...1G}. The distances to the stars studied in this work are obtained from the catalogue provided by \citet{2021AJ....161..147B}. The limiting magnitude of the sources for the distance calculation is $\sim21$ in G-band. The uncertainties in the obtained parallax values vary from 0.02-0.03 milliarcsecond (mas) to 0.07 mas for sources with G < 15 mag and G = 17 mag, respectively. The uncertainties in the parallax values are $\sim$ 0.5 mas for the fainter sources with G$\sim$20 mag  \citep{2021A&A...649A...2L}. 
\begin{figure*}
\centering
\includegraphics[height=11.5cm, width=8.5cm]{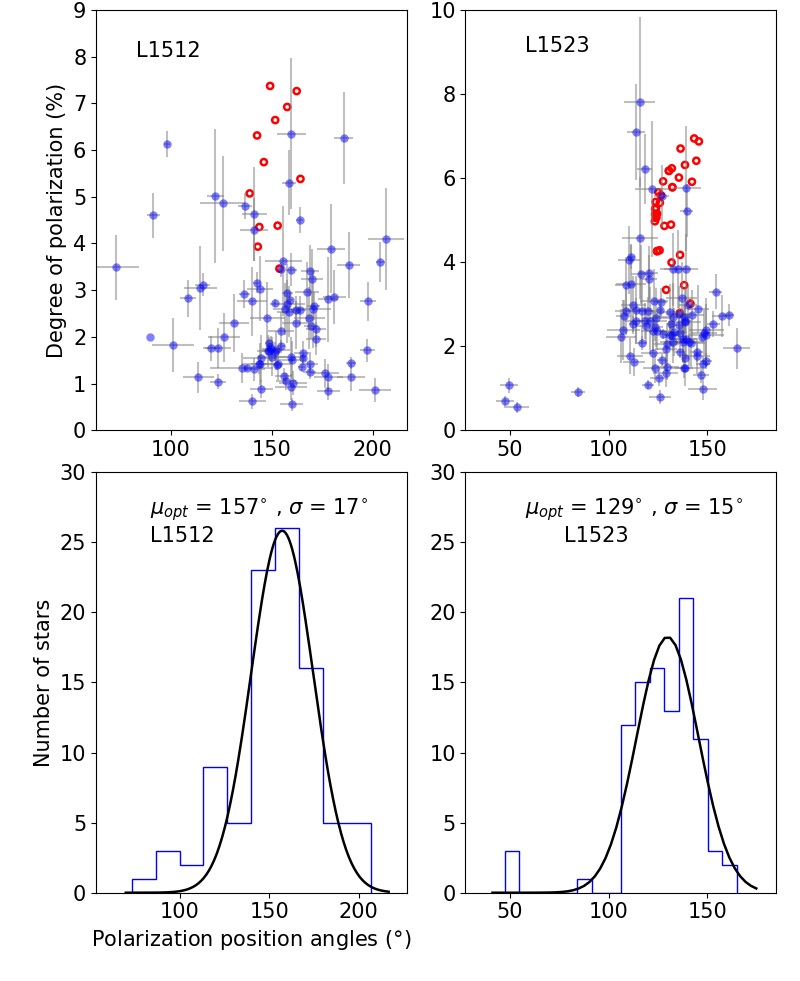}
\includegraphics[height=11.5cm, width=8.5cm]{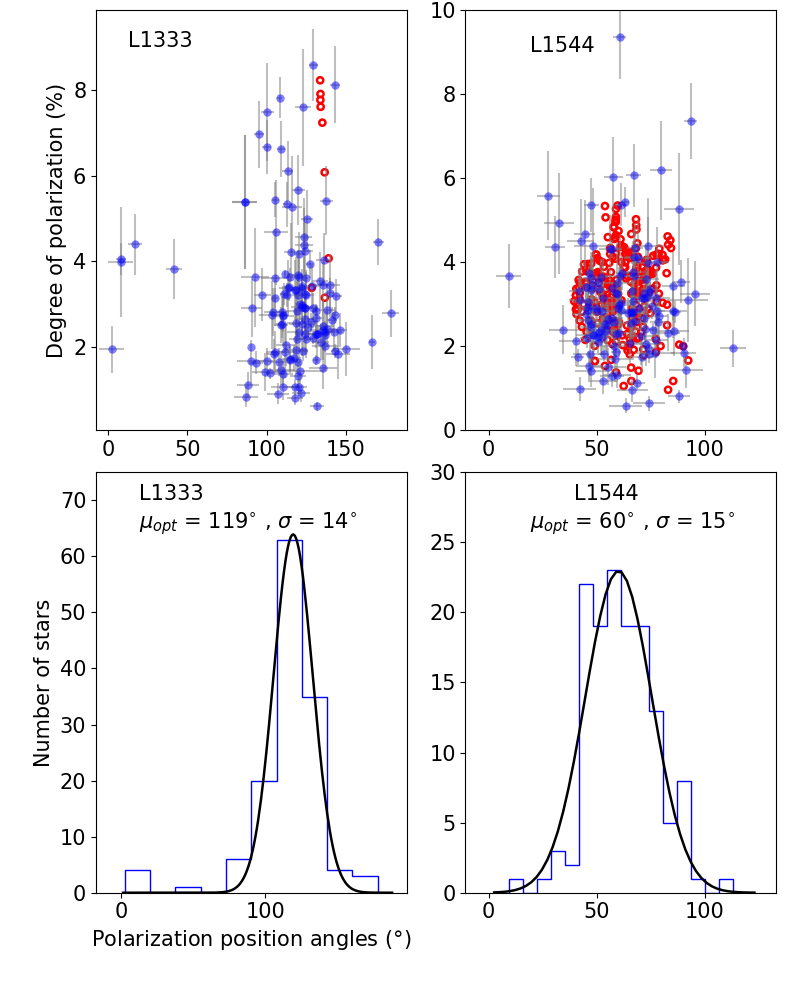}
\includegraphics[height=11.5cm, width=8.5cm]{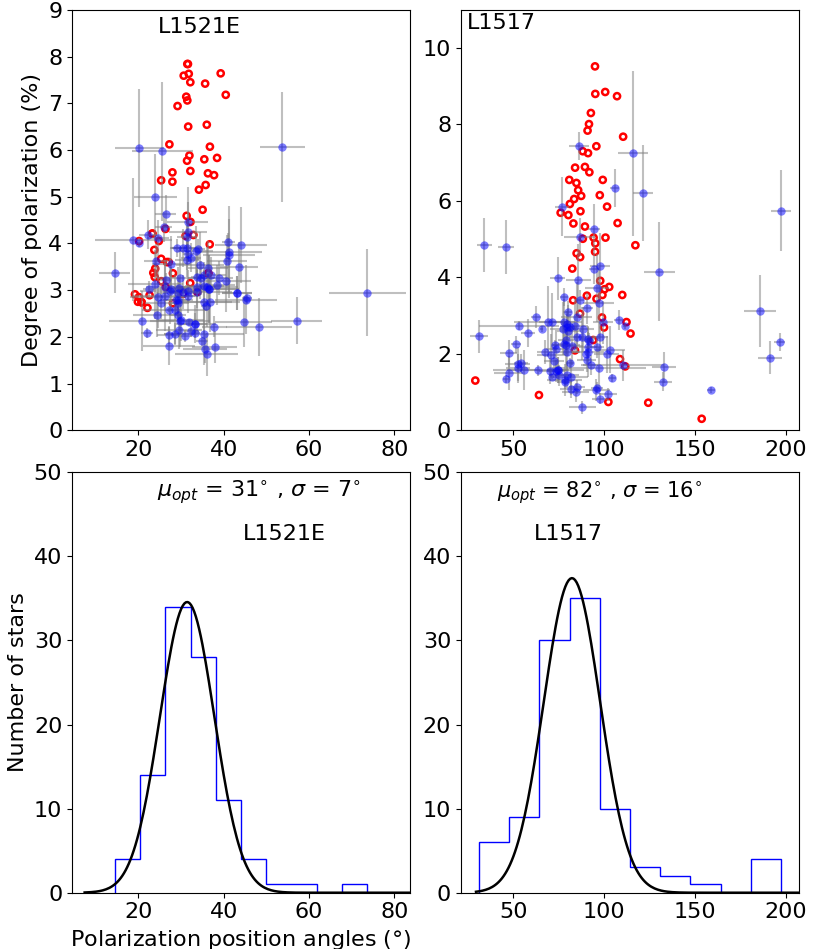}
\includegraphics[height=11.5cm, width=8.5cm]{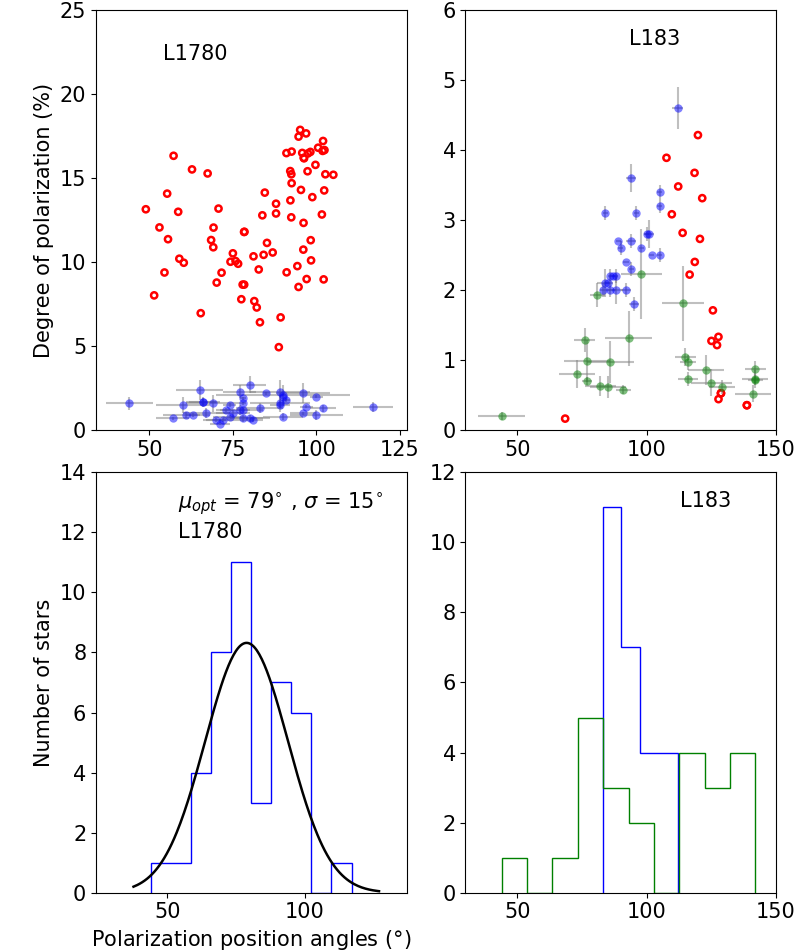}
\caption{\label{fig:hist_12_13}{\bf Top:} Variation of the degree of polarization as a function of polarization position angle for L1512, L1523, L1333, L1544, L1521E, L1517, L1780 and L183. The blue and red symbols show our R-band and the {\it Planck} polarization results, respectively. Green filled circles and histogram show the H-polarization results from \citet{2012ApJ...748...18C}. {\bf Bottom:} Histogram of polarization angles from optical measurements.}\end{figure*}
\section{Results}
The degree of polarization (P$_\text{op}$\%) and the polarization position angle ($\theta_{op}$ in degrees) of the stars projected on the molecular clouds from R-band observations are presented in Fig. \ref{fig:hist_12_13}. We plotted  in the figure the data only for which the ratio of the P$_\text{op}$\% and the error in the degree of polarization (e$_{p}$) greater than 3. 
The number of stars thus finally observed towards each cloud are listed in column 6 of Table \ref{tab:core_archive}. The mean value of P$_\text{op}$\% and the standard deviation obtained from the stars projected on each cloud are given in column 7 of Table \ref{tab:core_archive}. In lower panels of Fig. \ref{fig:hist_12_13} for each cloud, we also show the histogram of the $\theta_{op}$. Based on a Gaussian fit to the data, the mean values of the $\theta_{op}$ for the stars projected on each cloud are obtained and listed in column 8 of Table \ref{tab:core_archive}. The polarization results obtained from the \textit{Planck} data for all the eight clouds are also shown in Fig. \ref{fig:hist_12_13}. The \textit{Planck} polarization results are selected from the regions where the 250 $\mu$m intensity is greater than the standard deviation above the background intensity by a factor of three. The mean values of the polarization position angles ($\theta_{plk}$) and corresponding standard deviations are given in column 9 of Table \ref{tab:core_archive}.


\begin{table*}
\centering
\caption{Basic information of the eight clouds studied here with results from the R-band and the Planck polarization measurements.}\label{tab:core_archive}
\renewcommand{\arraystretch}{1.15}
\setlength{\tabcolsep}{4pt}
\begin{tabular}{lccllclll} \hline
Cloud   &RA (J2000)     &Dec (J2000)    & Distance    &Ref. &\# of Stars&          P$\pm$e$_{p}$ &$\theta_{op}\pm$e$_{op}$    &     $\theta_{Plk}\pm$e$_{plk}$\\
Id      &(h:m:s)   & (d:m:s)  & (pc)        &     &Obs. & (\%)              &($^{\circ}$)&($^{\circ}$)\\ 
 (1)    &(2)            &(3)            & (4)         &(5)  &(6)&(7) &(8) & (9)\\ \hline\hline
 L1333  & 02:26:04 & +75:28:30  & 170         &5    &137&  $3.0\pm1.3$ & 119$\pm$14  & 135$\pm$3\\
 L1521E	& 	04:29:21 & +26:14:19 & 145$^{+12}_{-16}$ &6&98& $3.1\pm0.8$ & 31$\pm$7   & 30$\pm$6\\ 
 L1517	& 04:55:45 & +30:33:00  & 160$\pm$3   &7    &99 &  $2.6\pm1.4$ & 82$\pm$16   & 93$\pm$19\\
 L1512  & 05:04:09 & +32:43:09 & 140$\pm$20  & 1, 2&94 &  $2.5\pm1.3$ & 157$\pm$17  & 151$\pm$8\\
 L1544  & 	05:04:17 & +25:10:48   & 130$\pm$6   &8    &136&  $3.0\pm1.3$ & 60$\pm$15   & 61$\pm$11\\
 L1523  & 	05:06:14 & +31:42:20  & 140         &3, 4 &97 &  $2.7\pm1.3$ & 129$\pm$15  & 131$\pm$7 \\
 L1780  & 	15:39:42 & -07:10:00   &  74         &9, 10&41 &  $1.4\pm0.6$ & 79$\pm$15   & 84$\pm$15 \\ 
 L183   & 15:54:12 & -02:49:42 & 115 & 9, 10 &26 &  $2.6\pm0.6$ & $94\pm8$ &$120\pm14$\\      
\hline
\end{tabular}

1. \cite{1978ApJ...224..857E}, 2. \cite{1989ApJS...71...89B}, 3. \cite{1983ApJ...264..517M}, 4. \cite{ 1992ApJ...394..196W}, 5. \cite{1998AJ....115..274O}, 6. \cite{2019A&A...624A...6Y}, 7. \cite{2018AJ....156...58B}, 8. \cite{2020A&A...633A..51Z}, 9. \cite{2019ApJ...879..125Z}, 10. \cite{2018MNRAS.476.4442N}.
\renewcommand{\arraystretch}{1.1}
\end{table*}


\begin{figure}
\includegraphics[height=6cm, width=8.5cm]{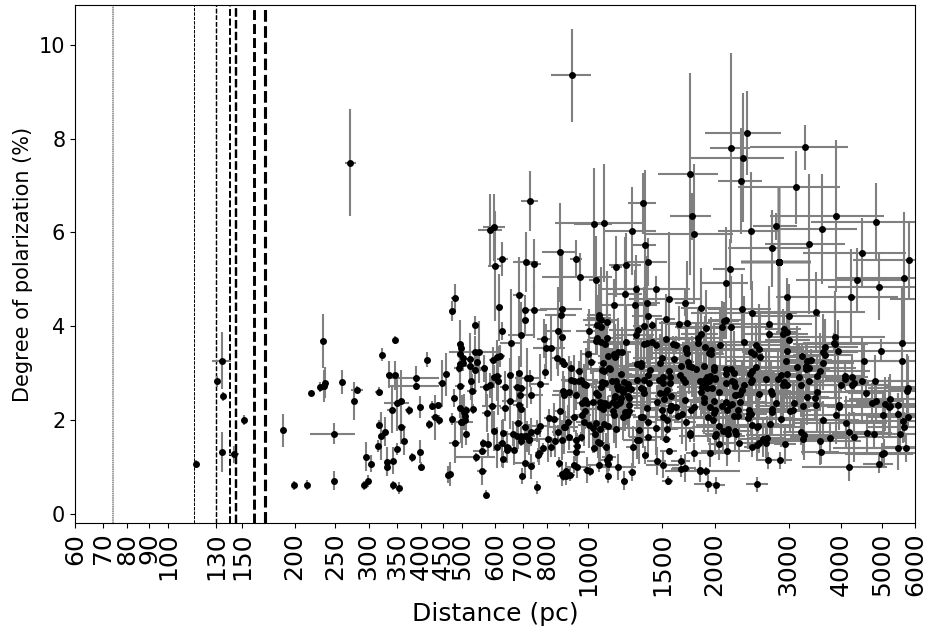}
\caption{\normalsize{The variation of the fraction of polarization (P\%) versus distance. The dotted lines show the known distances of all the eight clouds.}}\label{fig:dist_all}
\end{figure}

\subsection{The magnetic field geometry and the cloud morphology}
The magnetic field geometry of a cloud is traced using the polarisation measurements of the stars that are background to the cloud. With the \textit{Gaia} EDR3 data, we can ascertain whether the projected stars for which we have polarization measurements are foreground or background to the molecular clouds. If the observed star is background, the starlight will pass through the cloud and the asymmetric dust grains present in the clouds that are aligned with their minor axis parallel to the ambient magnetic field interact with the starlight. The starlight becomes polarized due to selective extinction and the polarization position angle gives the orientation of the projected magnetic field which is a projected component of the magnetic field in the plane of the sky \citep{1951ApJ...114..206D, 2007JQSRT.106..225L}. We obtained the distances of the stars having R-band polarization measurements from the \citet{2021AJ....161..147B} catalogue by making a search around each star with a search radius of 1$^{\prime\prime}$. For the majority of the stars, we obtained a \textit{GAIA} EDR3 counterpart well within our search radius. The P$_\text{op}$\% vs. distance plot for the stars observed by us are presented in Fig. \ref{fig:dist_all}. For all the eight clouds, almost all the observed stars are found to lie beyond the distances quoted in column 4 of Table \ref{tab:core_archive}. This implies that the $\theta_{op}$ obtained from the R-band polarization measurements are tracing the projected component of the magnetic fields of the clouds in the plane of the sky. For the reason that all the eight clouds are located very close to us ($\lesssim200$ pc), we did not correct for any foreground contribution to the measured polarization. In this context, previous magnetic field studies by \cite{2017MNRAS.464.2403S,2018MNRAS.476.4782S} also reported no significant change in the polarization properties due to line-of-sight contribution for the clouds which are at distances closer than 500 pc. 

\begin{figure*}
\includegraphics[height=5.5cm,width=12cm]{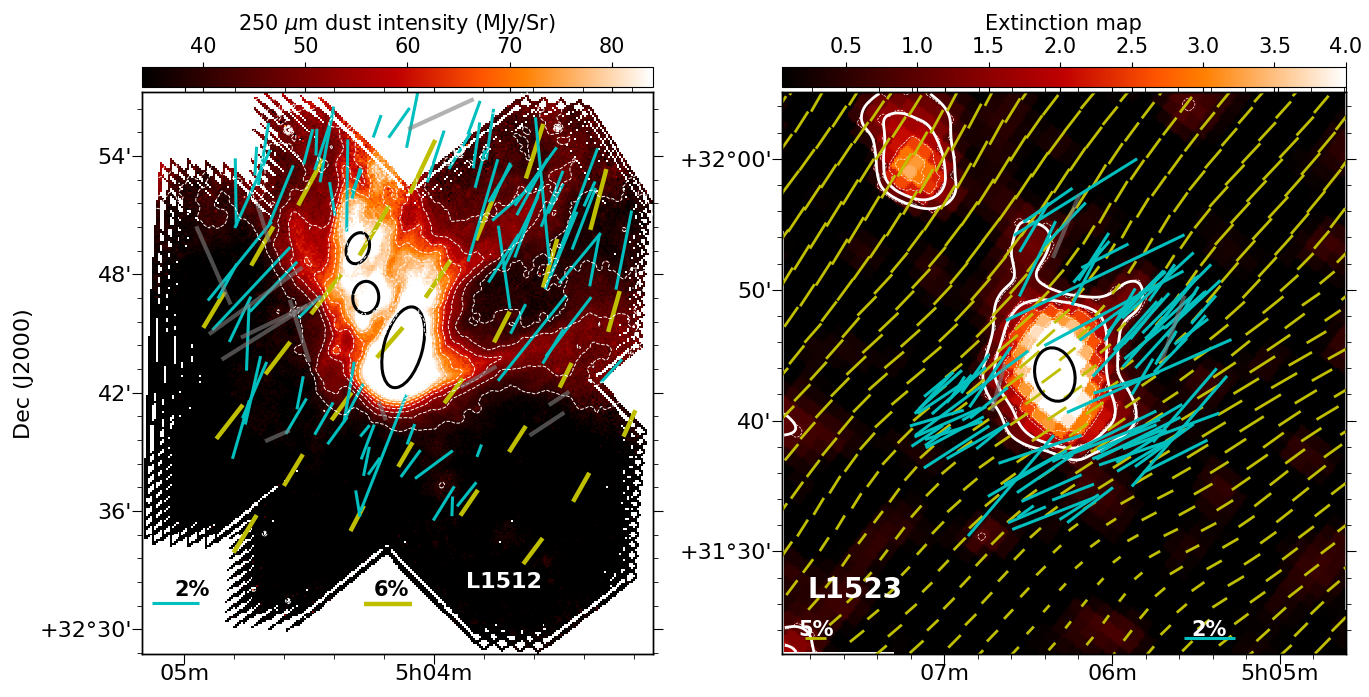} 
\includegraphics[height=5.4cm,width=5.5cm]{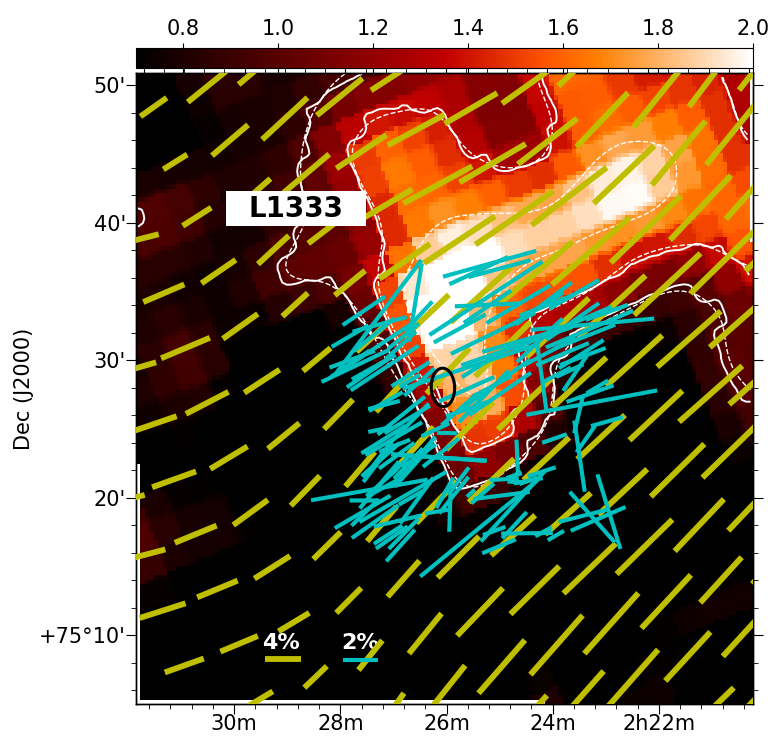}
\includegraphics[height=5.5cm,width=5.6cm]{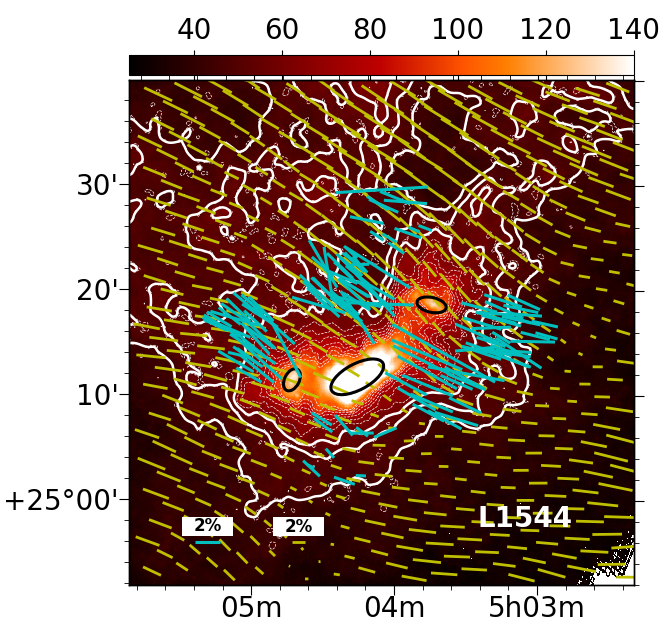}
\includegraphics[height=5.5cm, width=5.8cm]{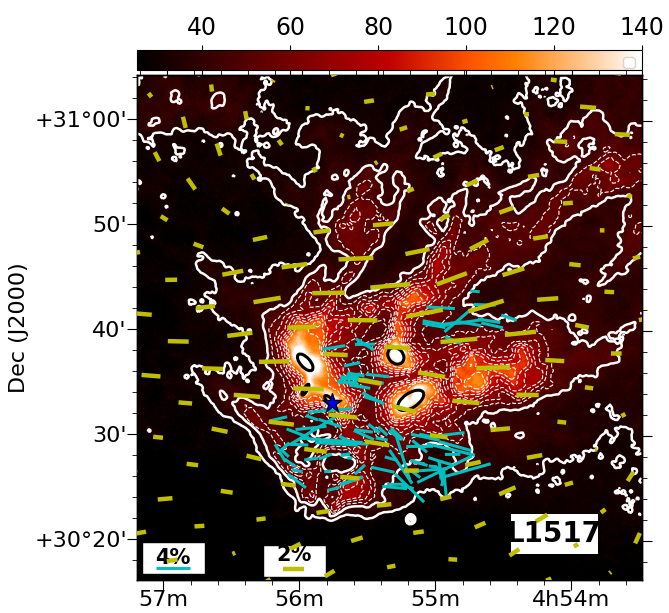}
\includegraphics[height=5.5cm, width=5.6cm]{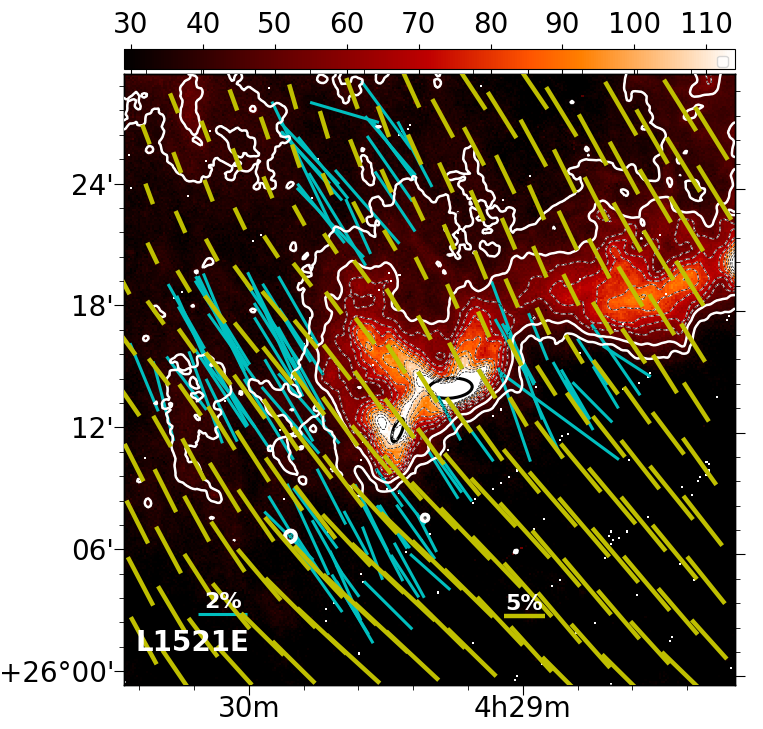}
\includegraphics[height=5.5cm,width=5.5cm]{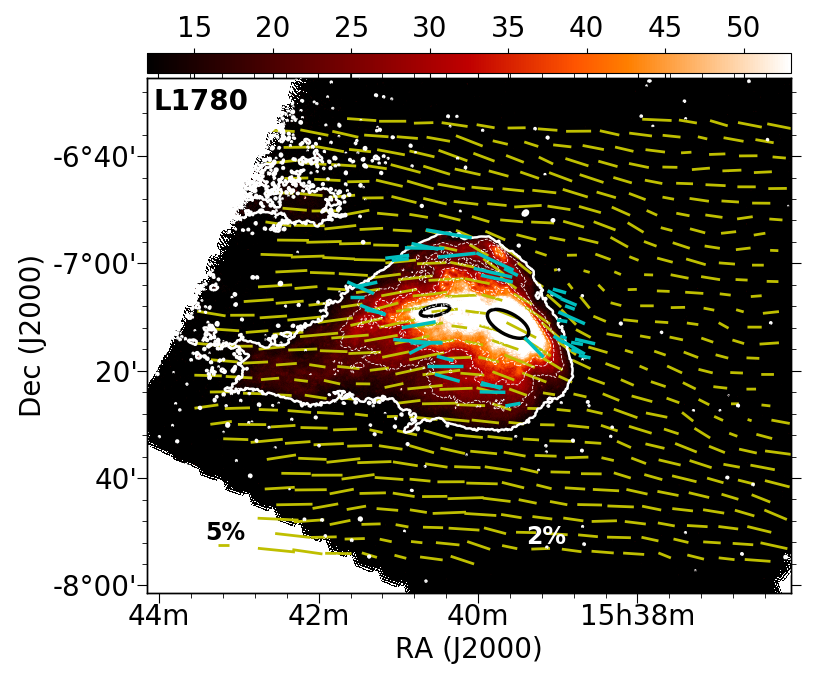}
\includegraphics[height=5.5cm,width=5.5cm]{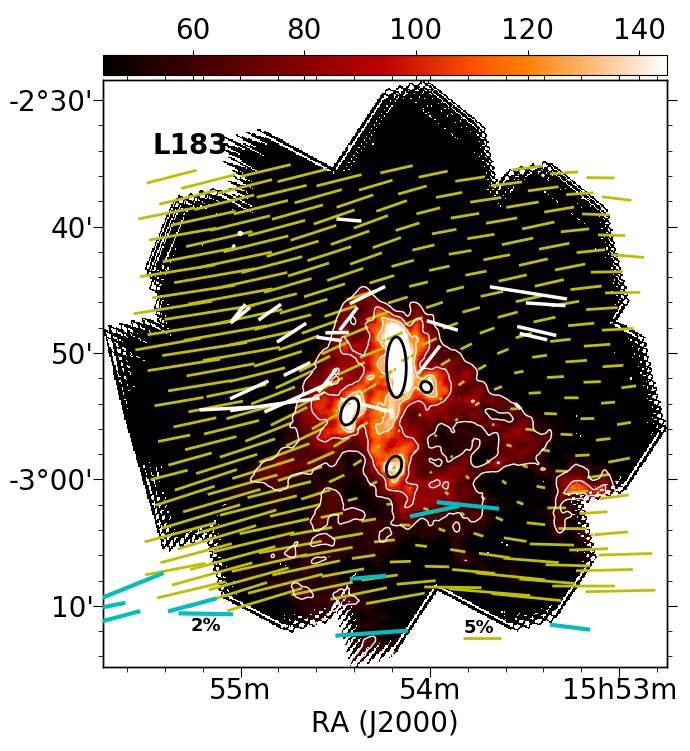}

\caption{\normalsize{Optical R-band polarization vectors in cyan and \textit{Planck} polarization measurements in yellow color overplotted on \textit{Herschel} 250 $\mu$m image of L1512, L1544, L1517, L1521E, L1780 and L183 clouds and on \textit{Dobashi} extinction map of L1523 and L1333 clouds. White vectors in bottom right last panel are H-band polarization observations for L183 \citep{2012ApJ...748...18C}. Distribution of the extracted cores identified using the \textit{Astrodendro} algorithm shown by black ellipses.}}\label{fig:3cld_morp_1}
\end{figure*}

The projected magnetic field geometries traced by the R-band polarization results (\br) are presented on the \textit{Herschel} 250 $\mu$m images containing the clouds L1512, L1523, L1333, L1521E, L1517, L1544, L1780 and L183 in Fig. \ref{fig:3cld_morp_1}. Since L1523 and L1333 were not observed with the \textit{Herschel} telescope, we overplotted the polarization results on the \textit{Dobashi} extinction map for these clouds as shown in Fig. \ref{fig:l12_l13_el} (second and third panel). The values of $\theta_\text{op}$ that are less than twice the standard deviation about the mean are plotted in the respective panels of Fig. \ref{fig:3cld_morp_1} using the lines drawn in cyan and those greater than twice that value are shown using the lines drawn in green. The projected magnetic field traced using the \textit{Planck} polarization vectors (\bp) are represented using the vectors drawn in yellow in all the panels of Fig. \ref{fig:3cld_morp_1}. The length of the vectors corresponds to the values of P$_\text{op}$ and the direction corresponds to the $\theta_\text{op}$ measured from the north increasing towards the east. 

The optical polarization traces the field lines towards the outer low-density regions of the cloud along particular line-of-sight \citep{2010ApJ...723..146F,2011ApJ...734...63S,2021A&A...655A..76S} and the \textit{Planck} polarization is more efficient in tracing the field orientations towards the denser parts of the cloud \citep{2016A&A...586A.138P}. Therefore, ideally we should use R-band polarization measurements to trace magnetic field geometry towards the low-density extended regions of the eight clouds studied here and the \textit{Planck} polarization measurements to trace the field lines towards the high-density cores of the clouds. However, as can be seen from the Fig. \ref{fig:3cld_morp_1}, the R-band polarization measurements are for sources that are located around the immediate neighbourhood of the clouds while the \textit{Planck} polarization measurements are available for the full extent of the clouds. 
It is worth noticing here that the sizes of the dense cores are typically $\sim$3-4$^{'}$ and the resolution of $\textit{Planck}$ data we used here is $\sim$ 8 arcmin. Therefore, the \textit{Planck} data traces the large-scale magnetic field towards the centre of the core, not the core scale magnetic fields. Since the polarized emission from dust grains given by \textit{Planck} data is weighted by column density \citep{2016A&A...586A.138P} and we have considered the mean \textit{Planck} polarization angle as the average of nearest vectors within an 8 arcmin region around the core centre, the $\theta_{plk}$ still represents the large scale magnetic fields around the core in this paper.
The advantage of using the R-band polarization of the stars is that as we know the distances of the individual stars and the majority of those stars towards all the eight clouds are located at the background, we are certain that the measured polarization is actually due to the dust grains belonging to the cloud and hence traces the field geometry towards the clouds. Thus, the projected magnetic field directions traced by the R-band polarization measurements can be compared with those from the \textit{Planck} measurements towards the low-density outer regions of the cloud. This is because the polarization detected in R-band is caused by the dust grains due to selective absorption which also emit thermally and causes the polarization detected by the \textit{Planck}. Hence the polarization measurements made in the R-band and the \textit{Planck} should be consistent towards the low-density parts of the cloud. Such agreements between the polarization measurements made in the optical and in the sub- or milli-meter wavelengths are seen in other studies also \citep[e.g., ][]{2009MNRAS.398..394W, 2016A&A...596A..93S, 2019ApJ...871L..15G}. In addition to this, since polarization in R-band gives field geometry with relatively higher spatial resolution in comparison to the \textit{Planck} polarization measurements, the R-band results are used to examine field distortions at smaller scales. Below we discuss the results obtained for individual clouds.

\subsubsection{L1512}
With the distance of 140 pc (shown with dotted line in Fig. \ref{fig:dist_all}), it is clear that the majority of the stars observed by us in the direction of L1512 are background sources and hence are tracing the projected magnetic field geometry of the cloud as shown in Fig. \ref{fig:3cld_morp_1} (Left top). The orientations of the \br~ and the \bp~ are in excellent agreement (both showing an orientation of 158\degr). The consistency of the \br~ and the \bp~ is even higher when we compare the $\theta_{op}$ within two standard deviation about the mean as shown by the vectors drawn in cyan. Since the optical polarization traces the field lines towards the outer low-density regions of the cloud and the \textit{Planck} polarization is more efficient in tracing the field orientations towards the denser parts of the cloud, the agreement seen here suggests that the cloud is threaded by the same magnetic field from outer low-density medium to the high density regions of L1512. The projected field is aligned along the morphology of the cloud which is elongated in the northern direction. There is a curvature in the field lines along the northwest and southeast direction. The field lines are found to be tracing the two filaments seen to the western parts of the cloud. The degree of polarization (measured from both R-band and the \textit{Planck}) is relatively lower towards the region with sharp edges as compared to the diffuse region on the opposite side. This could be due to the lack of aligned dust grains which may happen due to different grain size distribution or collisional disalignment with the gas \citep{2021AJ....161..149S}. It could also be due to a tangled magnetic field causing depolarization \citep{2015A&A...576A.104P}. The other reason might be that the magnetic field lines could be bending along the line of sight with a smaller component projected in the plane-of-the-sky.

\subsubsection{L1523}
The L1523 cloud shows a diffused tail towards the north-western parts of the cloud. The \br~ (130\degr) and the \bp~(132\degr) are found to be in excellent agreement in L1523 also indicating that the inner cloud magnetic field is inherited from the envelope magnetic field. Here again, the agreement is more pronounced as we consider the $\theta_{op}$ within two standard deviation about the mean as shown by the vectors drawn in cyan in Fig. \ref{fig:3cld_morp_1} (top right). Both \br~ and \bp~show highly ordered distribution.  The magnetic field orientation between the region to the south-eastern and the north-western parts of the cloud shows a small variation. It is at the position of the cloud where the change may be occurring. However, more data towards the north-eastern parts of the cloud is required to make a definitive statement. Whether the change in the field orientation is caused because of the formation of the cloud or the formation of the cloud occurred because of the bend in the field created by the two components of the magnetic field is unclear. In the case of a weaker magnetic field, gas dynamics is not controlled by the field and hence forced to follow the motion of the gas. However, as the field strength increases, the field lines dominate the gas dynamics helping the clouds to remain coherent \citep{2015MNRAS.451.3340K}.  

\subsubsection{L1333}
The region, L1333, a nearby cloud in Cassiopeia at 180 pc \citep{1998AJ....115..274O}, is considered to be an extreme environment of star-formation where a few low-mass stars are present in the filamentary cloud. The cloud presents an extended morphology towards the north-west and high density part towards the south-east. The field lines from \bp~ (135 \degr) are well correlated with the \br~ (119 \dgr) although there is more variation present in \br~ towards the south-west side accounting for higher dispersion in the polarization angles. The field lines are again aligned along the direction of extended emission towards north-west as seen in \textit{Dobashi} extinction map. It harbours a starless core at the southern part of the cloud shown by black ellipse which is well studied using multiple molecular lines \citep{2011ApJ...734...60L,2007ApJ...664..928S,2004ApJS..152...81P}. The large-scale magnetic field is perpendicular to the orientation of the core.

\subsubsection{L1544}
The cloud L1544, one of the dark clouds hosting a starless core in Taurus \citep{1983ApJ...264..517M,Heyeretal1987}, with unusual chemical richness \citep{2003ApJ...583..789L,2014A&A...569A..27B}, offers a key laboratory to test physical conditions and stability of gas in starless cores. The \br~ (60\degr) shown in cyan vectors are found to be in excellent agreement with the \bp~ (61\degr) implying that the field lines inside the cloud are preserved with the fields from the envelope of the cloud. Both \br~ and \bp~ are mostly oriented perpendicular to the high density emission structures seen towards the cloud. The JCMT-SCUBA polarization maps of 850 $\mu$m thermal emission from dust in L1544 were obtained by \citet{2000ApJ...537L.135W}. The mean position angle of the magnetic field inferred is found to be at an offset of 23\degr~ with the \br~and $\sim37$\degr~with \bp. It is interesting to note that towards the southern tip of the cloud both optical and the \textit{Planck} polarization values show a change in the field direction from $\sim$60\degr~ to $\sim$90\degr. The degree of polarization shows a drop at this location consistently in measurements made in both the techniques. The magnetic fields orientation towards the diffuse extended region is found to be more parallel to the elongated morphology of the cloud in the north eastern direction. More details on the magnetic fields using optical polarization data will be discussed in Kumar \& Soam 2022 (under preparation).

\subsubsection{L1517}
The L1517 cloud is relatively isolated only to be disturbed by the nearby pre-main sequence stars, AB Aurigae and SB Aurigae which are physically associated with the cloud \citep{Heyeretal1987}. The cores are at different evolutionary stages with detected \nthp emission in selected cores.
Fig. \ref{fig:3cld_morp_1} shows that there is an excellent agreement between the field lines in the inner parts of the clouds (93\degr) to the field lines in the outer envelope (83\degr) of the cloud. \citet{2006MNRAS.369.1445K} mapped the linear polarization emission from L1517B (one of the five cores in the cloud) at 850 $\mu$m. The weighted mean of position angle was estimated as 106\degr~ which is at an offset of 13\degr~ with the \bp~ and at an offset of 23\degr~ from the \br~. \citet{2006MNRAS.369.1445K} also reported a shift in the position angles from the north (84\degr) to the middle (94\degr) and to the southern parts (156\degr) of L1517B. It is evident from the Fig. \ref{fig:3cld_morp_1}, that there are changes in the large-scale magnetic field orientation around the cloud as found towards the north-east and south-west corners of the figure where field rotations are most likely causing a reduction in the degree of polarization possibly due to depolarization. \citet{2011A&A...533A..34H} have made a detailed study of L1517 in molecular lines and 1.2 mm dust continuum emission. They identified four filaments and five starless cores embedded in these filaments that are at different evolutionary stages. These filamentary structures stretch parallel to the direction of the extended emission and also to the direction of the magnetic fields. The magnetic fields towards the south of the cloud near AB Aur shown by blue star in Fig. \ref{fig:3cld_morp_1}) show more dis-oriented vectors with relatively large dispersion which could be because of the effects from this Herbig Ae star. Opposite to that, the vectors in the north of cloud show more ordered pattern. 

\subsubsection{L1521E}
The cloud L1521E \citep{1999ApJS..123..233L} is considered to be chemically and dynamically young starless core in a larger cloud L1521 in Taurus constellation.
The field orientations inferred from both the techniques are in excellent agreement implying that the magnetic field lines are preserved at core as well envelope scales in L1521E also. The field orientations follow the extended cloud structure of the cloud. The H$_{2}$ density at the peak position of the core is estimated to be 1.3-5.6$\times$10$^{5}$ cm$^{-3}$. However, although the H$_{2}$ density is such high enough to excite the inversion transitions of NH$_{3}$, these lines are found to be very faint in L1521E \citep{2002ApJ...565..359H}. \citet{2004A&A...414L..53T} compared the emission of two molecules, C$^{18}$O and N$_{2}$H$^{+}$ which are known to have different depletion behaviours. While C$^{18}$O gets heavily depleted at densities of a few 10$^{4}$ cm$^{-3}$, N$_{2}$H$^{+}$ remains at gas phase up to densities of $\sim$10$^{6}$ cm$^{-3}$. Thus, the ratio between the integrated intensities of C$^{18}$O and N$_{2}$H$^{+}$ is a good indicator of depletion which is found to be 3.4 in L1521E implying that this core is chemically less processed and therefore youngest core known with a contraction age of $\lesssim 1.5\times10^{5}$ yr \citep{2004A&A...414L..53T}. However, this age is not compatible with the models of core formation via ambipolar diffusion, as they require evolutionary times of the order of 1 Myr. In that case, a more dynamical core formation scenario is required to create such a highly contracted core on a short time scale as estimated by using chemical models \citep{2004A&A...414L..53T}. The sharp edges to the southern side and a more extended distribution of material to the opposite direction could be a result of some external compression acted from the southern direction along the field lines that might have created the cloud.

\subsubsection{L1780}
The L1780 is a cometary shaped high cloud located at higher latitudes. For the case of L1780, the \br~ (77\degr) and the \bp~ (84\degr) agree very well not only with respect to the mean orientation but also to the overall field morphology traced with the two techniques. The field lines remarkably follow the cloud structure. The cloud presents a cometary morphology with a broad high density part to the west and a diffuse narrow part to the east. The structure shows a curvature from east to west which is strikingly traced by the field lines as well. Towards the head, the degree of polarization is significantly lower compared to the other regions as traced by both the techniques.

\subsubsection{L183}
The morphology of L183 is directed towards south in form of V-shaped structure as shown by outer contours in Herschel emission (see Fig. \ref{fig:3cld_morp_1}). The R-band measurements made by us for cloud L183 are for the background stars lying towards the southern parts of the cloud while the H-band polarization measurements made by \citet{2012ApJ...748...18C} are for the background sources lying to the northern parts of the cloud. The magnetic fields inferred from the H-band polarization measurements made by \citet{2012ApJ...748...18C} are shown using the vectors drawn in white. Though the field orientations inferred from both R-band and H-band are in good agreement with the \textit{Planck} polarization measurements, there are differences seen between the H-band and the \textit{Planck} magnetic field orientations towards the western parts of the cloud where a reduction in the degree of polarization is seen. There could be two components of magnetic field orientation in this case which might be causing a depolarization resulting in the reduction of polarization. The maps generated using the R-, H- and \textit{Planck} present a smooth and ordered field geometry over the entire field containing the cloud. The magnetic field is predominantly perpendicular to the north-south elongation of the cloud. The weighted mean position angle of the magnetic field, averaged over the 16 positions observed by \citet{2000ApJ...537L.135W} in 850 $\mu$m is 46\degr. This represents the magnetic field geometry of the high density inner regions of the cloud which is at an offset of 74\degr, 57\degr and 48\degr~with respect to the magnetic fields inferred by the \textit{Planck}, H-band and R-band polarization measurements. This is the only cloud where we find the sharp edged northern and extended diffuse emission to the southern parts of the cloud are perpendicular to the magnetic field orientation. \cite{2020ApJ...900..181K} has carried out the comparative study of the magnetic field structure of the cloud in optical, near-infrared wavelengths with sub-millimetre polarization using Pol-2 at JCMT. The authors have mapped the whole cloud using optical polarization. The mean value of magnetic field orientation is around $\sim$ 87$^{\circ}$ using optical observations which is well consistent with our reported value. 

\section{Discussion}
\begin{table}
\centering
\caption{Position angle of extended structure, magnetic field using R-band optical polarization and Planck polarization measurements.}\label{tab:offsets_env}
\renewcommand{\arraystretch}{0.9}
\begin{tabular}{lccccc}
Cloud Name	&  $\theta_{op}$     & $\theta_{plk}$      &  $\theta_{env}$  &  $\theta_{plk}$-$\theta_{env}$ &$\theta_{op}$-$\theta_{env}$ \\
    & ($^{\circ}$)   &   ($^{\circ}$)  & ($^{\circ}$)  & ($^{\circ}$) & ($^{\circ}$)   \\ 
\hline
  (1) &(2)&(3)&(4)&(5)&(6)\\ \hline\hline
L1333  &119 &	135 &	128 &	9  &	7 \\
L1521E & 31  &	30  &	43  &	12  &	13 \\
L1517  &82  &	93  &	136 &	54  &	43 \\
L1512  &157 &	151 &	148 &	9  &	3 \\
L1544  & 60  &	61 &	36  &	24  &	25 \\
L1523  & 129 &	131 &	137 &	8   &	 6 \\
L1780  &79  &	84  &	100 &	21  &	16 \\
L183   & 94  &	120 &	26 &	68  &	94 \\
\hline
\end{tabular}
\renewcommand{\arraystretch}{1}
\end{table}

We have studied the large-scale magnetic field structure of the clouds using \textit{Planck} polarization and its small scale variations using optical polarization at the outer periphery of the cloud. The cloud magnetic fields are found to be parallel to inter-cloud magnetic fields in all the clouds. One common factor in all these selected clouds is their observed morphology. All the clouds show relatively bright rim on one side and the extended diffuse structure on the other side which can be noticed by the change in intensity along the elongation. We are interested in the direction of small intensity gradient in the cloud's extended structure which is defined by $\theta_{env}$. We visually selected directions from the brightest core in the map till the outer isocontours and calculated the gradient within the outer intensity isocontours. Hence, $\theta_{env}$ was extracted by taking the slowest change of intensity out of many selected directions.
 Table \ref{tab:offsets_env} shows the comparison between the magnetic field orientation (optical and \textit{Planck}) with the structure of the cloud ($\theta_{env}$). Our results suggest that the magnetic field lines in six clouds are almost oriented along the diffuse extended emission which is opposite the sharp edge of the cloud. One of the clouds, L183 which do not shows parallel orientation, showed almost same value of gradient in all the directions but different orientations. \cite{2016ApJ...824...85K} investigated a sample of starless cores using \textit{Akari} emission maps and molecular line observations to conclude that the external feedback from the stars or the isotropic interstellar radiation field might play a crucial role in the evolution of the cores. In that study, L1512 and L1517B have been studied in detail using 3-160 $\mu$m emission and they are found to be affected by the presence of star in their vicinity. This could be responsible for the asymmetric shape of the far-infrared emission in L1512 and L1517.
The fact that the magnetic fields in this study are along the elongation of cloud structure in majority of the clouds may suggest a link to the feedback or effect from the ISRF. \cite{2011MNRAS.414.2511V} has suggested that the scenario of cloud formation is driven by colliding flows which can be guided by magnetic field and thereby facilitate the formation of filaments and fragmentation to the clumps/cores. The origin of these flows is still unknown. One possible explanation for the B-field aligned along the extended emission of the clouds is that the feedback from the stars nearby to the clouds might have stripped the clouds shown here.

\subsection{Magnetic field strength and cloud stability}
The Davis-Chandrasekhar-Fermi (DCF) method \citep{1951ApJ...114..206D,1953ApJ...118..113C} is an effective means of estimating magnetic field strength by taking into account the dispersion in polarization angles and the non-thermal gas motions which are responsible to distort the magnetic field structure at smaller scales and also measure mass-to-flux ratio.

We used the following equation to calculate the strength of the plane-of-the-sky component of magnetic field:
\begin{align}
\frac{B_{pos}}{( \mu G)}= 9.3 \times \sqrt{\frac{n_{H_{2}}}{cm^{-3}}} \times  \frac{\Delta v}{km s^{-1}} \times \Big(\frac{\Delta  \phi}{1^{\circ}}\Big)^{-1}
\end{align}
where $\Delta v$ is the FWHM obtained from the Gaussian fitting to the CO spectra and $\Delta$ $\phi$ is the dispersion in polarization angle in degrees.

We estimated the magnetic field strength for all the clouds by taking dispersion in polarization angles from the optical polarization data and velocity dispersion from the available molecular line data of $^{13}$CO for all the sources from \cite{2004ApJS..153..523L}. Using 2$\times$10$^{3}$ cm$^{-3}$ as the average number density for all the clouds and the dispersion in polarization angle from table \ref{tab:core_archive}, the strength has been calculated as $\sim$ 27 $\mu$G, $\sim$ 18 $\mu$G, $\sim$ 21 $\mu$G, $\sim$ 26 $\mu$G,$\sim$ 23 $\mu$G and $\sim$ 78 $\mu$G for L1333, L1512, L1523, L1544, L1517 and L1521E, respectively. The magnetic field strength for high-latitude clouds, L1780 and L183 were estimated as 11 $\mu$G and 42 $\mu$G by \cite{2016A&A...588A..45N}. Assuming typical uncertainty in the velocity dispersion and the polarization angles, the uncertainty in the magnetic field strength is estimated to be $\sim$ 0.5 B$_{pos}$.

The relative importance of the magnetic fields with respect to the self-gravity can be understood by calculating the mass-to-flux ratio which can tell whether or not magnetic fields are strong enough to support a molecular cloud against gravitational collapse \citep{1976ApJ...210..326M}. This can be quantified by a parameter \citep{2004Ap&SS.292..225C,1978PASJ...30..671N}, $\lambda$ = 7.6$\times$ 10$^{-21}$ N$_{{H}_{2}}$/B$_{pos}$ where N$_{H_{2}}$ is the column density in units of cm$^{-2}$ and B$_{pos}$ is the magnetic field strength in $\mu$G. The mass-to-flux ratio has been calculated for all the sources to discuss the magnetic criticality/stability. If $\lambda$ > 1, the cloud envelope is magnetically supercritical implying that the magnetic fields cannot support enough the cloud against gravitational collapse. Whereas if $\lambda$ < 1, it is magnetically sub-critical implying the magnetic fields can prevent the gravitational collapse of the cloud. We took the extinction values from GAIA DR3 measurements for the stars observed in optical polarization observations. Using an average value of extinction for each cloud, we estimated the column density for all the clouds as 
1.95, 1.96, 1.95, 2.56, 2.22, 1.69 ($\times$ 10$^{21}$) cm$^{-2}$ using the relation as N$_{H_{2}}$ (cm$^{-2}$) = 0.93$\times$ 10$^{21}$ A$_{v}$ (mag) \citep{1978ApJ...224..132B}. We estimated the values of mass-to-flux ratio as 0.55, 0.83, 0.71, 0.75, 0.73, 0.16 for the clouds L1333, L1512, L1523, L1544, L1517 and L1521E, respectively.
By adopting the column density as 1$\times$ 10$^{21}$ \citep{2015A&A...584A..92M}, we estimated $\lambda$ to be 0.7 and 0.18 for L1780 and L183, respectively. Since both L1780 and L183 are high latitude clouds and are at similar galactic latitude, we used the same column density for L183 also. Considering the uncertainty in the values of magnetic field strength, all the clouds are approaching the critical value except L1333, L1521E and L183. This implies that the magnetic fields are marginally supporting the cloud envelope against the self-gravity and are likely to play a significant role in governing the cloud dynamics. The very low values of L1333, L1521E and L183 indicate that the magnetic fields are strong enough to dominate the gas dynamics in the cloud envelope and can delay the gravitational collapse.

\subsection{Relative orientation between the magnetic field and the core orientations}

The star-forming cores or starless cores are considered to be the consequences of fragmentation of the clouds. Magneto-hydrodynamic simulations carried out by various authors \cite{2020MNRAS.497.4196S,2009ApJ...695..248H} showed that the there is a significant change of magnetic field orientation (from low to high field strengths) in the formation of clouds. As the relative orientation of filaments and B-field are correlated in nearby cloud complexes \citep{2017A&A...607A...2S}, the cores, being the nested structures, will have effect from the magnetization of the cloud. In this regard, the clump shapes and the orientations are important indicators of clump dynamics and fate of star formation. The core-background magnetic field can provide important clues about the dynamical importance of magnetic field at the cloud scale in comparison to stellar feedback or turbulent motions.
In the previous section, we discussed the magnetic field orientation with respect to the distribution of the material around the clouds. In this section, we investigate the relative orientation between the core major axis and the ambient and the core-scale magnetic fields. Since the R-band polarization measurements trace the envelope magnetic fields or the fields at the periphery of the clouds, we define this as the ambient ($\theta_{op}$) magnetic fields. The core-scale magnetic field defined by $\theta_{plk}$ is calculated by taking an average of the nearest three \textit{Planck} polarization vectors around the inner regions of the cloud. The $\theta_{plk}$ thus obtained are listed in the column 6 of Table \ref{tab:core_pa}.  
The positions of the cores associated with the selected clouds and their properties are obtained using the python package \textit{Astrodendro} algorithm \citep{2008ApJ...679.1338R} on \textit{Herschel} 250 $\mu$m emission maps. The dendrogram method identifies emission features at successive isocontours in emission maps which are called "leaves" and then find the intensity values at which they merge with the neighbouring structures (branches and trunks). We supplied an emission threshold above which all the structures are identified and contour intervals which decides the boundary between the distinct structures. The initial threshold and the contour step size were selected as a multiple of the $\sigma$, the rms of the intensity map. In our analysis, we are interested in the dense cores which are identified as leaves. 


 For each core, the dendrogram method estimated the core positions, the major and the minor axes, the effective radius of the core, and the position angle in degrees which is measured counter-clockwise from the positive x-axis. The core major and minor axes are calculated from the intensity-weighted second moment in the direction of the major axis or perpendicular to the major axis.  The derived properties of the cores are presented in Table \ref{tab:core_pa}. A total of 19 cores were identified in six clouds studied here. The right ascension and the declination of the extracted cores are given in the columns 2 and 3. The column 4 gives the aspect ratios of the cores. The orientation of the major axis measured from the north increasing to the east is given in column 5. The sources having aspect ratio higher than 0.8 were removed from the analysis to consider only elongated cores \citep{2020MNRAS.494.1971C}. The smallest structure that can be considered as real is of size $\sim$ 0.75 FWHM  \citep{2017A&A...599A.109M}. We selected only those sources where the derived radius is larger than 2-3 pixels with the beam size of 250 $\mu$m intensity map. In addition to these, we considered only those sources lying within the asymmetric structure of the cloud (shown by white contours in fig. \ref{fig:3cld_morp_1} (right panel)). 
A total of 19 cores are identified in six clouds studied here, five cores in L1517, four in L183, three each in L1512 and L1544, two each in L1521E and L1780. We adopted the positions and major axis orientation of the cores in L1523 and L1333 from \cite{1999ApJS..123..233L} as there are no \textit{Herschel} maps available. The extracted cores in each cloud are shown in the panels of Fig. \ref{fig:3cld_morp_1}. In L1512, the major axis of the three extracted cores are almost parallel to each other and parallel to the ambient magnetic field. Interestingly, the core major axis orientation is along the elongated morphology of the cloud. In contrast, in L183, the major axes of three of the four cores are oriented parallel to each other, but perpendicular to the ambient magnetic field.

\begin{figure}
\centering
\includegraphics[height=6cm, width=9cm]{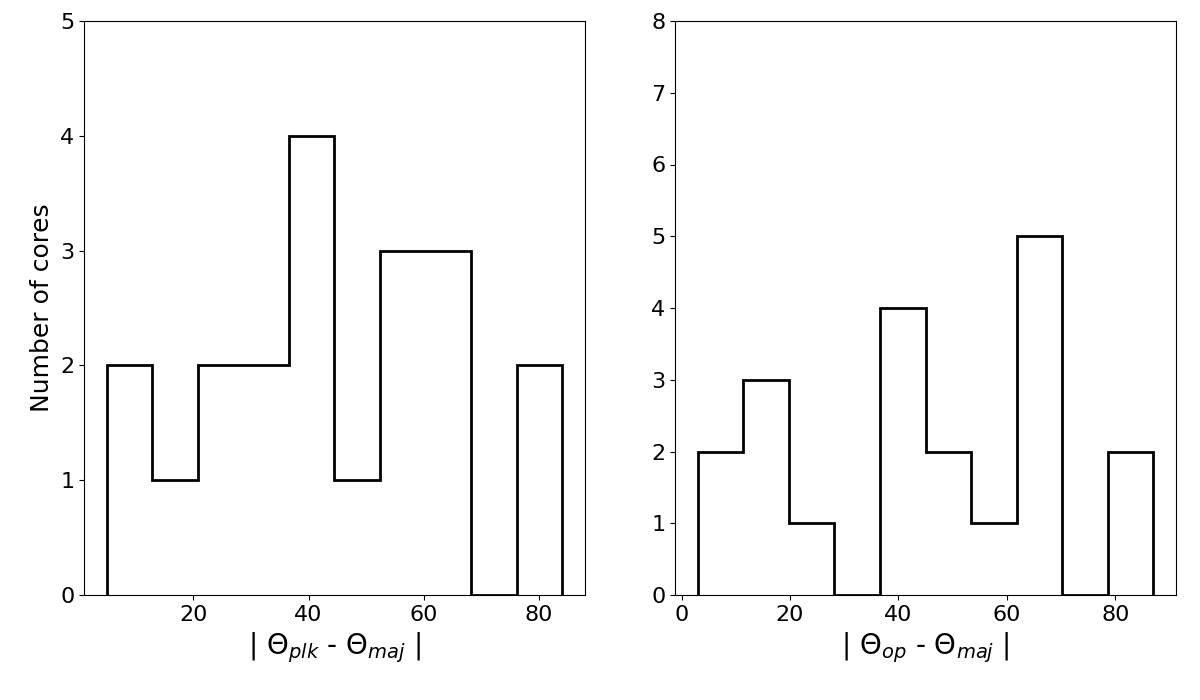}
\caption{{\bf Left:} \normalsize{Distribution of the difference of inner magnetic field and position angle of major axis. {\bf Right:} Difference of outer magnetic field and position angle of major axis. }}\label{fig:diff}
\end{figure}

\begin{table*}
\caption{Position angle of cores, cloud magnetic field using R-band optical polarization and core-scale magnetic field \textit{Planck} polarization measurements.}\label{tab:core_pa}
\renewcommand{\arraystretch}{1.1}
\begin{tabular}{p{0.5cm}p{1.3cm}p{1.3cm}p{0.5cm}p{0.5cm}p{0.4cm}p{0.5cm}p{0.5cm}p{1.8cm}} \hline
Core    & RA & Dec &  (b/a)    &  $\theta_{maj}^{\ddagger}$    & $\theta_{plk}^{c}$ & $\Delta\theta^{plk}_{maj}$  & $\Delta\theta^{opt}_{maj}$ & Cores studied  \\
 Id     &  (h:m:s)  & (d:m:s)  & &  ($^{\circ}$) & ($^{\circ}$)  & ($^{\circ}$) & ($^{\circ}$)  & in literature  \\ 
  (1) &(2)&(3)&(4)&(5) & (6) & (7) & (8) & (9)\\ \hline\hline
\multicolumn{9}{c}{{\bf L1333}, $\theta_{op}$ = 119$^{\circ}\pm$15$^{\circ}$}\\ 
C1 & 02:26:04 & +75:28:30  &	0.18 & 2$\dagger$ &	135	&  117 & 133 & L1333 (1) \\  \hline
\multicolumn{9}{c}{{\bf L1521E}, $\theta_{op}$ = 32$^{\circ}\pm$7$^{\circ}$}\\
C1 & 04:29:28 & +26:11:60 &	0.35 &	159 &	37	&   58 &  51  \\
C2 & 04:29:16 &   +26:14:05 & 0.45 &	 94 &	 31  &	63 &	 64 & L1521E (1)	 \\ \hline
\multicolumn{9}{c}{{\bf L1517}, $\theta_{op}$ = 83$^{\circ}\pm$14$^{\circ}$ }\\
C1 & 04:55:11 & +30:33:19	& 0.49 	& 123 &	 79 &  	44 & 	 41 & L1517B (1) \\
C2 & 04:55:47 & +30:33:14	& 0.38 	& 42 &	 86  &	44 &     41 & L1517A (1)\\
C3 & 04:55:57 &	+30:34:17 	& 0.27 	& 150 &	 87 & 	63 &	 66 & L1517C (1)\\
C4 & 04:55:58 & +30:36:54 	& 0.47 	& 41 &	 90  &	49 &	 42 \\
C5 & 04:55:18 &	+30:37:34	& 0.79 	& 41 &	 82 & 	41 &	 42  \\ \hline 
\multicolumn{9}{c}{{\bf L1512}, $\theta_{op}$ = 158$^{\circ}\pm$16$^{\circ}$}\\
C1 & 05:04:07	& +32:44:19 & 	 0.47 &  165 & 	 142 & 	23 &  7  & L1512-1 (1) \\ 
C2 & 05:04:16 	& +32:46:50 &	 0.80 &	 176 &	 143 & 	33 &  18  \\
C3 & 05:04:18 	& +32:49:20 &	 0.70 &	 161 &	 147 & 	14 &  3 \\ \hline
\multicolumn{9}{c}{{\bf L1544}, $\theta_{op}$ = 60$^{\circ}\pm$16$^{\circ}$}\\
C1 & 05:04:16 & +25:11:44 & 0.45 &   118  & 	60 &  	58 &	58 & L1544 (2)\\  
C2 & 05:04:43 & +25:11:25 &  0.62 &  153 & 	69 & 	84 &	 87 & L1544-E (2)\\
C3 & 05:03:45 & +25:18:37 &  0.50 &  77 &   40 & 	37 &	 17  & L1544-W (2) \\ \hline 
\multicolumn{9}{c}{{\bf L1523}, $\theta_{op}$ = 130$^{\circ}\pm$14$^{\circ}$}\\
C1  &  05:06:20  &	 +31:43:33  &	 0.73  &13$\dagger$ & 126 & 	67 &  64 & L1523 (1)	 \\ \hline
\multicolumn{9}{c}{{\bf L1780}, $\theta_{op}$ = 79$^{\circ}\pm$15$^{\circ}$}\\
C1 & 15:39:37 &  -07:11:20 & 	 0.46 &	 61 & 	 68 & 	7  & 17 \\ 
C2 & 15:40:32 &	 -07:08:52 & 0.33 &	 102 &	 97  &	5 &	 23  \\ \hline  
\multicolumn{9}{c}{{\bf L183}, $\theta_{op}$ = 93$^{\circ}\pm$8$^{\circ}$}\\
C1 & 15:54:11	& -02:59:00 &	 0.66 &	 160 & 	 132  &	28 & 	 67 \\ 
C2 & 15:54:25 & -02:54:38 &	 0.57 &	 158 &	 122  &	36 &	 65 \\
C3 & 15:54:10	& -02:51:08	&    0.33 &  180 &   120  & 60 &	 87 & L183B (1)\\
C4 & 15:54:01 & -02:52:42 &	 0.85 &	 42 &	 126  &	84 &	 51  \\ \hline  

\end{tabular}

 $^{\ddagger}$ Position angle of the major axis of the cores.\\
 $\Delta\theta^{pla}_{maj}$=$\left| \theta_{pla} -\theta_{maj} \right|$, 
 $\Delta\theta^{opt}_{maj}$=$\left| \theta_{opt} -\theta_{maj} \right|$
 
\renewcommand{\arraystretch}{1}

$^{\dagger}$ Value of major axes have been taken from \cite{1999ApJS..123..233L}.
 1. \cite{1999ApJ...526..788L}, 2. \cite{ 1998ApJ...504..900T}
\end{table*}

In Fig. \ref{fig:diff}, we show the distribution of the angle offset between the direction of core major axes with respect to the magnetic field from \textit{Planck}, |$\theta_{plk} - \theta_{maj}$| in left panel and the magnetic field from R-band polarization, |$\theta_{op} - \theta_{maj}$| in right panel. If the core orientation has a significant effect due to the large-scale magnetic field, we expect the core major axis to be perpendicular to the magnetic field lines with a significant effect on the gas motions. Contrary to this, in the Fig. \ref{fig:diff}, we find that all the cores extracted show a random orientation with respect to the outer (R-band) as well inner ($\textit{Planck}$) magnetic field. We quantified our inference using a statistical technique, the \textit{Rayleigh circular statistic} which gives a measure of relative alignment of structures with respect to the magnetic field and to obtain a statistical strength of that alignment \citep{2018MNRAS.474.1018J}. 
We define projected Rayleigh statistic (PRS) as
\begin{equation}
    \textit{Z} = \sum_{i=1}^{n}{cos  \phi_{i}}/\sqrt{n/2},
\end{equation}
where $\phi_{i}$=2$\theta_{i}$ and $\theta_{i}$ is the offset between the magnetic field and core major axes.
If Z << 0, it is considered to be parallel alignment and if Z >> 0, it is considered to be a perpendicular alignment. If Z$\approx$ 0, the distribution is either random or preferentially lies around 45$^{\circ}$. We have estimated Z to be 0.05 and -0.15 for \textit{Planck} and R-band data, respectively. The offsets in both the cases are neither parallel nor perpendicular owing to the values close to zero and hence, follow random distribution. 
The same analysis was done by \cite{2018MNRAS.474.1018J} on the nearby Gould belt regions and Vela cloud to test the preferential alignment of column density structures with magnetic field and the values vary approximately from -5 to 5 to show transition from parallel to perpendicular alignment.

This result implies that the core orientation might not have a significant effect from the large-scale magnetic field as per the scenario of magnetic field dominated low mass star formation \citep{1987ARA&A..25...23S,1999ASIC..540..305M}. There might be other factors that can contribute to shaping the core morphology such as feedback effects or the ram pressure as suggested by \cite{2018ApJ...865...34C}. Although we expect that the cores tend to align their major axis perpendicular to the local magnetic field where the flow of material will be preferential to favour the collapse mechanism, the environment can be a crucial effect towards the morphology of the cores. \cite{2018ApJ...865...34C} showed that the environment also plays an important role in shaping the prestellar cores through ram pressure and magnetic pressure. The turbulence created by stellar feedback may affect the orientation of the cores' minor axis \citep{2016ApJ...824...85K} or the local magnetization within the cloud \citep{2020MNRAS.494.1971C}.  %

\section{Conclusions}
In this work, we presented our R-band polarization measurements of over 500 stars projected on eight small, isolated clouds. Of these, we took observations for six of these clouds, namely, L1512, L1523, L1333, L1521E, L1517 and L1544. These sources were chosen on the basis of their cloud morphology from \textit{Herschel} 250 $\mu$m emission or \textit{Dobashi} extinction map having sharp edges on one side and an extended emission on the opposite direction to trace the plane-of-the-sky magnetic field geometry of these clouds. Two more clouds, L1780 and L183, showing this emission were added to the sample from the literature. The main goals of the study are (a) to investigate the relationship between the observed magnetic field orientation with respect to the elongated morphology seen in the clouds; (b) to investigate the relationship between the magnetic field orientations inside the clouds inferred from the \textit{Planck} polarization and the envelope-scale field inferred from the optical polarization measurements of background starlight; (c) to examine the relative offsets between the core major axes orientation with respect to both inner and outer magnetic field directions. The results obtained from this study are given below:
\begin{enumerate}
    \item Based on the distribution of the degree of polarization and the position angles of the stars projected on all the eight clouds (measured by us in R-band and taken from archival data) with respect to the distances of the stars obtained from the \textit{Gaia} EDR3, we show evidence that all the stars are found to be located at distance farther than $\sim$200 pc.
    
    \item In all the eight clouds studied here, the magnetic fields inferred from the R-band polarization and the \textit{Planck} polarization measurements are in excellent agreement, suggesting that it is the same dust population which is possibly responsible for the production of polarization that is detected in the two techniques. 
    
    \item In seven out of eight clouds studied, the projected magnetic field is aligned with the direction of diffuse emission of cloud. The one cloud, L183, has its diffuse emission oriented perpendicular to the ambient magnetic field.
    
    \item The magnetic field strength estimated with DCF method in all the clouds ranges from 11-78 $\mu$G. With the column density and the derived field strength, we calculated the mass-to-flux ratio, $\lambda$ with the values ranging from 0.16-0.83. Except L1333, L1521E, and L183 all the clouds show magnetically critical state suggesting that the magnetic pressure are almost supporting the gravity in the cloud envelope.
    
    \item A total of 19 cores are extracted from the six clouds. The major axes of the three cores found in L1512 from \textit{Herschel} 250 $\mu$m emission are not only found to be parallel to each other but also to both the projected magnetic fields and the direction of diffuse emission of the cloud. In contrast, for three out of four cores in L183, their major axes are found to be parallel to each other and also to the cloud elongation, but almost perpendicular to the orientation of the ambient magnetic field.
    
    \item The orientation of the core major axis with respect to the (projected) ambient and the inner magnetic fields is randomly distributed.
\end{enumerate}

\section*{Acknowledgements}
We thank all the supporting staff at 104-cm Sampurnanand Telescope, ARIES, Nainital who had provided assistance in carrying out these observations. This research has made use of the SIMBAD database, operated at CDS, Strasbourg, France. We also acknowledge the use of NASA’s SkyView facility (http://skyview.gsfc.nasa.gov) located at NASA Goddard Space Flight Center. This work has made use of data from the following sources: (1) European Space Agency (ESA) mission \textit {Gaia} (\url{https://www.cosmos.esa.int/gaia}), processed by the {\it Gaia} Data Processing and Analysis Consortium (DPAC, \url{https://www.cosmos.esa.int/web/gaia/dpac/consortium}). Funding for the DPAC has been provided by national institutions, in particular the institutions participating in the {\it Gaia} Multilateral Agreement; (2) the \textit{Planck} Legacy Archive (PLA) contains all public products originating from the \textit{Planck} mission, and we take the opportunity to thank ESA/\textit{Planck} and the \textit{Planck} collaboration for the same; (3) the \textit {Herschel} SPIRE images from \textit {Herschel} Science Archive (HSA). \textit{Herschel} is an ESA space observatory with science instruments provided by European-led Principal Investigator consortia and with important participation from NASA. C.W.L is supported by the Basic Science Research Program (2019R1A2C1010851) through the National Research Foundation of Korea. Some of the results in this paper have been derived using the healpy and HEALPix package. This paper is based on one of the chapters of thesis by E. Sharma on "Investigation of the evolution of dark clouds" (\url{http://hdl.handle.net/2248/7861}).

\section*{Data Availability}
The Gaia distances of the stars are available at \url{https://vizier.u-strasbg.fr/viz-bin/VizieR-3?-source=I/350/gaiaedr3}. The Herschel maps are available at Herschel science archive (\url{http://archives.esac.esa.int/hsa/whsa/}). The optical polarization data is given in the appendices below.



\bibliographystyle{mnras}
\bibliography{example} 

\begin{thebibliography}{}
\makeatletter
\relax
\def\mn@urlcharsother{\let\do\@makeother \do\$\do\&\do\#\do\^\do\_\do\%\do\~}
\def\mn@doi{\begingroup\mn@urlcharsother \@ifnextchar [ {\mn@doi@}
  {\mn@doi@[]}}
\def\mn@doi@[#1]#2{\def\@tempa{#1}\ifx\@tempa\@empty \href
  {http://dx.doi.org/#2} {doi:#2}\else \href {http://dx.doi.org/#2} {#1}\fi
  \endgroup}
\def\mn@eprint#1#2{\mn@eprint@#1:#2::\@nil}
\def\mn@eprint@arXiv#1{\href {http://arxiv.org/abs/#1} {{\tt arXiv:#1}}}
\def\mn@eprint@dblp#1{\href {http://dblp.uni-trier.de/rec/bibtex/#1.xml}
  {dblp:#1}}
\def\mn@eprint@#1:#2:#3:#4\@nil{\def\@tempa {#1}\def\@tempb {#2}\def\@tempc
  {#3}\ifx \@tempc \@empty \let \@tempc \@tempb \let \@tempb \@tempa \fi \ifx
  \@tempb \@empty \def\@tempb {arXiv}\fi \@ifundefined
  {mn@eprint@\@tempb}{\@tempb:\@tempc}{\expandafter \expandafter \csname
  mn@eprint@\@tempb\endcsname \expandafter{\@tempc}}}

\bibitem[\protect\citeauthoryear{{Alves}, {Franco}  \& {Girart}}{{Alves}
  et~al.}{2008}]{2008A&A...486L..13A}
{Alves} F.~O.,  {Franco} G.~A.~P.,   {Girart} J.~M.,  2008, \mn@doi [\aap]
  {10.1051/0004-6361:200810091}, \href
  {http://adsabs.harvard.edu/abs/2008A&A...486L..13A} {486, L13}

\bibitem[\protect\citeauthoryear{{Alves}, {Frau}, {Girart}, {Franco}, {Santos}
  \& {Wiesemeyer}}{{Alves} et~al.}{2014}]{2014A&A...569L...1A}
{Alves} F.~O.,  {Frau} P.,  {Girart} J.~M.,  {Franco} G.~A.~P.,  {Santos}
  F.~P.,   {Wiesemeyer} H.,  2014, \mn@doi [\aap]
  {10.1051/0004-6361/201424678}, \href
  {http://adsabs.harvard.edu/abs/2014A&A...569L...1A} {569, L1}

\bibitem[\protect\citeauthoryear{{Andr{\'e}} et~al.,}{{Andr{\'e}}
  et~al.}{2010}]{2010A&A...518L.102A}
{Andr{\'e}} P.,  et~al., 2010, \mn@doi [\aap] {10.1051/0004-6361/201014666},
  \href {http://adsabs.harvard.edu/abs/2010A&A...518L.102A} {518, L102}

\bibitem[\protect\citeauthoryear{{Bailer-Jones}, {Rybizki}, {Fouesneau},
  {Mantelet}  \& {Andrae}}{{Bailer-Jones} et~al.}{2018}]{2018AJ....156...58B}
{Bailer-Jones} C.~A.~L.,  {Rybizki} J.,  {Fouesneau} M.,  {Mantelet} G.,
  {Andrae} R.,  2018, \mn@doi [\aj] {10.3847/1538-3881/aacb21}, \href
  {http://adsabs.harvard.edu/abs/2018AJ....156...58B} {156, 58}

\bibitem[\protect\citeauthoryear{{Bailer-Jones}, {Rybizki}, {Fouesneau},
  {Demleitner}  \& {Andrae}}{{Bailer-Jones} et~al.}{2021}]{2021AJ....161..147B}
{Bailer-Jones} C.~A.~L.,  {Rybizki} J.,  {Fouesneau} M.,  {Demleitner} M.,
  {Andrae} R.,  2021, \mn@doi [\aj] {10.3847/1538-3881/abd806}, \href
  {https://ui.adsabs.harvard.edu/abs/2021AJ....161..147B} {161, 147}

\bibitem[\protect\citeauthoryear{{Bailey}, {Basu}  \& {Caselli}}{{Bailey}
  et~al.}{2017}]{2017A&A...601A..18B}
{Bailey} N.~D.,  {Basu} S.,   {Caselli} P.,  2017, \mn@doi [\aap]
  {10.1051/0004-6361/201628273}, \href
  {https://ui.adsabs.harvard.edu/abs/2017A&A...601A..18B} {601, A18}

\bibitem[\protect\citeauthoryear{{Ballesteros-Paredes}, {Klessen}, {Mac Low}
  \& {Vazquez-Semadeni}}{{Ballesteros-Paredes}
  et~al.}{2007}]{2007prpl.conf...63B}
{Ballesteros-Paredes} J.,  {Klessen} R.~S.,  {Mac Low} M.~M.,
  {Vazquez-Semadeni} E.,  2007, in {Reipurth} B.,  {Jewitt} D.,   {Keil} K.,
  eds, Protostars and Planets V. p.~63 (\mn@eprint {arXiv} {astro-ph/0603357})

\bibitem[\protect\citeauthoryear{{Banerjee}, {V{\'a}zquez-Semadeni},
  {Hennebelle}  \& {Klessen}}{{Banerjee} et~al.}{2009}]{2009MNRAS.398.1082B}
{Banerjee} R.,  {V{\'a}zquez-Semadeni} E.,  {Hennebelle} P.,   {Klessen} R.~S.,
   2009, \mn@doi [\mnras] {10.1111/j.1365-2966.2009.15115.x}, \href
  {https://ui.adsabs.harvard.edu/abs/2009MNRAS.398.1082B} {398, 1082}

\bibitem[\protect\citeauthoryear{{Basu} \& {Mouschovias}}{{Basu} \&
  {Mouschovias}}{1995}]{1995ApJ...453..271B}
{Basu} S.,  {Mouschovias} T.~C.,  1995, \mn@doi [\apj] {10.1086/176387}, \href
  {https://ui.adsabs.harvard.edu/abs/1995ApJ...453..271B} {453, 271}

\bibitem[\protect\citeauthoryear{{Benson} \& {Myers}}{{Benson} \&
  {Myers}}{1983}]{1983ApJ...270..589B}
{Benson} P.~J.,  {Myers} P.~C.,  1983, \mn@doi [\apj] {10.1086/161151}, \href
  {http://adsabs.harvard.edu/abs/1983ApJ...270..589B} {270, 589}

\bibitem[\protect\citeauthoryear{{Benson} \& {Myers}}{{Benson} \&
  {Myers}}{1989}]{1989ApJS...71...89B}
{Benson} P.~J.,  {Myers} P.~C.,  1989, \mn@doi [\apjs] {10.1086/191365}, \href
  {http://adsabs.harvard.edu/abs/1989ApJS...71...89B} {71, 89}

\bibitem[\protect\citeauthoryear{{Bizzocchi}, {Caselli}, {Spezzano}  \&
  {Leonardo}}{{Bizzocchi} et~al.}{2014}]{2014A&A...569A..27B}
{Bizzocchi} L.,  {Caselli} P.,  {Spezzano} S.,   {Leonardo} E.,  2014, \mn@doi
  [\aap] {10.1051/0004-6361/201423858}, \href
  {https://ui.adsabs.harvard.edu/abs/2014A&A...569A..27B} {569, A27}

\bibitem[\protect\citeauthoryear{{Bohlin}, {Savage}  \& {Drake}}{{Bohlin}
  et~al.}{1978}]{1978ApJ...224..132B}
{Bohlin} R.~C.,  {Savage} B.~D.,   {Drake} J.~F.,  1978, \mn@doi [\apj]
  {10.1086/156357}, \href {http://adsabs.harvard.edu/abs/1978ApJ...224..132B}
  {224, 132}

\bibitem[\protect\citeauthoryear{{Chandrasekhar} \& {Fermi}}{{Chandrasekhar} \&
  {Fermi}}{1953}]{1953ApJ...118..113C}
{Chandrasekhar} S.,  {Fermi} E.,  1953, \mn@doi [\apj] {10.1086/145731}, \href
  {http://adsabs.harvard.edu/abs/1953ApJ...118..113C} {118, 113}

\bibitem[\protect\citeauthoryear{{Chen} \& {Ostriker}}{{Chen} \&
  {Ostriker}}{2014}]{2014ApJ...785...69C}
{Chen} C.-Y.,  {Ostriker} E.~C.,  2014, \mn@doi [\apj]
  {10.1088/0004-637X/785/1/69}, \href
  {https://ui.adsabs.harvard.edu/abs/2014ApJ...785...69C} {785, 69}

\bibitem[\protect\citeauthoryear{{Chen} \& {Ostriker}}{{Chen} \&
  {Ostriker}}{2015}]{2015ApJ...810..126C}
{Chen} C.-Y.,  {Ostriker} E.~C.,  2015, \mn@doi [\apj]
  {10.1088/0004-637X/810/2/126}, \href
  {https://ui.adsabs.harvard.edu/abs/2015ApJ...810..126C} {810, 126}

\bibitem[\protect\citeauthoryear{{Chen} \& {Ostriker}}{{Chen} \&
  {Ostriker}}{2018}]{2018ApJ...865...34C}
{Chen} C.-Y.,  {Ostriker} E.~C.,  2018, \mn@doi [\apj]
  {10.3847/1538-4357/aad905}, \href
  {https://ui.adsabs.harvard.edu/abs/2018ApJ...865...34C} {865, 34}

\bibitem[\protect\citeauthoryear{{Chen} et~al.,}{{Chen}
  et~al.}{2020}]{2020MNRAS.494.1971C}
{Chen} C.-Y.,  et~al., 2020, \mn@doi [\mnras] {10.1093/mnras/staa835}, \href
  {https://ui.adsabs.harvard.edu/abs/2020MNRAS.494.1971C} {494, 1971}

\bibitem[\protect\citeauthoryear{{Clemens}}{{Clemens}}{2012}]{2012ApJ...748...18C}
{Clemens} D.~P.,  2012, \mn@doi [\apj] {10.1088/0004-637X/748/1/18}, \href
  {http://cdsads.u-strasbg.fr/abs/2012ApJ...748...18C} {748, 18}

\bibitem[\protect\citeauthoryear{{Commer{\c{c}}on}, {Hennebelle}  \&
  {Henning}}{{Commer{\c{c}}on} et~al.}{2011}]{2011ApJ...742L...9C}
{Commer{\c{c}}on} B.,  {Hennebelle} P.,   {Henning} T.,  2011, \mn@doi [\apjl]
  {10.1088/2041-8205/742/1/L9}, \href
  {https://ui.adsabs.harvard.edu/abs/2011ApJ...742L...9C} {742, L9}

\bibitem[\protect\citeauthoryear{{Crutcher}}{{Crutcher}}{2004}]{2004Ap&SS.292..225C}
{Crutcher} R.~M.,  2004, \mn@doi [\apss] {10.1023/B:ASTR.0000045021.42255.95},
  \href {http://adsabs.harvard.edu/abs/2004Ap&SS.292..225C} {292, 225}

\bibitem[\protect\citeauthoryear{{Crutcher}}{{Crutcher}}{2012}]{2012ARA&A..50...29C}
{Crutcher} R.~M.,  2012, \mn@doi [\araa] {10.1146/annurev-astro-081811-125514},
  \href {http://adsabs.harvard.edu/abs/2012ARA&A..50...29C} {50, 29}

\bibitem[\protect\citeauthoryear{{Davis} \& {Greenstein}}{{Davis} \&
  {Greenstein}}{1951}]{1951ApJ...114..206D}
{Davis} Jr. L.,  {Greenstein} J.~L.,  1951, \mn@doi [\apj] {10.1086/145464},
  \href {http://adsabs.harvard.edu/abs/1951ApJ...114..206D} {114, 206}

\bibitem[\protect\citeauthoryear{{Elias}}{{Elias}}{1978}]{1978ApJ...224..857E}
{Elias} J.~H.,  1978, \mn@doi [\apj] {10.1086/156436}, \href
  {https://ui.adsabs.harvard.edu/abs/1978ApJ...224..857E} {224, 857}

\bibitem[\protect\citeauthoryear{{Franco}, {Alves}  \& {Girart}}{{Franco}
  et~al.}{2010}]{2010ApJ...723..146F}
{Franco} G.~A.~P.,  {Alves} F.~O.,   {Girart} J.~M.,  2010, \mn@doi [\apj]
  {10.1088/0004-637X/723/1/146}, \href
  {http://adsabs.harvard.edu/abs/2010ApJ...723..146F} {723, 146}

\bibitem[\protect\citeauthoryear{{Gaia Collaboration} et~al.,}{{Gaia
  Collaboration} et~al.}{2021}]{2021A&A...649A...1G}
{Gaia Collaboration} et~al., 2021, \mn@doi [\aap]
  {10.1051/0004-6361/202039657}, \href
  {https://ui.adsabs.harvard.edu/abs/2021A&A...649A...1G} {649, A1}

\bibitem[\protect\citeauthoryear{{Goodman}, {Bastien}, {Menard}  \&
  {Myers}}{{Goodman} et~al.}{1990}]{1990ApJ...359..363G}
{Goodman} A.~A.,  {Bastien} P.,  {Menard} F.,   {Myers} P.~C.,  1990, \mn@doi
  [\apj] {10.1086/169070}, \href
  {http://adsabs.harvard.edu/abs/1990ApJ...359..363G} {359, 363}

\bibitem[\protect\citeauthoryear{{G{\'o}rski}, {Hivon}, {Banday}, {Wandelt},
  {Hansen}, {Reinecke}  \& {Bartelmann}}{{G{\'o}rski}
  et~al.}{2005}]{2005ApJ...622..759G}
{G{\'o}rski} K.~M.,  {Hivon} E.,  {Banday} A.~J.,  {Wandelt} B.~D.,  {Hansen}
  F.~K.,  {Reinecke} M.,   {Bartelmann} M.,  2005, \mn@doi [\apj]
  {10.1086/427976}, \href {http://adsabs.harvard.edu/abs/2005ApJ...622..759G}
  {622, 759}

\bibitem[\protect\citeauthoryear{{Gu} \& {Li}}{{Gu} \&
  {Li}}{2019}]{2019ApJ...871L..15G}
{Gu} Q.,  {Li} H.-b.,  2019, \mn@doi [\apjl] {10.3847/2041-8213/aafdb1}, \href
  {https://ui.adsabs.harvard.edu/abs/2019ApJ...871L..15G} {871, L15}

\bibitem[\protect\citeauthoryear{{Hacar} \& {Tafalla}}{{Hacar} \&
  {Tafalla}}{2011}]{2011A&A...533A..34H}
{Hacar} A.,  {Tafalla} M.,  2011, \mn@doi [\aap] {10.1051/0004-6361/201117039},
  \href {http://adsabs.harvard.edu/abs/2011A&A...533A..34H} {533, A34}

\bibitem[\protect\citeauthoryear{{Hatchell}, {Richer}, {Fuller}, {Qualtrough},
  {Ladd}  \& {Chandler}}{{Hatchell} et~al.}{2005}]{2005A&A...440..151H}
{Hatchell} J.,  {Richer} J.~S.,  {Fuller} G.~A.,  {Qualtrough} C.~J.,  {Ladd}
  E.~F.,   {Chandler} C.~J.,  2005, \mn@doi [\aap]
  {10.1051/0004-6361:20041836}, \href
  {https://ui.adsabs.harvard.edu/abs/2005A&A...440..151H} {440, 151}

\bibitem[\protect\citeauthoryear{{Heitsch}, {Stone}  \& {Hartmann}}{{Heitsch}
  et~al.}{2009}]{2009ApJ...695..248H}
{Heitsch} F.,  {Stone} J.~M.,   {Hartmann} L.~W.,  2009, \mn@doi [\apj]
  {10.1088/0004-637X/695/1/248}, \href
  {http://adsabs.harvard.edu/abs/2009ApJ...695..248H} {695, 248}

\bibitem[\protect\citeauthoryear{{Heyer}, {Vrba}, {Snell}, {Schloerb}, {Strom},
  {Goldsmith}  \& {Strom}}{{Heyer} et~al.}{1987}]{Heyeretal1987}
{Heyer} M.~H.,  {Vrba} F.~J.,  {Snell} R.~L.,  {Schloerb} F.~P.,  {Strom}
  S.~E.,  {Goldsmith} P.~F.,   {Strom} K.~M.,  1987, \mn@doi [\apj]
  {10.1086/165678}, \href {http://adsabs.harvard.edu/abs/1987ApJ...321..855H}
  {321, 855}

\bibitem[\protect\citeauthoryear{{Heyer}, {Gong}, {Ostriker}  \&
  {Brunt}}{{Heyer} et~al.}{2008}]{2008ApJ...680..420H}
{Heyer} M.,  {Gong} H.,  {Ostriker} E.,   {Brunt} C.,  2008, \mn@doi [\apj]
  {10.1086/587510}, \href
  {https://ui.adsabs.harvard.edu/abs/2008ApJ...680..420H} {680, 420}

\bibitem[\protect\citeauthoryear{{Hiltner}}{{Hiltner}}{1949}]{1949Natur.163..283H}
{Hiltner} W.~A.,  1949, \mn@doi [\nat] {10.1038/163283a0}, \href
  {http://adsabs.harvard.edu/abs/1949Natur.163..283H} {163, 283}

\bibitem[\protect\citeauthoryear{{Hirota}, {Ito}  \& {Yamamoto}}{{Hirota}
  et~al.}{2002}]{2002ApJ...565..359H}
{Hirota} T.,  {Ito} T.,   {Yamamoto} S.,  2002, \mn@doi [\apj]
  {10.1086/324476}, \href
  {https://ui.adsabs.harvard.edu/abs/2002ApJ...565..359H} {565, 359}

\bibitem[\protect\citeauthoryear{{Inutsuka} \& {Miyama}}{{Inutsuka} \&
  {Miyama}}{1992}]{1992ApJ...388..392I}
{Inutsuka} S.-I.,  {Miyama} S.~M.,  1992, \mn@doi [\apj] {10.1086/171162},
  \href {https://ui.adsabs.harvard.edu/abs/1992ApJ...388..392I} {388, 392}

\bibitem[\protect\citeauthoryear{{Jow}, {Hill}, {Scott}, {Soler}, {Martin},
  {Devlin}, {Fissel}  \& {Poidevin}}{{Jow} et~al.}{2018}]{2018MNRAS.474.1018J}
{Jow} D.~L.,  {Hill} R.,  {Scott} D.,  {Soler} J.~D.,  {Martin} P.~G.,
  {Devlin} M.~J.,  {Fissel} L.~M.,   {Poidevin} F.,  2018, \mn@doi [\mnras]
  {10.1093/mnras/stx2736}, \href
  {https://ui.adsabs.harvard.edu/abs/2018MNRAS.474.1018J} {474, 1018}

\bibitem[\protect\citeauthoryear{{Karoly}, {Soam}, {Andersson}, {Coud{\'e}},
  {Bastien}, {Vaillancourt}  \& {Lee}}{{Karoly}
  et~al.}{2020}]{2020ApJ...900..181K}
{Karoly} J.,  {Soam} A.,  {Andersson} B.~G.,  {Coud{\'e}} S.,  {Bastien} P.,
  {Vaillancourt} J.~E.,   {Lee} C.~W.,  2020, \mn@doi [\apj]
  {10.3847/1538-4357/abad37}, \href
  {https://ui.adsabs.harvard.edu/abs/2020ApJ...900..181K} {900, 181}

\bibitem[\protect\citeauthoryear{{Kawada} et~al.,}{{Kawada}
  et~al.}{2007}]{2007PASJ...59S.389K}
{Kawada} M.,  et~al., 2007, \mn@doi [\pasj] {10.1093/pasj/59.sp2.S389}, \href
  {https://ui.adsabs.harvard.edu/abs/2007PASJ...59S.389K} {59, S389}

\bibitem[\protect\citeauthoryear{{Kim}, {Lee}, {Gopinathan}, {Jeong}  \&
  {Kim}}{{Kim} et~al.}{2016}]{2016ApJ...824...85K}
{Kim} G.,  {Lee} C.~W.,  {Gopinathan} M.,  {Jeong} W.-S.,   {Kim} M.-R.,  2016,
  \mn@doi [\apj] {10.3847/0004-637X/824/2/85}, \href
  {https://ui.adsabs.harvard.edu/abs/2016ApJ...824...85K} {824, 85}

\bibitem[\protect\citeauthoryear{{Kirk}, {Ward-Thompson}  \& {Crutcher}}{{Kirk}
  et~al.}{2006}]{2006MNRAS.369.1445K}
{Kirk} J.~M.,  {Ward-Thompson} D.,   {Crutcher} R.~M.,  2006, \mn@doi [\mnras]
  {10.1111/j.1365-2966.2006.10392.x}, \href
  {http://cdsads.u-strasbg.fr/abs/2006MNRAS.369.1445K} {369, 1445}

\bibitem[\protect\citeauthoryear{{Klessen} \& {Hennebelle}}{{Klessen} \&
  {Hennebelle}}{2010}]{2010A&A...520A..17K}
{Klessen} R.~S.,  {Hennebelle} P.,  2010, \mn@doi [\aap]
  {10.1051/0004-6361/200913780}, \href
  {https://ui.adsabs.harvard.edu/abs/2010A&A...520A..17K} {520, A17}

\bibitem[\protect\citeauthoryear{{K{\"o}rtgen} \& {Banerjee}}{{K{\"o}rtgen} \&
  {Banerjee}}{2015}]{2015MNRAS.451.3340K}
{K{\"o}rtgen} B.,  {Banerjee} R.,  2015, \mn@doi [\mnras]
  {10.1093/mnras/stv1200}, \href
  {https://ui.adsabs.harvard.edu/abs/2015MNRAS.451.3340K} {451, 3340}

\bibitem[\protect\citeauthoryear{{Kudoh} \& {Basu}}{{Kudoh} \&
  {Basu}}{2011}]{2011ApJ...728..123K}
{Kudoh} T.,  {Basu} S.,  2011, \mn@doi [\apj] {10.1088/0004-637X/728/2/123},
  \href {https://ui.adsabs.harvard.edu/abs/2011ApJ...728..123K} {728, 123}

\bibitem[\protect\citeauthoryear{{Launhardt} et~al.,}{{Launhardt}
  et~al.}{2013}]{2013A&A...551A..98L}
{Launhardt} R.,  et~al., 2013, \mn@doi [\aap] {10.1051/0004-6361/201220477},
  \href {http://adsabs.harvard.edu/abs/2013A&A...551A..98L} {551, A98}

\bibitem[\protect\citeauthoryear{{Lazarian}}{{Lazarian}}{2007}]{2007JQSRT.106..225L}
{Lazarian} A.,  2007, \mn@doi [\jqsrt] {10.1016/j.jqsrt.2007.01.038}, \href
  {http://adsabs.harvard.edu/abs/2007JQSRT.106..225L} {106, 225}

\bibitem[\protect\citeauthoryear{{Lee} \& {Myers}}{{Lee} \&
  {Myers}}{1999}]{1999ApJS..123..233L}
{Lee} C.~W.,  {Myers} P.~C.,  1999, \mn@doi [\apjs] {10.1086/313234}, \href
  {http://adsabs.harvard.edu/abs/1999ApJS..123..233L} {123, 233}

\bibitem[\protect\citeauthoryear{{Lee} \& {Myers}}{{Lee} \&
  {Myers}}{2011}]{2011ApJ...734...60L}
{Lee} C.~W.,  {Myers} P.~C.,  2011, \mn@doi [\apj]
  {10.1088/0004-637X/734/1/60}, \href
  {http://adsabs.harvard.edu/abs/2011ApJ...734...60L} {734, 60}

\bibitem[\protect\citeauthoryear{{Lee}, {Myers}  \& {Tafalla}}{{Lee}
  et~al.}{1999}]{1999ApJ...526..788L}
{Lee} C.~W.,  {Myers} P.~C.,   {Tafalla} M.,  1999, \mn@doi [\apj]
  {10.1086/308027}, \href
  {https://ui.adsabs.harvard.edu/abs/1999ApJ...526..788L} {526, 788}

\bibitem[\protect\citeauthoryear{{Lee}, {Evans}, {Shirley}  \&
  {Tatematsu}}{{Lee} et~al.}{2003}]{2003ApJ...583..789L}
{Lee} J.-E.,  {Evans} Neal~J. I.,  {Shirley} Y.~L.,   {Tatematsu} K.,  2003,
  \mn@doi [\apj] {10.1086/345428}, \href
  {https://ui.adsabs.harvard.edu/abs/2003ApJ...583..789L} {583, 789}

\bibitem[\protect\citeauthoryear{{Lee}, {Myers}  \& {Plume}}{{Lee}
  et~al.}{2004}]{2004ApJS..153..523L}
{Lee} C.~W.,  {Myers} P.~C.,   {Plume} R.,  2004, \mn@doi [\apjs]
  {10.1086/421996}, \href
  {https://ui.adsabs.harvard.edu/abs/2004ApJS..153..523L} {153, 523}

\bibitem[\protect\citeauthoryear{{Li}, {Blundell}, {Hedden}, {Kawamura},
  {Paine}  \& {Tong}}{{Li} et~al.}{2011}]{2011MNRAS.411.2067L}
{Li} H.-B.,  {Blundell} R.,  {Hedden} A.,  {Kawamura} J.,  {Paine} S.,   {Tong}
  E.,  2011, \mn@doi [\mnras] {10.1111/j.1365-2966.2010.17839.x}, \href
  {http://adsabs.harvard.edu/abs/2011MNRAS.411.2067L} {411, 2067}

\bibitem[\protect\citeauthoryear{{Li}, {Goodman}, {Sridharan}, {Houde}, {Li},
  {Novak}  \& {Tang}}{{Li} et~al.}{2014}]{2014prpl.conf..101L}
{Li} H.~B.,  {Goodman} A.,  {Sridharan} T.~K.,  {Houde} M.,  {Li} Z.~Y.,
  {Novak} G.,   {Tang} K.~S.,  2014, in {Beuther} H.,  {Klessen} R.~S.,
  {Dullemond} C.~P.,   {Henning} T.,  eds, Protostars and Planets VI. p.~101
  (\mn@eprint {arXiv} {1404.2024}),
  \mn@doi{10.2458/azu_uapress_9780816531240-ch005}

\bibitem[\protect\citeauthoryear{{Li}, {Jiang}, {Fan}, {Gu}  \& {Zhang}}{{Li}
  et~al.}{2017}]{2017NatAs...1E.158L}
{Li} H.-B.,  {Jiang} H.,  {Fan} X.,  {Gu} Q.,   {Zhang} Y.,  2017, \mn@doi
  [Nature Astronomy] {10.1038/s41550-017-0158}, \href
  {https://ui.adsabs.harvard.edu/abs/2017NatAs...1E.158L} {1, 0158}

\bibitem[\protect\citeauthoryear{{Lindegren} et~al.,}{{Lindegren}
  et~al.}{2021}]{2021A&A...649A...2L}
{Lindegren} L.,  et~al., 2021, \mn@doi [\aap] {10.1051/0004-6361/202039709},
  \href {https://ui.adsabs.harvard.edu/abs/2021A&A...649A...2L} {649, A2}

\bibitem[\protect\citeauthoryear{{McKee} \& {Ostriker}}{{McKee} \&
  {Ostriker}}{2007}]{2007ARA&A..45..565M}
{McKee} C.~F.,  {Ostriker} E.~C.,  2007, \mn@doi [\araa]
  {10.1146/annurev.astro.45.051806.110602}, \href
  {http://adsabs.harvard.edu/abs/2007ARA&A..45..565M} {45, 565}

\bibitem[\protect\citeauthoryear{{Miville-Desch{\^e}nes}
  et~al.,}{{Miville-Desch{\^e}nes} et~al.}{2017}]{2017A&A...599A.109M}
{Miville-Desch{\^e}nes} M.~A.,  et~al., 2017, \mn@doi [\aap]
  {10.1051/0004-6361/201628289}, \href
  {https://ui.adsabs.harvard.edu/abs/2017A&A...599A.109M} {599, A109}

\bibitem[\protect\citeauthoryear{{Montillaud} et~al.,}{{Montillaud}
  et~al.}{2015}]{2015A&A...584A..92M}
{Montillaud} J.,  et~al., 2015, \mn@doi [\aap] {10.1051/0004-6361/201424063},
  \href {https://ui.adsabs.harvard.edu/abs/2015A&A...584A..92M} {584, A92}

\bibitem[\protect\citeauthoryear{{Mouschovias} \& {Ciolek}}{{Mouschovias} \&
  {Ciolek}}{1999}]{1999ASIC..540..305M}
{Mouschovias} T.~C.,  {Ciolek} G.~E.,  1999, in {Lada} C.~J.,  {Kylafis} N.~D.,
   eds,  NATO Advanced Study Institute (ASI) Series C Vol. 540, The Origin of
  Stars and Planetary Systems. p.~305

\bibitem[\protect\citeauthoryear{{Mouschovias} \& {Spitzer}}{{Mouschovias} \&
  {Spitzer}}{1976}]{1976ApJ...210..326M}
{Mouschovias} T.~C.,  {Spitzer} Jr. L.,  1976, \mn@doi [\apj] {10.1086/154835},
  \href {http://adsabs.harvard.edu/abs/1976ApJ...210..326M} {210, 326}

\bibitem[\protect\citeauthoryear{{Myers} \& {Benson}}{{Myers} \&
  {Benson}}{1983}]{1983RMxAA...7..238M}
{Myers} P.~C.,  {Benson} P.~J.,  1983, \rmxaa, \href
  {http://adsabs.harvard.edu/abs/1983RMxAA...7..238M} {7, 238}

\bibitem[\protect\citeauthoryear{{Myers}, {Linke}  \& {Benson}}{{Myers}
  et~al.}{1983}]{1983ApJ...264..517M}
{Myers} P.~C.,  {Linke} R.~A.,   {Benson} P.~J.,  1983, \mn@doi [\apj]
  {10.1086/160619}, \href
  {https://ui.adsabs.harvard.edu/abs/1983ApJ...264..517M} {264, 517}

\bibitem[\protect\citeauthoryear{{Nakamura} \& {Li}}{{Nakamura} \&
  {Li}}{2011}]{2011ApJ...740...36N}
{Nakamura} F.,  {Li} Z.-Y.,  2011, \mn@doi [\apj] {10.1088/0004-637X/740/1/36},
  \href {https://ui.adsabs.harvard.edu/abs/2011ApJ...740...36N} {740, 36}

\bibitem[\protect\citeauthoryear{{Nakano} \& {Nakamura}}{{Nakano} \&
  {Nakamura}}{1978}]{1978PASJ...30..671N}
{Nakano} T.,  {Nakamura} T.,  1978, \pasj, \href
  {http://adsabs.harvard.edu/abs/1978PASJ...30..671N} {30, 671}

\bibitem[\protect\citeauthoryear{{Neha}, {Maheswar}, {Soam}, {Lee}  \&
  {Tej}}{{Neha} et~al.}{2016}]{2016A&A...588A..45N}
{Neha} S.,  {Maheswar} G.,  {Soam} A.,  {Lee} C.~W.,   {Tej} A.,  2016, \mn@doi
  [\aap] {10.1051/0004-6361/201526845}, \href
  {https://ui.adsabs.harvard.edu/abs/2016A&A...588A..45N} {588, A45}

\bibitem[\protect\citeauthoryear{{Neha}, {Maheswar}, {Soam}  \& {Lee}}{{Neha}
  et~al.}{2018}]{2018MNRAS.476.4442N}
{Neha} S.,  {Maheswar} G.,  {Soam} A.,   {Lee} C.~W.,  2018, \mn@doi [\mnras]
  {10.1093/mnras/sty485}, \href
  {https://ui.adsabs.harvard.edu/abs/2018MNRAS.476.4442N} {476, 4442}

\bibitem[\protect\citeauthoryear{{Nielbock} et~al.,}{{Nielbock}
  et~al.}{2012}]{2012A&A...547A..11N}
{Nielbock} M.,  et~al., 2012, \mn@doi [\aap] {10.1051/0004-6361/201219139},
  \href {https://ui.adsabs.harvard.edu/abs/2012A&A...547A..11N} {547, A11}

\bibitem[\protect\citeauthoryear{{Nutter}, {Stamatellos}  \&
  {Ward-Thompson}}{{Nutter} et~al.}{2009}]{2009MNRAS.396.1851N}
{Nutter} D.,  {Stamatellos} D.,   {Ward-Thompson} D.,  2009, \mn@doi [\mnras]
  {10.1111/j.1365-2966.2009.14837.x}, \href
  {https://ui.adsabs.harvard.edu/abs/2009MNRAS.396.1851N} {396, 1851}

\bibitem[\protect\citeauthoryear{{Obayashi}, {Kun}, {Sato}, {Yonekura}  \&
  {Fukui}}{{Obayashi} et~al.}{1998}]{1998AJ....115..274O}
{Obayashi} A.,  {Kun} M.,  {Sato} F.,  {Yonekura} Y.,   {Fukui} Y.,  1998,
  \mn@doi [\aj] {10.1086/300172}, \href
  {https://ui.adsabs.harvard.edu/abs/1998AJ....115..274O} {115, 274}

\bibitem[\protect\citeauthoryear{{Palau} et~al.,}{{Palau}
  et~al.}{2013}]{2013ApJ...762..120P}
{Palau} A.,  et~al., 2013, \mn@doi [\apj] {10.1088/0004-637X/762/2/120}, \href
  {https://ui.adsabs.harvard.edu/abs/2013ApJ...762..120P} {762, 120}

\bibitem[\protect\citeauthoryear{{Park}, {Lee}  \& {Myers}}{{Park}
  et~al.}{2004}]{2004ApJS..152...81P}
{Park} Y.-S.,  {Lee} C.~W.,   {Myers} P.~C.,  2004, \mn@doi [\apjs]
  {10.1086/382506}, \href {http://adsabs.harvard.edu/abs/2004ApJS..152...81P}
  {152, 81}

\bibitem[\protect\citeauthoryear{{Planck Collaboration} et~al.,}{{Planck
  Collaboration} et~al.}{2015}]{2015A&A...576A.104P}
{Planck Collaboration} et~al., 2015, \mn@doi [\aap]
  {10.1051/0004-6361/201424082}, \href
  {http://adsabs.harvard.edu/abs/2015A&A...576A.104P} {576, A104}

\bibitem[\protect\citeauthoryear{{Planck Collaboration} et~al.,}{{Planck
  Collaboration} et~al.}{2016}]{2016A&A...586A.138P}
{Planck Collaboration} et~al., 2016, \mn@doi [\aap]
  {10.1051/0004-6361/201525896}, \href
  {http://adsabs.harvard.edu/abs/2016A&A...586A.138P} {586, A138}

\bibitem[\protect\citeauthoryear{{Poidevin} et~al.,}{{Poidevin}
  et~al.}{2014}]{2014ApJ...791...43P}
{Poidevin} F.,  et~al., 2014, \mn@doi [\apj] {10.1088/0004-637X/791/1/43},
  \href {https://ui.adsabs.harvard.edu/abs/2014ApJ...791...43P} {791, 43}

\bibitem[\protect\citeauthoryear{{Rautela}, {Joshi}  \& {Pandey}}{{Rautela}
  et~al.}{2004}]{2004BASI...32..159R}
{Rautela} B.~S.,  {Joshi} G.~C.,   {Pandey} J.~C.,  2004, Bulletin of the
  Astronomical Society of India, \href
  {https://ui.adsabs.harvard.edu/abs/2004BASI...32..159R} {32, 159}

\bibitem[\protect\citeauthoryear{{Rosolowsky}, {Pineda}, {Kauffmann}  \&
  {Goodman}}{{Rosolowsky} et~al.}{2008}]{2008ApJ...679.1338R}
{Rosolowsky} E.~W.,  {Pineda} J.~E.,  {Kauffmann} J.,   {Goodman} A.~A.,  2008,
  \mn@doi [\apj] {10.1086/587685}, \href
  {https://ui.adsabs.harvard.edu/abs/2008ApJ...679.1338R} {679, 1338}

\bibitem[\protect\citeauthoryear{{Saha}, {Gopinathan}, {Sharma}, {Won Lee},
  {Ghosh}  \& {Kim}}{{Saha} et~al.}{2021}]{2021A&A...655A..76S}
{Saha} P.,  {Gopinathan} M.,  {Sharma} E.,  {Won Lee} C.,  {Ghosh} T.,   {Kim}
  S.,  2021, \mn@doi [\aap] {10.1051/0004-6361/202039948}, \href
  {https://ui.adsabs.harvard.edu/abs/2021A&A...655A..76S} {655, A76}

\bibitem[\protect\citeauthoryear{{Schmidt}, {Elston}  \& {Lupie}}{{Schmidt}
  et~al.}{1992}]{1992AJ....104.1563S}
{Schmidt} G.~D.,  {Elston} R.,   {Lupie} O.~L.,  1992, \mn@doi [aj]
  {10.1086/116341}, \href {http://adsabs.harvard.edu/abs/1992AJ....104.1563S}
  {104, 1563}

\bibitem[\protect\citeauthoryear{{Seifried}, {Walch}, {Weis}, {Reissl},
  {Soler}, {Klessen}  \& {Joshi}}{{Seifried}
  et~al.}{2020}]{2020MNRAS.497.4196S}
{Seifried} D.,  {Walch} S.,  {Weis} M.,  {Reissl} S.,  {Soler} J.~D.,
  {Klessen} R.~S.,   {Joshi} P.~R.,  2020, \mn@doi [\mnras]
  {10.1093/mnras/staa2231}, \href
  {https://ui.adsabs.harvard.edu/abs/2020MNRAS.497.4196S} {497, 4196}

\bibitem[\protect\citeauthoryear{{Sharma} et~al.,}{{Sharma}
  et~al.}{2020}]{2020A&A...639A.133S}
{Sharma} E.,  et~al., 2020, \mn@doi [\aap] {10.1051/0004-6361/202037438}, \href
  {https://ui.adsabs.harvard.edu/abs/2020A&A...639A.133S} {639, A133}

\bibitem[\protect\citeauthoryear{{Shu}, {Adams}  \& {Lizano}}{{Shu}
  et~al.}{1987}]{1987ARA&A..25...23S}
{Shu} F.~H.,  {Adams} F.~C.,   {Lizano} S.,  1987, \mn@doi [\araa]
  {10.1146/annurev.aa.25.090187.000323}, \href
  {http://adsabs.harvard.edu/abs/1987ARA&A..25...23S} {25, 23}

\bibitem[\protect\citeauthoryear{{Soam}, {Maheswar}, {Bhatt}, {Lee}  \&
  {Ramaprakash}}{{Soam} et~al.}{2013}]{2013MNRAS.432.1502S}
{Soam} A.,  {Maheswar} G.,  {Bhatt} H.~C.,  {Lee} C.~W.,   {Ramaprakash} A.~N.,
   2013, \mn@doi [\mnras] {10.1093/mnras/stt576}, \href
  {http://adsabs.harvard.edu/abs/2013MNRAS.432.1502S} {432, 1502}

\bibitem[\protect\citeauthoryear{{Soam}, {Maheswar}, {Lee}, {Dib}, {Bhatt},
  {Tamura}  \& {Kim}}{{Soam} et~al.}{2015}]{2015A&A...573A..34S}
{Soam} A.,  {Maheswar} G.,  {Lee} C.~W.,  {Dib} S.,  {Bhatt} H.~C.,  {Tamura}
  M.,   {Kim} G.,  2015, \mn@doi [\aap] {10.1051/0004-6361/201322536}, \href
  {http://adsabs.harvard.edu/abs/2015A&A...573A..34S} {573, A34}

\bibitem[\protect\citeauthoryear{{Soam}, {Lee}, {Maheswar}, {Kim}, {Neha}  \&
  {Kim}}{{Soam} et~al.}{2017}]{2017MNRAS.464.2403S}
{Soam} A.,  {Lee} C.~W.,  {Maheswar} G.,  {Kim} G.,  {Neha} S.,   {Kim} M.-R.,
  2017, \mn@doi [\mnras] {10.1093/mnras/stw2454}, \href
  {https://ui.adsabs.harvard.edu/abs/2017MNRAS.464.2403S} {464, 2403}

\bibitem[\protect\citeauthoryear{{Soam}, {Maheswar}, {Lee}, {Neha}  \&
  {Kim}}{{Soam} et~al.}{2018}]{2018MNRAS.476.4782S}
{Soam} A.,  {Maheswar} G.,  {Lee} C.~W.,  {Neha} S.,   {Kim} K.-T.,  2018,
  \mn@doi [\mnras] {10.1093/mnras/sty517}, \href
  {https://ui.adsabs.harvard.edu/abs/2018MNRAS.476.4782S} {476, 4782}

\bibitem[\protect\citeauthoryear{{Soam} et~al.,}{{Soam}
  et~al.}{2021}]{2021AJ....161..149S}
{Soam} A.,  et~al., 2021, \mn@doi [\aj] {10.3847/1538-3881/abdd3b}, \href
  {https://ui.adsabs.harvard.edu/abs/2021AJ....161..149S} {161, 149}

\bibitem[\protect\citeauthoryear{{Sohn}, {Lee}, {Park}, {Lee}, {Myers}  \&
  {Lee}}{{Sohn} et~al.}{2007}]{2007ApJ...664..928S}
{Sohn} J.,  {Lee} C.~W.,  {Park} Y.-S.,  {Lee} H.~M.,  {Myers} P.~C.,   {Lee}
  Y.,  2007, \mn@doi [\apj] {10.1086/519159}, \href
  {http://adsabs.harvard.edu/abs/2007ApJ...664..928S} {664, 928}

\bibitem[\protect\citeauthoryear{{Soler} \& {Hennebelle}}{{Soler} \&
  {Hennebelle}}{2017}]{2017A&A...607A...2S}
{Soler} J.~D.,  {Hennebelle} P.,  2017, \mn@doi [\aap]
  {10.1051/0004-6361/201731049}, \href
  {https://ui.adsabs.harvard.edu/abs/2017A&A...607A...2S} {607, A2}

\bibitem[\protect\citeauthoryear{{Soler} et~al.,}{{Soler}
  et~al.}{2016}]{2016A&A...596A..93S}
{Soler} J.~D.,  et~al., 2016, \mn@doi [\aap] {10.1051/0004-6361/201628996},
  \href {https://ui.adsabs.harvard.edu/abs/2016A&A...596A..93S} {596, A93}

\bibitem[\protect\citeauthoryear{{Sugitani} et~al.,}{{Sugitani}
  et~al.}{2010}]{2010ApJ...716..299S}
{Sugitani} K.,  et~al., 2010, \mn@doi [\apj] {10.1088/0004-637X/716/1/299},
  \href {https://ui.adsabs.harvard.edu/abs/2010ApJ...716..299S} {716, 299}

\bibitem[\protect\citeauthoryear{{Sugitani} et~al.,}{{Sugitani}
  et~al.}{2011}]{2011ApJ...734...63S}
{Sugitani} K.,  et~al., 2011, \mn@doi [\apj] {10.1088/0004-637X/734/1/63},
  \href {http://adsabs.harvard.edu/abs/2011ApJ...734...63S} {734, 63}

\bibitem[\protect\citeauthoryear{{Tafalla} \& {Santiago}}{{Tafalla} \&
  {Santiago}}{2004}]{2004A&A...414L..53T}
{Tafalla} M.,  {Santiago} J.,  2004, \mn@doi [\aap]
  {10.1051/0004-6361:20031766}, \href
  {http://adsabs.harvard.edu/abs/2004A&A...414L..53T} {414, L53}

\bibitem[\protect\citeauthoryear{{Tafalla}, {Mardones}, {Myers}, {Caselli},
  {Bachiller}  \& {Benson}}{{Tafalla} et~al.}{1998}]{1998ApJ...504..900T}
{Tafalla} M.,  {Mardones} D.,  {Myers} P.~C.,  {Caselli} P.,  {Bachiller} R.,
  {Benson} P.~J.,  1998, \mn@doi [\apj] {10.1086/306115}, \href
  {http://adsabs.harvard.edu/abs/1998ApJ...504..900T} {504, 900}

\bibitem[\protect\citeauthoryear{{V{\'a}zquez-Semadeni}, {Banerjee},
  {G{\'o}mez}, {Hennebelle}, {Duffin}  \& {Klessen}}{{V{\'a}zquez-Semadeni}
  et~al.}{2011}]{2011MNRAS.414.2511V}
{V{\'a}zquez-Semadeni} E.,  {Banerjee} R.,  {G{\'o}mez} G.~C.,  {Hennebelle}
  P.,  {Duffin} D.,   {Klessen} R.~S.,  2011, \mn@doi [\mnras]
  {10.1111/j.1365-2966.2011.18569.x}, \href
  {https://ui.adsabs.harvard.edu/abs/2011MNRAS.414.2511V} {414, 2511}

\bibitem[\protect\citeauthoryear{{Vrba}, {Strom}  \& {Strom}}{{Vrba}
  et~al.}{1976}]{1976AJ.....81..958V}
{Vrba} F.~J.,  {Strom} S.~E.,   {Strom} K.~M.,  1976, \mn@doi [\aj]
  {10.1086/111976}, \href
  {https://ui.adsabs.harvard.edu/abs/1976AJ.....81..958V} {81, 958}

\bibitem[\protect\citeauthoryear{{Ward-Thompson}, {Scott}, {Hills}  \&
  {Andre}}{{Ward-Thompson} et~al.}{1994}]{1994MNRAS.268..276W}
{Ward-Thompson} D.,  {Scott} P.~F.,  {Hills} R.~E.,   {Andre} P.,  1994,
  \mn@doi [\mnras] {10.1093/mnras/268.1.276}, \href
  {https://ui.adsabs.harvard.edu/abs/1994MNRAS.268..276W} {268, 276}

\bibitem[\protect\citeauthoryear{{Ward-Thompson}, {Kirk}, {Crutcher},
  {Greaves}, {Holland}  \& {Andr{\'e}}}{{Ward-Thompson}
  et~al.}{2000}]{2000ApJ...537L.135W}
{Ward-Thompson} D.,  {Kirk} J.~M.,  {Crutcher} R.~M.,  {Greaves} J.~S.,
  {Holland} W.~S.,   {Andr{\'e}} P.,  2000, \mn@doi [\apjl] {10.1086/312764},
  \href {http://cdsads.u-strasbg.fr/abs/2000ApJ...537L.135W} {537, L135}

\bibitem[\protect\citeauthoryear{{Ward-Thompson}, {Sen}, {Kirk}  \&
  {Nutter}}{{Ward-Thompson} et~al.}{2009}]{2009MNRAS.398..394W}
{Ward-Thompson} D.,  {Sen} A.~K.,  {Kirk} J.~M.,   {Nutter} D.,  2009, \mn@doi
  [\mnras] {10.1111/j.1365-2966.2009.15159.x}, \href
  {http://adsabs.harvard.edu/abs/2009MNRAS.398..394W} {398, 394}

\bibitem[\protect\citeauthoryear{{Wu}, {Zhou}  \& {Evans}}{{Wu}
  et~al.}{1992}]{1992ApJ...394..196W}
{Wu} Y.,  {Zhou} S.,   {Evans} Neal~J. I.,  1992, \mn@doi [\apj]
  {10.1086/171571}, \href
  {https://ui.adsabs.harvard.edu/abs/1992ApJ...394..196W} {394, 196}

\bibitem[\protect\citeauthoryear{{Yan}, {Zhang}, {Xu}, {Guo}, {Macquart},
  {Tang}  \& {Walsh}}{{Yan} et~al.}{2019}]{2019A&A...624A...6Y}
{Yan} Q.-Z.,  {Zhang} B.,  {Xu} Y.,  {Guo} S.,  {Macquart} J.-P.,  {Tang}
  Z.-H.,   {Walsh} A.~J.,  2019, \mn@doi [\aap] {10.1051/0004-6361/201834337},
  \href {https://ui.adsabs.harvard.edu/abs/2019A&A...624A...6Y} {624, A6}

\bibitem[\protect\citeauthoryear{Zonca, Singer, Lenz, Reinecke, Rosset, Hivon
  \& Gorski}{Zonca et~al.}{2019}]{Zonca2019}
Zonca A.,  Singer L.,  Lenz D.,  Reinecke M.,  Rosset C.,  Hivon E.,   Gorski
  K.,  2019, \mn@doi [Journal of Open Source Software] {10.21105/joss.01298},
  4, 1298

\bibitem[\protect\citeauthoryear{{Zucker}, {Speagle}, {Schlafly}, {Green},
  {Finkbeiner}, {Goodman}  \& {Alves}}{{Zucker}
  et~al.}{2019}]{2019ApJ...879..125Z}
{Zucker} C.,  {Speagle} J.~S.,  {Schlafly} E.~F.,  {Green} G.~M.,  {Finkbeiner}
  D.~P.,  {Goodman} A.~A.,   {Alves} J.,  2019, \mn@doi [\apj]
  {10.3847/1538-4357/ab2388}, \href
  {https://ui.adsabs.harvard.edu/abs/2019ApJ...879..125Z} {879, 125}

\bibitem[\protect\citeauthoryear{{Zucker}, {Speagle}, {Schlafly}, {Green},
  {Finkbeiner}, {Goodman}  \& {Alves}}{{Zucker}
  et~al.}{2020}]{2020A&A...633A..51Z}
{Zucker} C.,  {Speagle} J.~S.,  {Schlafly} E.~F.,  {Green} G.~M.,  {Finkbeiner}
  D.~P.,  {Goodman} A.,   {Alves} J.,  2020, \mn@doi [\aap]
  {10.1051/0004-6361/201936145}, \href
  {https://ui.adsabs.harvard.edu/abs/2020A&A...633A..51Z} {633, A51}

\makeatother
\end{thebibliography}



\newpage
\appendix
\section{Tables}
\begin{center}
\begin{supertabular}{lllll} \hline
\renewcommand{\arraystretch}{2.0}
ID & RA  & Dec  &  P $\pm$ e$_{p}$    & $\theta_{op}\pm$ e$_{op}$\\
   & ($^{\circ}$)   & ($^{\circ}$)   &  (\%) & ($^{\circ}$)   \\               
\hline
\multicolumn{5}{c}{\bf L1333}\\\hline
1&	 36.635574&	75.557785 &	2.3$\pm$0.1&	137$\pm$1\\
2&	 36.832424&	75.556046 &	8.1$\pm$0.9&	143$\pm$3\\
3&	 36.900402&	75.564400 &	2.9$\pm$0.5&	124$\pm$5\\
4&	 36.706692&	75.534019 &	2.2$\pm$0.1&	129$\pm$2\\
5&	 36.824272&	75.548027 &	3.1$\pm$0.2&	105$\pm$2\\
6&	 36.721371&	75.522621 &	2.3$\pm$0.2&	132$\pm$3\\
7&	 36.851704&	75.535210 &	2.9$\pm$0.3&	129$\pm$3\\
8&	 36.815571&	75.524368 &	4.4$\pm$0.9&	124$\pm$6\\
9&	 36.968765&	75.533386 &	2.4$\pm$0.6&	120$\pm$7\\
10&	 36.673347&	75.487823 &	2.5$\pm$0.8&	123$\pm$8\\
11&	 36.630009&	75.478310 &	2.3$\pm$0.7&	132$\pm$8\\
12&	 36.678047&	75.479797 &	1.3$\pm$0.2&	120$\pm$5\\
13&	 36.798355&	75.495140 &	4.6$\pm$0.9&	123$\pm$6\\
14&	 36.851345&	75.496162 &	2.8$\pm$0.7&	119$\pm$7\\
15&	 36.955891&	75.505592 &	2.7$\pm$0.7&	110$\pm$7\\
16&	 36.878319&	75.489311 &	2.9$\pm$0.2&	124$\pm$2\\
17&	 36.939865&	75.497597 &	4.2$\pm$0.7&	115$\pm$4\\
18&	 36.966125&	75.499939 &	2.6$\pm$0.5&	124$\pm$5\\
19&	 36.497143&	75.416504 &	1.1$\pm$0.3&	89$\pm$7\\
20&	 36.552902&	75.420486 &	1.7$\pm$0.3&	115$\pm$5\\
21&	 36.792648&	75.454453 &	2.2$\pm$0.4&	119$\pm$5\\
22&	 36.801079&	75.450340 &	1.8$\pm$0.3&	105$\pm$5\\
23&	 36.717266&	75.437553 &	3.4$\pm$0.5&	129$\pm$4\\
24&	 36.688969&	75.432884 &	3.0$\pm$0.4&	121$\pm$3\\
25&	 36.687901&	75.425781 &	3.5$\pm$0.5&	135$\pm$4\\
26&	 36.526684&	75.392769 &	2.6$\pm$0.2&	128$\pm$2\\
27&	 36.536018&	75.386826 &	5.4$\pm$1.6&	86$\pm$8\\
28&	 36.536018&	75.386826 &	5.4$\pm$1.6&	86$\pm$8\\
29&	 36.818443&	75.420723 &	1.4$\pm$0.1&	102$\pm$3\\
30&	 36.753849&	75.409576 &	3.7$\pm$0.6&	120$\pm$4\\
31&	 36.633022&	75.388786 &	2.3$\pm$0.5&	124$\pm$6\\
32&	 36.642246&	75.385384 &	3.2$\pm$0.3&	140$\pm$3\\
33&	 36.749104&	75.400932 &	1.9$\pm$0.6&	111$\pm$9\\
34&	 36.802547&	75.408531 &	2.9$\pm$0.7&	122$\pm$7\\
35&	 36.739025&	75.388939 &	2.5$\pm$0.3&	126$\pm$4\\
36&	 36.682495&	75.379448 &	2.4$\pm$0.5&	137$\pm$6\\
37&	 36.665897&	75.374657 &	2.9$\pm$0.6&	138$\pm$6\\
38&	 36.816051&	75.384026 &	2.9$\pm$0.1&	131$\pm$1\\
39&	 36.835003&	75.382202 &	2.4$\pm$0.5&	143$\pm$5\\
40&	 36.828247&	75.380699 &	2.8$\pm$0.2&	143$\pm$2\\
41&	 36.441269&	75.358467 &	1.9$\pm$0.2&	143$\pm$2\\
42&	 36.642090&	75.385460 &	2.5$\pm$0.6&	136$\pm$6\\
43&	 36.464394&	75.341240 &	2.6$\pm$0.3&	141$\pm$3\\
44&	 36.718288&	75.378220 &	1.5$\pm$0.5&	136$\pm$9\\
45&	 36.507980&	75.344017 &	2.0$\pm$0.6&	150$\pm$9\\
46&	 36.482780&	75.326973 &	2.8$\pm$0.6&	178$\pm$5\\
47&	 36.543869&	75.321381 &	1.4$\pm$0.4&	99$\pm$8\\
48&	 36.667274&	75.338501 &	1.6$\pm$0.1&	120$\pm$2\\
49&	 36.642097&	75.334679 &	3.2$\pm$0.4&	144$\pm$3\\
50&	 36.810066&	75.356255 &	2.3$\pm$0.3&	131$\pm$4\\
51&	 36.708378&	75.339081 &	5.7$\pm$0.8&	120$\pm$4\\
52&	 36.821564&	75.353394 &	3.4$\pm$0.7&	140$\pm$6\\
53&	 36.804974&	75.349754 &	1.9$\pm$0.6&	118$\pm$8\\
54&	 36.869255&	75.347252 &	6.7$\pm$0.6&	100$\pm$3\\
55&	 36.892963&	75.334496 &	1.7$\pm$0.5&	90$\pm$8\\
56&	 36.733772&	75.309822 &	2.8$\pm$0.8&	103$\pm$8\\
57&	 36.704391&	75.294312 &	2.4$\pm$0.7&	135$\pm$8\\
58&	 36.889622&	75.319733 &	2.1$\pm$0.4&	134$\pm$5\\
59&	 36.900982&	75.320618 &	3.6$\pm$0.4&	120$\pm$3\\
60&	 36.840511&	75.308975 &	3.2$\pm$0.5&	125$\pm$5\\
61&	 36.778320&	75.293205 &	1.9$\pm$0.2&	123$\pm$3\\
62&	 36.760612&	75.289764 &	2.2$\pm$0.5&	125$\pm$6\\
63&	 36.718964&	75.280144 &	2.4$\pm$0.4&	138$\pm$5\\
64&	 36.160011&	75.387260 &	1.9$\pm$0.6&	2$\pm$8\\
65&	 36.042805&	75.369087 &	1.6$\pm$0.3&	108$\pm$5\\
66&	 36.107712&	75.358253 &	1.1$\pm$0.3&	111$\pm$8\\
67&	 36.175201&	75.364693 &	0.9$\pm$0.2&	107$\pm$7\\
68&	 36.222542&	75.360321 &	1.6$\pm$0.3&	93$\pm$5\\
69&	 36.237003&	75.340622 &	3.2$\pm$0.6&	97$\pm$5\\
70&	 36.180279&	75.328026 &	5.4$\pm$0.8&	138$\pm$4\\
71&	 36.326046&	75.343208 &	1.7$\pm$0.1&	131$\pm$2\\
72&	 36.036568&	75.296249 &	0.8$\pm$0.1&	118$\pm$5\\
73&	 36.382717&	75.343605 &	0.6$\pm$0.1&	132$\pm$4\\
74&	 36.175095&	75.304520 &	2.0$\pm$0.6&	137$\pm$8\\
75&	 36.115463&	75.294716 &	2.9$\pm$0.7&	91$\pm$7\\
76&	 36.272186&	75.297707 &	1.4$\pm$0.3&	110$\pm$6\\
77&	 36.343594&	75.301308 &	8.6$\pm$0.8&	129$\pm$3\\
78&	 36.249485&	75.279053 &	2.0$\pm$0.5&	112$\pm$6\\
79&	 35.723125&	75.319908 &	4.4$\pm$0.7&	17$\pm$5\\
80&	 35.775055&	75.311264 &	3.4$\pm$0.7&	113$\pm$6\\
81&	 35.802952&	75.313507 &	3.8$\pm$0.7&	42$\pm$5\\
82&	 36.107559&	75.358276 &	1.7$\pm$0.3&	100$\pm$5\\
83&	 35.847786&	75.315231 &	2.8$\pm$0.6&	104$\pm$6\\
84&	 36.175217&	75.364716 &	0.9$\pm$0.2&	122$\pm$6\\
85&	 36.172031&	75.356987 &	0.8$\pm$0.2&	87$\pm$7\\
86&	 35.977787&	75.291695 &	1.1$\pm$0.2&	120$\pm$5\\
87&	 36.036797&	75.296349 &	1.1$\pm$0.2&	118$\pm$5\\
88&	 35.719463&	75.561485 &	2.5$\pm$0.3&	110$\pm$4\\
89&	 35.877468&	75.583862 &	3.2$\pm$0.5&	120$\pm$4\\
90&	 35.790482&	75.547256 &	7.0$\pm$0.8&	95$\pm$3\\
91&	 35.805134&	75.541405 &	5.3$\pm$0.5&	113$\pm$3\\
92&	 36.025848&	75.573044 &	3.9$\pm$0.3&	127$\pm$2\\
93&	 36.011715&	75.570824 &	3.2$\pm$0.2&	111$\pm$2\\
94&	 35.886318&	75.550804 &	3.6$\pm$0.2&	125$\pm$1\\
95&	 35.829212&	75.532265 &	3.7$\pm$0.1&	111$\pm$1\\
96&	 36.057571&	75.566589 &	3.4$\pm$0.2&	122$\pm$2\\
97&	 35.937981&	75.543114 &	6.1$\pm$0.7&	113$\pm$3\\
98&	 36.150185&	75.572212 &	4.2$\pm$0.5&	121$\pm$3\\
99&	 35.846653&	75.511169 &	3.0$\pm$0.1&	122$\pm$1\\
100& 35.839558&	75.508217 &	2.8$\pm$0.4&	110$\pm$4\\
101& 36.047020&	75.528419 &	3.5$\pm$0.7&	134$\pm$5\\
102& 35.937199&	75.511421 &	3.2$\pm$0.3&	113$\pm$2\\
103&	 36.000168&	75.514542 &	3.3$\pm$0.7&	118$\pm$6\\
104&	 36.131355&	75.528450 &	4.7$\pm$1.2&	106$\pm$7\\
105&	 35.902927&	75.491394 &	1.7$\pm$0.3&	115$\pm$5\\
106&	 36.158512&	75.521553 &	6.6$\pm$0.7&	109$\pm$3\\
107&	 35.964561&	75.486557 &	1.7$\pm$0.3&	117$\pm$5\\
108&	 36.240715&	75.615578 &	3.6$\pm$1.1&	105$\pm$9\\
109&	 36.288883&	75.621521 &	5.4$\pm$0.4&	106$\pm$2\\
110&	 36.342354&	75.613274 &	3.6$\pm$0.2&	115$\pm$2\\
111&	 36.307533&	75.579750 &	7.6$\pm$1.4&	123$\pm$5\\
112&	 36.302315&	75.571709 &	3.6$\pm$1.2&	93$\pm$9\\
113&	 36.300140&	75.569557 &	3.2$\pm$0.2&	124$\pm$2\\
114&	 36.273445&	75.537132 &	5.3$\pm$1.2&	116$\pm$6\\
115&	 36.435184&	75.543159 &	3.3$\pm$0.8&	118$\pm$7\\
116&	 36.651173&	75.574608 &	4.5$\pm$0.5&	171$\pm$3\\
117&	 35.726604&	75.429031 &	1.9$\pm$0.5&	106$\pm$6\\
118&	 35.770763&	75.471237 &	1.4$\pm$0.3&	121$\pm$16\\
119&	 35.806629&	75.463112 &	2.8$\pm$0.4&	110$\pm$5\\
120&	 35.798588&	75.452263 &	7.5$\pm$1.1&	101$\pm$4\\
121&	 35.877754&	75.488167 &	2.4$\pm$0.4&	146$\pm$5\\
122&	 35.829231&	75.408340 &	1.8$\pm$0.6&	145$\pm$8\\
123&	 35.863377&	75.437630 &	2.1$\pm$0.6&	167$\pm$2\\
124&	 35.860695&	75.387650 &	4.0$\pm$0.4&	8$\pm$3\\
125&	 35.980408&	75.409248 &	1.5$\pm$0.3&	109$\pm$9\\
126&	 36.039444&	75.484955 &	4.0$\pm$1.3&	8$\pm$8\\
127&	 36.241055&	75.463737 &	4.2$\pm$0.7&	124$\pm$5\\
128&	 36.269241&	75.474220 &	5.0$\pm$0.7&	126$\pm$4\\
129&	 36.302826&	75.503593 &	3.4$\pm$0.4&	122$\pm$4\\
130&	 36.282207&	75.472626 &	2.6$\pm$0.5&	118$\pm$5\\
131&	 36.319000&	75.460953 &	4.0$\pm$0.6&	136$\pm$4\\
132&	 36.392452&	75.496414 &	7.8$\pm$0.5&	108$\pm$2\\
133&	 36.402100&	75.469627 &	3.4$\pm$0.3&	114$\pm$3\\
134&	 36.399605&	75.445602 &	2.5$\pm$0.4&	109$\pm$5\\
135&	 36.437878&	75.444946 &	2.7$\pm$0.3&  131$\pm$3\\\hline
\multicolumn{5}{c}{\bf L1521E}\\\hline
1&	 67.481590&	26.287052 &	4.6$\pm$0.4&	26$\pm$2\\
2&	 67.476410&	26.272148 &	5.0$\pm$0.9&	24$\pm$5\\
3&	 67.527367&	26.288368 &	4.2$\pm$0.3&	22$\pm$2\\
4&	 67.528519&	26.285912 &	4.3$\pm$0.3&	26$\pm$2\\
5&	 67.560616&	26.291574 &	3.0$\pm$0.2&	30$\pm$2\\
6&	 67.518944&	26.267017 &	3.2$\pm$0.2&	26$\pm$1\\
7&	 67.552635&	26.275476 &	3.6$\pm$0.4&	24$\pm$3\\
8&	 67.519592&	26.227743 &	2.9$\pm$0.5&	29$\pm$5\\
9&	 67.542175&	26.225372 &	4.0$\pm$0.1&	20$\pm$1\\
10&	 67.598625&	26.241295 &	3.0$\pm$0.2&	23$\pm$1\\
11&	 67.580704&	26.221781 &	3.4$\pm$0.4&	14$\pm$4\\
12&	 67.352890&	26.136698 &	1.8$\pm$0.4&	38$\pm$5\\
13&	 67.339035&	26.116142 &	1.8$\pm$0.4&	27$\pm$6\\
14&	 67.349022&	26.116617 &	2.7$\pm$0.7&	35$\pm$7\\
15&	 67.369499&	26.120598 &	2.1$\pm$0.1&	32$\pm$2\\
16&	 67.389580&	26.127825 &	3.1$\pm$0.5&	26$\pm$5\\
17&	 67.426331&	26.144903 &	2.6$\pm$0.0&	27$\pm$1\\
18&	 67.335449&	26.083839 &	2.2$\pm$0.6&	48$\pm$8\\
19&	 67.379005&	26.107500 &	3.6$\pm$0.6&	28$\pm$5\\
20&	 67.403671&	26.115255 &	2.1$\pm$0.2&	29$\pm$3\\
21&	 67.354576&	26.085054 &	2.6$\pm$0.8&	29$\pm$8\\
22&	 67.388283&	26.098785 &	2.3$\pm$0.6&	21$\pm$8\\
23&	 67.440826&	26.127062 &	3.1$\pm$0.9&	36$\pm$7\\
24&	 67.358475&	26.080219 &	2.1$\pm$0.1&	33$\pm$1\\
25&	 67.447426&	26.117859 &	4.1$\pm$0.4&	25$\pm$3\\
26&	 67.430061&	26.104013 &	2.5$\pm$0.2&	29$\pm$2\\
27&	 67.462196&	26.114037 &	3.7$\pm$0.4&	31$\pm$3\\
28&	 67.464890&	26.112408 &	2.8$\pm$0.1&	45$\pm$1\\
29&	 67.447563&	26.098097 &	2.3$\pm$0.2&	33$\pm$3\\
30&	 67.373581&	26.056162 &	2.8$\pm$0.7&	45$\pm$7\\
31&	 67.410774&	26.075914 &	1.6$\pm$0.5&	36$\pm$7\\
32&	 67.412727&	26.072020 &	3.1$\pm$0.7&	24$\pm$6\\
33&	 67.435532&	26.075167 &	3.1$\pm$0.5&	35$\pm$5\\
34&	 67.408218&	26.056572 &	2.9$\pm$0.2&	32$\pm$2\\
35&	 67.403999&	26.050295 &	3.0$\pm$0.4&	31$\pm$3\\
36&	 67.373482&	26.152620 &	1.9$\pm$0.1&	35$\pm$1\\
37&	 67.458282&	26.297251 &	3.3$\pm$0.6&	30$\pm$5\\
38&	 67.468719&	26.287107 &	3.8$\pm$0.2&	32$\pm$1\\
39&	 67.481529&	26.287107 &	3.9$\pm$0.2&	34$\pm$2\\
40&	 67.455322&	26.271738 &	3.3$\pm$0.2&	34$\pm$1\\
41&	 67.476395&	26.272152 &	3.7$\pm$0.3&	32$\pm$2\\
42&	 67.480774&	26.266171 &	3.0$\pm$0.5&	32$\pm$4\\
43&	 67.455139&	26.249735 &	3.9$\pm$0.2&	29$\pm$1\\
44&	 67.527328&	26.288523 &	4.2$\pm$0.6&	32$\pm$4\\
45&	 67.528503&	26.285934 &	4.1$\pm$0.3&	31$\pm$2\\
46&	 67.448624&	26.232401 &	3.0$\pm$0.7&	27$\pm$7\\
47&	 67.518898&	26.267067 &	3.3$\pm$0.1&	35$\pm$1\\
48&	 67.552650&	26.275578 &	2.8$\pm$0.3&	29$\pm$3\\
49&	 67.447762&	26.217007 &	3.8$\pm$1.0&	41$\pm$8\\
50&	 67.466057&	26.223522 &	3.9$\pm$0.4&	31$\pm$3\\
51&	 67.464447&	26.207251 &	3.6$\pm$0.6&	41$\pm$5\\
52&	 67.528458&	26.232956 &	3.8$\pm$0.5&	33$\pm$4\\
53&	 67.519531&	26.227768 &	3.3$\pm$0.4&	37$\pm$3\\
54&	 67.484222&	26.207764 &	3.9$\pm$0.5&	31$\pm$4\\
55&	 67.501228&	26.207205 &	3.1$\pm$0.5&	34$\pm$4\\
56&	 67.559822&	26.236757 &	2.7$\pm$0.6&	37$\pm$7\\
57&	 67.527458&	26.214884 &	2.9$\pm$0.8&	26$\pm$8\\
58&	 67.157364&	26.257944 &	2.3$\pm$0.5&	57$\pm$6\\
59&	 67.167305&	26.259041 &	3.5$\pm$0.7&	34$\pm$6\\
60&	 67.188629&	26.257345 &	2.3$\pm$0.4&	33$\pm$5\\
61&	 67.238670&	26.277580 &	2.1$\pm$0.1&	22$\pm$1\\
62&	 67.176338&	26.236416 &	2.0$\pm$0.1&	31$\pm$2\\
63&	 67.257874&	26.266144 &	6.0$\pm$1.3&	20$\pm$6\\
64&	 67.263107&	26.268164 &	2.7$\pm$0.2&	29$\pm$2\\
65&	 67.256348&	26.263693 &	2.8$\pm$0.3&	29$\pm$3\\
66&	 67.255142&	26.262440 &	2.7$\pm$0.3&	25$\pm$4\\
67&	 67.212830&	26.238308 &	2.1$\pm$0.4&	30$\pm$5\\
68&	 67.188499&	26.224091 &	2.3$\pm$0.5&	45$\pm$7\\
69&	 67.203880&	26.230350 &	2.4$\pm$0.5&	30$\pm$5\\
70&	 67.200729&	26.221565 &	2.3$\pm$0.6&	30$\pm$7\\
71&	 67.218590&	26.212732 &	6.1$\pm$1.2&	54$\pm$5\\
72&	 67.229462&	26.203793 &	2.8$\pm$0.7&	25$\pm$7\\
73&	 67.259682&	26.213427 &	4.1$\pm$1.3&	19$\pm$9\\
74&	 67.379425&	26.458372 &	3.5$\pm$0.6&	36$\pm$5\\
75&	 67.352814&	26.426369 &	3.0$\pm$0.5&	28$\pm$4\\
76&	 67.417671&	26.460176 &	2.9$\pm$0.9&	74$\pm$9\\
77&	 67.375343&	26.433187 &	3.2$\pm$0.7&	40$\pm$6\\
78&	 67.372414&	26.429434 &	3.0$\pm$0.7&	36$\pm$7\\
79&	 67.372414&	26.429434 &	3.0$\pm$0.7&	36$\pm$7\\
80&	 67.375771&	26.414431 &	3.3$\pm$0.4&	35$\pm$3\\
81&	 67.370667&	26.387484 &	3.0$\pm$0.8&	34$\pm$8\\
82&	 67.449181&	26.426167 &	3.5$\pm$0.5&	44$\pm$4\\
83&	 67.436790&	26.414555 &	3.3$\pm$0.4&	39$\pm$3\\
84&	 67.453255&	26.422728 &	2.9$\pm$0.4&	43$\pm$4\\
85&	 67.453255&	26.422728 &	2.9$\pm$0.4&	43$\pm$4\\
86&	 67.447006&	26.415834 &	3.7$\pm$0.8&	41$\pm$6\\
87&	 67.454948&	26.414957 &	6.0$\pm$1.5&	26$\pm$7\\
88&	 67.396027&	26.382881 &	4.0$\pm$0.2&	41$\pm$1\\
89&	 67.404182&	26.366709 &	2.5$\pm$0.5&	24$\pm$6\\
90&	 67.423317&	26.372372 &	4.5$\pm$0.7&	32$\pm$5\\
91&	 67.438324&	26.377171 &	3.1$\pm$0.4&	39$\pm$4\\
92&	 67.429680&	26.369631 &	4.0$\pm$0.8&	44$\pm$6\\
93&	 67.319107&	26.209984 &	2.0$\pm$0.2&	27$\pm$3\\
94&	 67.296135&	26.196365 &	2.7$\pm$0.1&	36$\pm$1\\
95&	 67.318672&	26.163237 &	1.7$\pm$0.3&	36$\pm$5\\
96&	 67.311264&	26.158535 &	2.0$\pm$0.7&	35$\pm$9\\
97&	 67.324203&	26.163578 &	2.3$\pm$0.1&	32$\pm$1\\
98&	 67.369453&	26.120644 &	2.2$\pm$0.2&	38$\pm$3\\ 
\\ \\
\hline
\multicolumn{5}{c}{\bf L1517}\\\hline
1&	 73.722855&	30.501905 &	3.3$\pm$0.9&	97$\pm$8\\
2&	 73.714233&	30.483135 &	4.0$\pm$0.6&	74$\pm$4\\
3&	 73.722137&	30.477777 &	5.8$\pm$0.8&	77$\pm$4\\
4&	 73.770859&	30.500072 &	1.3$\pm$0.4&	78$\pm$9\\
5&	 73.769554&	30.497047 &	3.5$\pm$0.2&	78$\pm$2\\
6&	 73.719040&	30.469999 &	1.8$\pm$0.3&	90$\pm$5\\
7&	 73.696213&	30.443743 &	6.3$\pm$0.5&	106$\pm$2\\
8&	 73.779846&	30.482695 &	2.3$\pm$0.2&	51$\pm$3\\
9&	 73.730705&	30.456091 &	1.4$\pm$0.3&	75$\pm$6\\
10&	 73.750389&	30.461250 &	1.8$\pm$0.3&	73$\pm$6\\
11&	 73.733681&	30.446335 &	7.2$\pm$2.2&	116$\pm$8\\
12&	 73.741402&	30.446253 &	1.9$\pm$0.4&	191$\pm$7\\
13&	 73.755302&	30.450823 &	1.0$\pm$0.2&	84$\pm$7\\
14&	 73.740608&	30.436319 &	2.3$\pm$0.2&	197$\pm$3\\
15&	 73.781036&	30.456249 &	1.4$\pm$0.5&	82$\pm$10\\
16&	 73.761192&	30.445122 &	4.2$\pm$0.4&	94$\pm$3\\
17&	 73.736732&	30.426329 &	1.1$\pm$0.1&	159$\pm$2\\
18&	 73.733017&	30.424347 &	3.1$\pm$1.0&	186$\pm$9\\
19&	 73.814430&	30.450090 &	4.8$\pm$0.7&	34$\pm$4\\
20&	 73.779900&	30.482515 &	3.2$\pm$0.4&	91$\pm$4\\
21&	 73.821487&	30.502167 &	2.0$\pm$0.3&	47$\pm$5\\
22&	 73.825714&	30.500570 &	1.8$\pm$0.2&	54$\pm$4\\
23&	 73.847931&	30.506588 &	1.5$\pm$0.4&	47$\pm$8\\
24&	 73.863907&	30.510509 &	1.7$\pm$0.4&	52$\pm$7\\
25&	 73.849792&	30.493942 &	1.5$\pm$0.4&	79$\pm$7\\
26&	 73.781036&	30.456135 &	2.1$\pm$0.5&	74$\pm$6\\
27&	 73.883133&	30.506731 &	2.2$\pm$0.5&	82$\pm$6\\
28&	 73.811386&	30.466928 &	5.7$\pm$1.1&	197$\pm$5\\
29&	 73.882698&	30.500095 &	2.6$\pm$0.1&	66$\pm$1\\
30&	 73.832230&	30.462818 &	1.6$\pm$0.3&	74$\pm$6\\
31&	 73.788635&	30.439535 &	1.6$\pm$0.4&	53$\pm$6\\
32&	 73.904160&	30.487429 &	3.9$\pm$0.9&	86$\pm$7\\
33&	 73.847229&	30.445190 &	2.8$\pm$0.6&	78$\pm$6\\
34&	 73.797928&	30.416260 &	4.8$\pm$0.7&	46$\pm$4\\
35&	 73.802299&	30.418278 &	1.6$\pm$0.3&	75$\pm$6\\
36&	 73.916710&	30.468847 &	1.1$\pm$0.1&	96$\pm$3\\
37&	 73.823593&	30.422194 &	2.0$\pm$0.7&	91$\pm$9\\
38&	 73.908470&	30.463478 &	0.6$\pm$0.2&	88$\pm$7\\
39&	 73.825531&	30.417072 &	3.0$\pm$0.9&	85$\pm$9\\
40&	 73.889534&	30.445328 &	1.6$\pm$0.4&	133$\pm$7\\
41&	 73.905228&	30.427259 &	2.7$\pm$0.1&	111$\pm$1\\
42&	 73.881660&	30.644161 &	2.5$\pm$0.4&	58$\pm$4\\
43&	 73.858200&	30.594833 &	2.7$\pm$0.7&	84$\pm$7\\
44&	 73.866859&	30.585754 &	2.7$\pm$0.3&	79$\pm$3\\
45&	 73.897896&	30.596272 &	2.1$\pm$0.5&	80$\pm$7\\
46&	 73.903664&	30.592806 &	1.6$\pm$0.3&	73$\pm$5\\
47&	 73.909607&	30.590330 &	2.2$\pm$0.3&	96$\pm$4\\
48&	 73.913071&	30.583788 &	2.2$\pm$0.4&	91$\pm$5\\
49&	 73.865555&	30.558216 &	3.0$\pm$0.3&	63$\pm$3\\
50&	 73.865150&	30.556414 &	2.2$\pm$0.1&	73$\pm$2\\
51&	 73.885841&	30.644678 &	1.1$\pm$0.2&	81$\pm$6\\
52&	 73.897606&	30.577805 &	1.8$\pm$0.1&	81$\pm$2\\
53&	 73.945488&	30.538900 &	5.3$\pm$0.7&	94$\pm$4\\
54&	 73.945183&	30.518826 &	2.8$\pm$0.7&	71$\pm$7\\
55&	 73.978325&	30.527084 &	1.5$\pm$0.3&	70$\pm$6\\
56&	 74.040062&	30.525234 &	2.1$\pm$0.6&	103$\pm$8\\
57&	 73.917236&	30.493732 &	1.6$\pm$0.1&	64$\pm$25\\
58&	 73.980095&	30.511055 &	3.7$\pm$0.9&	96$\pm$7\\
59&	 73.942177&	30.491476 &	7.4$\pm$0.4&	86$\pm$5\\
60&	 73.933052&	30.486034 &	5.1$\pm$0.5&	86$\pm$9\\
61&	 73.977020&	30.506422 &	6.2$\pm$1.3&	121$\pm$6\\
62&	 73.948669&	30.489664 &	1.7$\pm$0.3&	93$\pm$4\\
63&	 73.916710&	30.469006 &	0.9$\pm$0.2&	102$\pm$5\\
64&	 74.004707&	30.502562 &	2.8$\pm$0.9&	69$\pm$19\\
65&	 73.953369&	30.465736 &	1.7$\pm$0.4&	110$\pm$23\\
66&	 74.003746&	30.484390 &	2.4$\pm$0.6&	88$\pm$7\\
67&	 73.905197&	30.427328 &	2.9$\pm$0.3&	108$\pm$3\\
68&	 73.979866&	30.461498 &	1.6$\pm$0.5&	97$\pm$26\\
69&	 74.027351&	30.481510 &	1.3$\pm$0.2&	132$\pm$5\\
70&	 73.949005&	30.442228 &	0.8$\pm$0.1&	97$\pm$4\\
71&	 74.028297&	30.471123 &	1.3$\pm$0.1&	46$\pm$2\\
72&	 74.036407&	30.470858 &	2.5$\pm$0.4&	31$\pm$5\\
73&	 74.018234&	30.455214 &	1.6$\pm$0.5&	56$\pm$9\\
74&	 73.937569&	30.414524 &	1.4$\pm$0.1&	104$\pm$2\\
75&	 74.004890&	30.442953 &	1.3$\pm$0.2&	79$\pm$4\\
76&	 74.005760&	30.426935 &	2.7$\pm$0.1&	53$\pm$22\\
77&	 73.678719&	30.728910 &	1.1$\pm$0.2&	85$\pm$4\\
78&	 73.660820&	30.707541 &	2.6$\pm$0.6&	88$\pm$6\\
79&	 73.615730&	30.687328 &	2.0$\pm$0.6&	71$\pm$8\\
80&	 73.652657&	30.682257 &	2.7$\pm$0.3&	81$\pm$3\\
81&	 73.646904&	30.675013 &	2.7$\pm$0.7&	77$\pm$7\\
82&	 73.742035&	30.702744 &	3.4$\pm$0.3&	87$\pm$2\\
83&	 73.695595&	30.680260 &	4.3$\pm$0.9&	98$\pm$6\\
84&	 73.687630&	30.677610 &	2.6$\pm$0.7&	80$\pm$7\\
85&	 73.710358&	30.682419 &	4.2$\pm$1.3&	130$\pm$8\\
86&	 73.705215&	30.678539 &	2.3$\pm$0.2&	92$\pm$3\\
87&	 73.727158&	30.677591 &	2.4$\pm$0.4&	98$\pm$4\\
88&	 73.751411&	30.678389 &	3.1$\pm$0.3&	80$\pm$3\\
89&	 73.754189&	30.674540 &	2.8$\pm$0.3&	80$\pm$3\\
90&	 73.719017&	30.651396 &	2.1$\pm$0.3&	68$\pm$4\\
91&	 73.881645&	30.644178 &	2.4$\pm$0.4&	91$\pm$4\\
92&	 73.885849&	30.644735 &	1.4$\pm$0.1&	71$\pm$1\\
93&	 73.939240&	30.638893 &	2.8$\pm$0.6&	99$\pm$6\\
94&	 73.905014&	30.612480 &	2.0$\pm$0.5&	101$\pm$7\\
95&	 73.866867&	30.585754 &	2.2$\pm$0.3&	78$\pm$3\\
96&	 73.897934&	30.596298 &	1.0$\pm$0.3&	96$\pm$7\\
97&	 73.909630&	30.590324 &	2.2$\pm$0.6&	79$\pm$7\\
98&	 73.889381&	30.577280 &	2.4$\pm$0.4&	79$\pm$5\\
99&	 73.913101&	30.583809 &	2.4$\pm$0.2&	85$\pm$2\\\hline
\multicolumn{5}{c}{\bf L1512}\\\hline
1&	 76.051712&	32.684940 &	0.9$\pm$0.3&	201$\pm$8\\
2&	 76.050499&	32.649193 &	5.3$\pm$0.7&	159$\pm$4\\
3&	 76.066269&	32.658588 &	1.6$\pm$0.4&	159$\pm$7\\
4&	 76.069061&	32.668003 &	0.6$\pm$0.2&	140$\pm$7\\
5&	 76.080231&	32.692600 &	1.7$\pm$0.4&	152$\pm$7\\
6&	 76.084976&	32.700874 &	2.6$\pm$0.8&	171$\pm$9\\
7&	 76.096680&	32.668175 &	1.4$\pm$0.5&	144$\pm$9\\
8&	 76.110268&	32.678352 &	1.6$\pm$0.4&	150$\pm$7\\
9&	 76.134186&	32.700123 &	1.4$\pm$0.2&	169$\pm$4\\
10&	 76.138077&	32.680393 &	1.7$\pm$0.3&	150$\pm$4\\
11&	 76.156876&	32.663719 &	1.1$\pm$0.3&	114$\pm$8\\
12&	 76.179893&	32.702320 &	2.6$\pm$0.6&	157$\pm$6\\
13&	 76.181557&	32.702168 &	3.0$\pm$0.6&	167$\pm$6\\
14&	 76.193573&	32.668980 &	2.6$\pm$0.3&	164$\pm$3\\
15&	 76.077141&	32.879276 &	1.0$\pm$0.2&	157$\pm$5\\
16&	 76.078087&	32.876869 &	1.4$\pm$0.1&	165$\pm$2\\
17&	 76.086960&	32.875275 &	3.9$\pm$1.0&	179$\pm$7\\
18&	 76.107506&	32.910011 &	3.2$\pm$0.6&	170$\pm$5\\
19&	 76.110176&	32.912781 &	2.6$\pm$0.2&	162$\pm$2\\
20&	 76.101997&	32.863926 &	1.4$\pm$0.1&	189$\pm$2\\
21&	 76.117981&	32.911945 &	1.1$\pm$0.3&	178$\pm$7\\
22&	 76.126297&	32.907944 &	1.5$\pm$0.1&	165$\pm$2\\
23&	 76.153435&	32.867275 &	2.7$\pm$0.8&	158$\pm$9\\
24&	 76.171211&	32.901875 &	2.7$\pm$0.3&	171$\pm$3\\
25&	 76.179031&	32.895840 &	1.2$\pm$0.3&	176$\pm$7\\
26&	 76.171562&	32.843739 &	1.7$\pm$0.2&	197$\pm$4\\
27&	 76.182274&	32.872364 &	3.6$\pm$1.2&	156$\pm$9\\
28&	 76.199074&	32.869507 &	2.9$\pm$0.6&	181$\pm$6\\
29&	 75.954376&	32.651176 &	0.6$\pm$0.1&	160$\pm$6\\
30&	 75.977623&	32.659569 &	2.8$\pm$0.4&	108$\pm$4\\
31&	 75.966957&	32.614342 &	1.3$\pm$0.3&	141$\pm$5\\
32&	 75.981514&	32.604176 &	0.8$\pm$0.2&	178$\pm$6\\
33&	 75.989944&	32.606697 &	1.7$\pm$0.3&	148$\pm$6\\
34&	 75.999878&	32.639458 &	2.0$\pm$0.5&	126$\pm$8\\
35&	 76.022934&	32.665119 &	0.9$\pm$0.3&	160$\pm$8\\
36&	 76.023193&	32.640354 &	1.4$\pm$0.2&	144$\pm$4\\
37&	 76.059883&	32.638119 &	0.9$\pm$0.2&	145$\pm$6\\
38&	 76.067558&	32.642220 &	1.3$\pm$0.2&	169$\pm$3\\
39&	 76.076180&	32.606525 &	1.1$\pm$0.3&	189$\pm$7\\
40&	 75.821548&	32.718387 &	1.3$\pm$0.1&	138$\pm$1\\
41&	 75.864555&	32.755851 &	3.2$\pm$0.2&	142$\pm$2\\
42&	 75.880150&	32.739234 &	1.7$\pm$0.2&	149$\pm$4\\
43&	 75.874168&	32.695308 &	1.0$\pm$0.2&	123$\pm$4\\
44&	 75.901093&	32.746322 &	4.6$\pm$1.0&	141$\pm$6\\
45&	 75.886436&	32.673606 &	1.8$\pm$0.4&	123$\pm$7\\
46&	 75.915199&	32.762527 &	1.4$\pm$0.1&	153$\pm$2\\
47&	 75.955101&	32.713634 &	1.8$\pm$0.3&	120$\pm$4\\
48&	 75.808868&	32.820480 &	3.4$\pm$0.5&	169$\pm$4\\
49&	 75.843201&	32.868038 &	1.2$\pm$0.2&	156$\pm$4\\
50&	 75.838257&	32.812008 &	2.2$\pm$0.3&	169$\pm$4\\
51&	 75.848991&	32.847015 &	2.8$\pm$0.2&	159$\pm$2\\
52&	 75.852913&	32.813793 &	2.8$\pm$0.7&	140$\pm$7\\
53&	 75.861488&	32.846989 &	6.4$\pm$1.6&	160$\pm$7\\
54&	 75.877274&	32.864197 &	1.4$\pm$0.3&	153$\pm$6\\
55&	 75.869102&	32.825848 &	2.9$\pm$0.3&	158$\pm$3\\
56&	 75.880882&	32.863953 &	2.5$\pm$0.5&	159$\pm$6\\
57&	 75.890434&	32.870522 &	6.3$\pm$1.0&	186$\pm$5\\
58&	 75.897346&	32.882835 &	3.0$\pm$0.7&	144$\pm$6\\
59&	 75.900948&	32.874069 &	2.7$\pm$0.1&	151$\pm$1\\
60&	 75.904366&	32.846634 &	3.4$\pm$0.3&	160$\pm$3\\
61&	 75.912674&	32.835228 &	3.4$\pm$0.2&	154$\pm$1\\
62&	 75.920509&	32.860737 &	2.3$\pm$0.6&	131$\pm$7\\
63&	 75.929703&	32.813766 &	2.4$\pm$0.8&	148$\pm$9\\
64&	 75.929123&	32.806206 &	3.5$\pm$0.7&	188$\pm$6\\
65&	 75.950523&	32.854229 &	4.5$\pm$0.3&	164$\pm$2\\
66&	 75.932022&	32.878132 &	1.8$\pm$0.1&	155$\pm$2\\
67&	 75.943344&	32.918541 &	2.2$\pm$0.4&	172$\pm$5\\
68&	 75.933197&	32.877018 &	1.9$\pm$0.1&	149$\pm$2\\
69&	 75.959351&	32.932571 &	1.5$\pm$0.4&	160$\pm$7\\
70&	 75.956909&	32.909149 &	2.0$\pm$0.3&	172$\pm$5\\
71&	 75.953285&	32.889267 &	1.6$\pm$0.3&	165$\pm$4\\
72&	 75.973724&	32.904507 &	1.8$\pm$0.1&	149$\pm$1\\
73&	 75.992607&	32.935608 &	3.0$\pm$0.9&	115$\pm$8\\
74&	 75.998016&	32.899822 &	2.3$\pm$0.6&	162$\pm$7\\
75&	 76.021591&	32.930729 &	2.4$\pm$0.7&	169$\pm$8\\
76&	 76.034760&	32.928242 &	1.5$\pm$0.5&	145$\pm$9\\
77&	 76.056839&	32.925343 &	1.0$\pm$0.3&	160$\pm$7\\
78&	 76.124908&	32.772536 &	1.3$\pm$0.3&	135$\pm$15\\
79&	 76.142113&	32.788318 &	4.1$\pm$1.1&	207$\pm$9\\
80&	 76.134727&	32.752444 &	2.8$\pm$0.4&	198$\pm$4\\
81&	 76.147842&	32.793490 &	1.8$\pm$0.6&	101$\pm$10\\
82&	 76.162682&	32.800082 &	6.1$\pm$0.3&	98$\pm$1\\
83&	 76.160995&	32.789722 &	2.9$\pm$0.3&	137$\pm$3\\
84&	 76.160347&	32.760299 &	3.1$\pm$0.2&	116$\pm$4\\
85&	 76.162483&	32.754821 &	5.0$\pm$1.4&	122$\pm$4\\
86&	 76.178909&	32.777843 &	4.9$\pm$1.0&	126$\pm$11\\
87&	 76.183281&	32.788123 &	4.8$\pm$0.3&	137$\pm$3\\
88&	 76.194901&	32.811313 &	4.3$\pm$0.7&	141$\pm$7\\
89&	 76.186783&	32.746787 &	2.8$\pm$0.9&	178$\pm$5\\
90&	 76.194557&	32.762157 &	2.1$\pm$0.4&	155$\pm$7\\
91&	 76.207397&	32.794330 &	1.7$\pm$0.2&	152$\pm$3\\
92&	 76.199966&	32.748481 &	4.6$\pm$0.5&	91$\pm$3\\
93&	 76.221397&	32.807666 &	3.6$\pm$0.4&	204$\pm$2\\
94&	 76.241035&	32.813396 &	3.5$\pm$0.7&	73$\pm$11\\\hline

\multicolumn{5}{c}{\bf L1544}\\\hline
1&	 76.057228&	25.108257 &	1.2$\pm$0.3&	53$\pm$7\\
2&	 76.089081&	25.118418 &	2.1$\pm$0.6&	41$\pm$8\\
3&	 76.127922&	25.128275 &	1.9$\pm$0.2&	59$\pm$2\\
4&	 76.125549&	25.118435 &	1.8$\pm$0.2&	53$\pm$3\\
5&	 76.113373&	25.074015 &	1.0$\pm$0.3&	42$\pm$8\\
6&	 76.058960&	25.038523 &	0.8$\pm$0.2&	88$\pm$5\\
7&	 76.087753&	25.030443 &	1.7$\pm$0.4&	71$\pm$6\\
8&	 76.146156&	25.052807 &	0.6$\pm$0.2&	63$\pm$8\\
9&	 76.145523&	25.049646 &	1.8$\pm$0.5&	47$\pm$8\\
10&	 76.016464&	25.107189 &	1.9$\pm$0.4&	113$\pm$6\\
11&	 76.057167&	25.106194 &	1.8$\pm$0.4&	90$\pm$6\\
12&	 76.235374&	25.302361 &	2.8$\pm$0.6&	45$\pm$7\\
13&	 76.251297&	25.310106 &	2.3$\pm$0.4&	49$\pm$5\\
14&	 76.260139&	25.309694 &	2.8$\pm$0.3&	50$\pm$3\\
15&	 76.237396&	25.291946 &	3.4$\pm$0.1&	57$\pm$1\\
16&	 76.212265&	25.276058 &	4.4$\pm$1.4&	48$\pm$9\\
17&	 76.275375&	25.300722 &	2.6$\pm$0.4&	46$\pm$5\\
18&	 76.245178&	25.283043 &	2.5$\pm$0.5&	47$\pm$5\\
19&	 76.281631&	25.295921 &	2.6$\pm$0.3&	55$\pm$3\\
20&	 76.193909&	25.243095 &	5.6$\pm$1.1&	28$\pm$5\\
21&	 76.281998&	25.287325 &	4.0$\pm$1.1&	74$\pm$8\\
22&	 76.234612&	25.260841 &	3.1$\pm$0.8&	46$\pm$7\\
23&	 76.290482&	25.290985 &	2.2$\pm$0.4&	52$\pm$5\\
24&	 76.237526&	25.257521 &	2.8$\pm$0.6&	52$\pm$6\\
25&	 76.245140&	25.255503 &	3.3$\pm$0.6&	42$\pm$5\\
26&	 76.308716&	25.285637 &	2.3$\pm$0.1&	59$\pm$2\\
27&	 76.296082&	25.272753 &	4.1$\pm$0.6&	67$\pm$4\\
28&	 76.304832&	25.275841 &	3.3$\pm$0.3&	59$\pm$2\\
29&	 76.256004&	25.247278 &	3.4$\pm$0.3&	51$\pm$2\\
30&	 76.315498&	25.276484 &	2.1$\pm$0.5&	65$\pm$6\\
31&	 76.249496&	25.238979 &	4.3$\pm$0.2&	57$\pm$1\\
32&	 76.242897&	25.227686 &	5.4$\pm$0.6&	48$\pm$3\\
33&	 76.254791&	25.225065 &	2.5$\pm$0.5&	54$\pm$5\\
34&	 76.261375&	25.228199 &	2.5$\pm$0.6&	55$\pm$7\\
35&	 76.273216&	25.229153 &	1.8$\pm$0.6&	77$\pm$9\\
36&	 76.242256&	25.207714 &	3.5$\pm$0.1&	51$\pm$1\\
37&	 76.279312&	25.219023 &	3.2$\pm$0.2&	59$\pm$2\\
38&	 76.256058&	25.197462 &	2.3$\pm$0.6&	60$\pm$7\\
39&	 76.062073&	25.373821 &	1.4$\pm$0.4&	91$\pm$8\\
40&	 76.071907&	25.377256 &	2.4$\pm$0.6&	52$\pm$7\\
41&	 76.076889&	25.369682 &	2.1$\pm$0.3&	51$\pm$3\\
42&	 76.067070&	25.357315 &	4.7$\pm$0.8&	44$\pm$5\\
43&	 76.087212&	25.360737 &	1.5$\pm$0.3&	46$\pm$6\\
44&	 76.035110&	25.325567 &	4.5$\pm$1.0&	43$\pm$6\\
45&	 76.072121&	25.336781 &	2.7$\pm$0.5&	57$\pm$5\\
46&	 76.068024&	25.315220 &	4.0$\pm$0.8&	78$\pm$6\\
47&	 76.117973&	25.328358 &	3.6$\pm$0.5&	48$\pm$4\\
48&	 76.065918&	25.289679 &	4.9$\pm$1.2&	33$\pm$7\\
49&	 76.030487&	25.368849 &	6.0$\pm$0.9&	58$\pm$4\\
50&	 76.078751&	25.393486 &	1.4$\pm$0.4&	48$\pm$7\\
51&	 76.062080&	25.373770 &	1.7$\pm$0.2&	41$\pm$3\\
52&	 76.071846&	25.377312 &	2.2$\pm$0.2&	45$\pm$3\\
53&	 76.076851&	25.369694 &	2.2$\pm$0.1&	51$\pm$1\\
54&	 76.067055&	25.357246 &	2.3$\pm$0.3&	51$\pm$4\\
55&	 76.097504&	25.371729 &	2.9$\pm$0.4&	46$\pm$4\\
56&	 76.030022&	25.333433 &	3.0$\pm$0.9&	60$\pm$8\\
57&	 76.087173&	25.360748 &	2.5$\pm$0.1&	47$\pm$1\\
58&	 76.035072&	25.325525 &	3.6$\pm$0.4&	52$\pm$3\\
59&	 76.126968&	25.374454 &	4.4$\pm$0.7&	31$\pm$5\\
60&	 76.095612&	25.354708 &	2.8$\pm$0.6&	67$\pm$6\\
61&	 76.127930&	25.370354 &	2.3$\pm$0.6&	48$\pm$8\\
62&	 76.118904&	25.365253 &	3.7$\pm$0.8&	9$\pm$6\\
63&	 76.072105&	25.336784 &	2.9$\pm$0.2&	60$\pm$2\\
64&	 76.025940&	25.311419 &	5.3$\pm$1.3&	88$\pm$7\\
65&	 76.071793&	25.327324 &	3.3$\pm$0.9&	47$\pm$8\\
66&	 76.068031&	25.315233 &	2.6$\pm$0.3&	58$\pm$4\\
67&	 76.117889&	25.328384 &	3.4$\pm$0.2&	45$\pm$2\\
68&	 76.087990&	25.312168 &	2.9$\pm$0.6&	49$\pm$6\\
69&	 76.097466&	25.315281 &	2.4$\pm$0.5&	54$\pm$6\\
70&	 76.091049&	25.306902 &	2.9$\pm$0.5&	46$\pm$4\\
71&	 76.137878&	25.331186 &	3.7$\pm$0.2&	46$\pm$1\\
72&	 76.085701&	25.302130 &	3.2$\pm$0.9&	53$\pm$8\\
73&	 76.065910&	25.289625 &	2.4$\pm$0.6&	35$\pm$7\\
74&	 76.138542&	25.324404 &	3.1$\pm$0.5&	42$\pm$5\\
75&	 76.126183&	25.313448 &	3.5$\pm$0.7&	48$\pm$5\\
76&	 76.134743&	25.309156 &	4.4$\pm$1.2&	74$\pm$8\\
77&	 75.914574&	25.233492 &	9.4$\pm$1.0&	61$\pm$3\\
78&	 75.875015&	25.201035 &	6.2$\pm$1.2&	80$\pm$5\\
79&	 75.943802&	25.230867 &	6.1$\pm$0.8&	67$\pm$3\\
80&	 75.876350&	25.193361 &	3.7$\pm$1.0&	61$\pm$8\\
81&	 75.877235&	25.189051 &	3.1$\pm$0.5&	67$\pm$5\\
82&	 75.949387&	25.220924 &	5.4$\pm$0.4&	63$\pm$2\\
83&	 75.891914&	25.187262 &	2.8$\pm$0.5&	66$\pm$5\\
84&	 75.943237&	25.202789 &	5.4$\pm$0.5&	61$\pm$3\\
85&	 75.877106&	25.151016 &	2.8$\pm$0.4&	64$\pm$3\\
86&	 75.942673&	25.182020 &	4.3$\pm$0.6&	68$\pm$4\\
87&	 75.980614&	25.184624 &	3.7$\pm$0.4&	62$\pm$3\\
88&	 75.935455&	25.154423 &	3.7$\pm$0.7&	67$\pm$5\\
89&	 75.891846&	25.129160 &	3.6$\pm$0.8&	73$\pm$6\\
90&	 75.961937&	25.160721 &	3.5$\pm$1.0&	60$\pm$8\\
91&	 75.912445&	25.127752 &	3.0$\pm$0.8&	61$\pm$8\\
92&	 75.940964&	25.139061 &	2.9$\pm$0.3&	61$\pm$3\\
93&	 75.774246&	25.314596 &	2.8$\pm$0.6&	66$\pm$6\\
94&	 75.779404&	25.304075 &	2.8$\pm$0.3&	78$\pm$3\\
95&	 75.737396&	25.280252 &	2.1$\pm$0.4&	78$\pm$6\\
96&	 75.768265&	25.294556 &	2.4$\pm$0.6&	86$\pm$7\\
97&	 75.810715&	25.317068 &	3.1$\pm$0.6&	72$\pm$5\\
98&	 75.808716&	25.302570 &	3.3$\pm$0.2&	66$\pm$2\\
99&	 75.734344&	25.261499 &	1.8$\pm$0.2&	74$\pm$3\\
100&	 75.783310&	25.287922 &	2.1$\pm$0.5&	73$\pm$7\\
101&	 75.778854&	25.285486 &	3.1$\pm$0.7&	78$\pm$6\\
102&	 75.754028&	25.261330 &	2.4$\pm$0.2&	76$\pm$2\\
103&	 75.822212&	25.296404 &	2.8$\pm$0.6&	85$\pm$6\\
104&	 75.852760&	25.306393 &	2.9$\pm$0.3&	75$\pm$3\\
105&	 75.814072&	25.285645 &	3.3$\pm$0.2&	75$\pm$2\\
106&	 75.822998&	25.290022 &	3.4$\pm$0.2&	75$\pm$2\\
107&	 75.758423&	25.252281 &	3.5$\pm$0.5&	89$\pm$4\\
108&	 75.831055&	25.274052 &	3.1$\pm$1.0&	92$\pm$9\\
109&	 75.847015&	25.280731 &	3.2$\pm$0.6&	73$\pm$6\\
110&	 75.813591&	25.258505 &	1.7$\pm$0.5&	59$\pm$9\\
111&	 75.782776&	25.233969 &	4.3$\pm$0.8&	56$\pm$5\\
112&	 75.791924&	25.236813 &	2.8$\pm$0.3&	86$\pm$3\\
113&	 75.816910&	25.248405 &	0.6$\pm$0.2&	74$\pm$7\\
114&	 75.787399&	25.231300 &	2.3$\pm$0.4&	83$\pm$5\\
115&	 75.804199&	25.237904 &	2.7$\pm$0.3&	72$\pm$3\\
116&	 75.856033&	25.265120 &	3.8$\pm$0.7&	67$\pm$6\\
117&	 75.831566&	25.251572 &	2.4$\pm$0.4&	72$\pm$4\\
118&	 75.815590&	25.235275 &	1.9$\pm$0.2&	74$\pm$2\\
119&	 75.791939&	25.221834 &	2.9$\pm$0.2&	70$\pm$2\\
120&	 75.806465&	25.226585 &	4.1$\pm$0.9&	69$\pm$6\\
121&	 75.843140&	25.240055 &	3.1$\pm$0.2&	71$\pm$2\\
122&	 75.854790&	25.239071 &	2.6$\pm$0.3&	79$\pm$4\\
123&	 75.859070&	25.240129 &	3.2$\pm$0.8&	95$\pm$7\\
124&	 75.848190&	25.234303 &	3.3$\pm$0.1&	72$\pm$1\\
125&	 75.982834&	25.473400 &	3.4$\pm$0.8&	86$\pm$5\\
126&	 76.000893&	25.480186 &	2.6$\pm$0.4&	68$\pm$3\\
127&	 76.024300&	25.492476 &	7.4$\pm$0.9&	94$\pm$3\\
128&	 75.947830&	25.432013 &	1.1$\pm$0.2&	69$\pm$4\\
129&	 76.021156&	25.466976 &	2.5$\pm$0.4&	66$\pm$4\\
130&	 76.031769&	25.471401 &	2.0$\pm$0.4&	58$\pm$6\\
131&	 75.967308&	25.429609 &	1.0$\pm$0.3&	66$\pm$7\\
132&	 75.978424&	25.424036 &	2.2$\pm$0.7&	71$\pm$7\\
133&	 76.050774&	25.444368 &	2.7$\pm$0.5&	79$\pm$4\\
134&	 76.010574&	25.416071 &	1.3$\pm$0.3&	59$\pm$6\\
135&	 76.032639&	25.419542 &	1.3$\pm$0.3&	57$\pm$6\\
136&	 75.998466&	25.400034 &	1.5$\pm$0.4&	55$\pm$8\\
\hline
\multicolumn{5}{c}{\bf L1523}\\\hline
1&	 76.510254&	31.877089 &	2.1$\pm$0.2&	117$\pm$2\\
2&	 76.527199&	31.875944 &	2.2$\pm$0.1&	147$\pm$1\\
3&	 76.524780&	31.849361 &	7.8$\pm$2.0&	116$\pm$8\\
4&	 76.530388&	31.869553 &	2.2$\pm$0.4&	139$\pm$5\\
5&	 76.554489&	31.951887 &	5.7$\pm$1.6&	122$\pm$9\\
6&	 76.537842&	31.850721 &	2.5$\pm$0.3&	153$\pm$3\\
7&	 76.568336&	31.914982 &	2.7$\pm$0.5&	157$\pm$5\\
8&	 76.569550&	31.917919 &	2.3$\pm$0.2&	136$\pm$2\\
9&	 76.595695&	31.941668 &	2.4$\pm$0.4&	125$\pm$6\\
10&	 76.616318&	31.913567 &	2.3$\pm$0.7&	149$\pm$9\\
11&	 76.611382&	31.873421 &	3.3$\pm$0.4&	154$\pm$3\\
12&	 76.362289&	31.808458 &	1.5$\pm$0.4&	139$\pm$8\\
13&	 76.369514&	31.820160 &	2.3$\pm$0.6&	133$\pm$7\\
14&	 76.380829&	31.780537 &	1.7$\pm$0.1&	127$\pm$2\\
15&	 76.390350&	31.807999 &	2.1$\pm$0.5&	132$\pm$7\\
16&	 76.405693&	31.815182 &	3.0$\pm$0.5&	140$\pm$4\\
17&	 76.430855&	31.839384 &	3.8$\pm$0.9&	133$\pm$6\\
18&	 76.418915&	31.760109 &	1.3$\pm$0.2&	147$\pm$4\\
19&	 76.455315&	31.829580 &	2.5$\pm$0.3&	136$\pm$3\\
20&	 76.447815&	31.797075 &	2.1$\pm$0.6&	142$\pm$7\\
21&	 76.451042&	31.766788 &	2.1$\pm$0.5&	129$\pm$7\\
22&	 76.380859&	31.780506 &	1.7$\pm$0.1&	139$\pm$2\\
23&	 76.390282&	31.808069 &	2.3$\pm$0.3&	131$\pm$3\\
24&	 76.393456&	31.805899 &	2.3$\pm$0.5&	148$\pm$6\\
25&	 76.402557&	31.822380 &	3.8$\pm$0.9&	135$\pm$7\\
26&	 76.405670&	31.815266 &	2.7$\pm$0.6&	142$\pm$6\\
27&	 76.408051&	31.784704 &	2.7$\pm$0.3&	161$\pm$3\\
28&	 76.412407&	31.800669 &	2.7$\pm$0.8&	132$\pm$8\\
29&	 76.424736&	31.842855 &	1.8$\pm$0.4&	145$\pm$7\\
30&	 76.412186&	31.750330 &	1.9$\pm$0.2&	136$\pm$2\\
31&	 76.418869&	31.760103 &	1.6$\pm$0.2&	148$\pm$4\\
32&	 76.436714&	31.760065 &	2.2$\pm$0.6&	138$\pm$7\\
33&	 76.455429&	31.829636 &	2.5$\pm$0.2&	132$\pm$2\\
34&	 76.447945&	31.797152 &	2.7$\pm$0.4&	138$\pm$4\\
35&	 76.456650&	31.825886 &	2.8$\pm$0.7&	136$\pm$7\\
36&	 76.451050&	31.766825 &	2.6$\pm$0.3&	139$\pm$3\\
37&	 76.464554&	31.788027 &	2.6$\pm$0.8&	138$\pm$8\\
38&	 76.480766&	31.805689 &	2.1$\pm$0.6&	141$\pm$8\\
39&	 76.487595&	31.802568 &	2.4$\pm$0.8&	150$\pm$9\\
40&	 76.496735&	31.810266 &	6.2$\pm$0.8&	118$\pm$4\\
41&	 76.488060&	31.760403 &	3.1$\pm$0.9&	137$\pm$8\\
42&	 76.501633&	31.798565 &	3.0$\pm$0.3&	126$\pm$3\\
43&	 76.390701&	31.686728 &	3.8$\pm$0.8&	139$\pm$6\\
44&	 76.398857&	31.653662 &	5.8$\pm$1.5&	139$\pm$7\\
45&	 76.411369&	31.654001 &	4.1$\pm$0.8&	110$\pm$6\\
46&	 76.404976&	31.621721 &	2.8$\pm$0.8&	117$\pm$8\\
47&	 76.421494&	31.635687 &	2.9$\pm$0.2&	126$\pm$2\\
48&	 76.444016&	31.724058 &	7.1$\pm$1.1&	114$\pm$5\\
49&	 76.454140&	31.723234 &	1.6$\pm$0.2&	149$\pm$3\\
50&	 76.455963&	31.657572 &	3.7$\pm$0.6&	120$\pm$5\\
51&	 76.449776&	31.624134 &	2.6$\pm$0.8&	121$\pm$9\\
52&	 76.455399&	31.634275 &	2.4$\pm$0.5&	122$\pm$6\\
53&	 76.460320&	31.644791 &	3.7$\pm$0.7&	117$\pm$5\\
54&	 76.478333&	31.715322 &	2.7$\pm$0.4&	108$\pm$4\\
55&	 76.468285&	31.625584 &	2.8$\pm$0.6&	109$\pm$6\\
56&	 76.477325&	31.646517 &	2.5$\pm$0.2&	113$\pm$2\\
57&	 76.477043&	31.636938 &	3.0$\pm$0.7&	112$\pm$6\\
58&	 76.500839&	31.644615 &	4.1$\pm$0.3&	111$\pm$2\\
59&	 76.511322&	31.647291 &	4.6$\pm$1.4&	116$\pm$9\\
60&	 76.532593&	31.559700 &	2.3$\pm$0.1&	128$\pm$1\\
61&	 76.556007&	31.602110 &	0.9$\pm$0.1&	85$\pm$3\\
62&	 76.548889&	31.557184 &	1.8$\pm$0.4&	122$\pm$6\\
63&	 76.558014&	31.576170 &	1.8$\pm$0.4&	111$\pm$7\\
64&	 76.575455&	31.579693 &	0.7$\pm$0.1&	47$\pm$5\\
65&	 76.590271&	31.581350 &	2.6$\pm$0.6&	114$\pm$6\\
66&	 76.587685&	31.548935 &	3.5$\pm$0.9&	111$\pm$8\\
67&	 76.607468&	31.630785 &	3.5$\pm$0.7&	109$\pm$5\\
68&	 76.600143&	31.599745 &	2.7$\pm$0.7&	124$\pm$7\\
69&	 76.620087&	31.604908 &	2.5$\pm$0.4&	119$\pm$4\\
70&	 76.626434&	31.599918 &	3.6$\pm$0.6&	120$\pm$4\\
71&	 76.626961&	31.547365 &	1.6$\pm$0.3&	113$\pm$6\\
72&	 76.628983&	31.551390 &	0.8$\pm$0.2&	126$\pm$6\\
73&	 76.650894&	31.584427 &	5.2$\pm$0.6&	140$\pm$3\\
74&	 76.669754&	31.709686 &	1.9$\pm$0.5&	165$\pm$7\\
75&	 76.682365&	31.722557 &	2.9$\pm$0.6&	146$\pm$6\\
76&	 76.683945&	31.689857 &	1.9$\pm$0.4&	129$\pm$5\\
77&	 76.682846&	31.644882 &	1.4$\pm$0.3&	129$\pm$6\\
78&	 76.704788&	31.726807 &	1.9$\pm$0.2&	145$\pm$3\\
79&	 76.709641&	31.702108 &	5.6$\pm$0.7&	127$\pm$4\\
80&	 76.706017&	31.676428 &	1.1$\pm$0.1&	120$\pm$3\\
81&	 76.700264&	31.641378 &	2.8$\pm$0.3&	120$\pm$3\\
82&	 76.728493&	31.707870 &	1.5$\pm$0.2&	138$\pm$4\\
83&	 76.730232&	31.709875 &	1.0$\pm$0.3&	148$\pm$7\\
84&	 76.728996&	31.705322 &	2.8$\pm$0.7&	131$\pm$7\\
85&	 76.728569&	31.679085 &	1.1$\pm$0.2&	49$\pm$5\\
86&	 76.735046&	31.652258 &	2.6$\pm$0.4&	118$\pm$5\\
87&	 76.736504&	31.654655 &	2.4$\pm$0.7&	107$\pm$8\\
88&	 76.732307&	31.633623 &	3.1$\pm$0.8&	123$\pm$7\\
89&	 76.752197&	31.698269 &	2.5$\pm$0.3&	124$\pm$4\\
90&	 76.745583&	31.664951 &	2.9$\pm$0.8&	114$\pm$8\\
91&	 76.756363&	31.706753 &	2.3$\pm$0.3&	132$\pm$4\\
92&	 76.759010&	31.713301 &	1.5$\pm$0.3&	130$\pm$6\\
93&	 76.768112&	31.700790 &	2.1$\pm$0.5&	138$\pm$7\\
94&	 76.761971&	31.670368 &	2.2$\pm$0.6&	106$\pm$8\\
95&	 76.760277&	31.646980 &	1.2$\pm$0.2&	126$\pm$4\\
96&	 76.781097&	31.678493 &	0.5$\pm$0.1&	53$\pm$6\\
97&	 76.776329&	31.647587 &	1.5$\pm$0.4&	124$\pm$6\\
\hline
\end{supertabular}
    \tabletail{Polarization results of background stars towards L1544.}\label{tab:core_L1544}
\end{center}



\bsp	
\label{lastpage}
\end{document}